\documentclass{aa}  

\usepackage{xcolor}
\usepackage{graphicx}
\usepackage[varg]{txfonts}
\usepackage{isotope}
\usepackage{siunitx}
\usepackage{ulem}
\usepackage{bm}
\usepackage{natbib}
\usepackage{url}
\usepackage{xspace}
\usepackage{placeins}
\usepackage{capt-of}
\usepackage{booktabs}
\bibpunct{(}{)}{;}{a}{}{,}

\usepackage[colorlinks=true,citecolor=blue,linkcolor=magenta,urlcolor=blue]{hyperref}
\usepackage[switch,columnwise]{lineno}

\newcommand{\msun}{$\text{M}_\odot$}
\newcommand{\teff}{$T_{\text{eff}}$}
\newcommand{\teffsed}{$T_{\text{eff}}^{\text{SED}}$}
\newcommand{\ebmv}{$E$\,$(B$\,$-$\,$V)$}
\newcommand{\emono}{$E$\,$(44$\,$-$\,$55)$}

\newcommand{\teffs}{effective temperatures }
\newcommand{\Cen}{$\omega$ Cen}
\newcommand{\ngc}{NGC\,6752}
\newcommand{\loggt}{log~$g$ $-$ \teff}

\begin{document}

\title{SHOTGLAS II. MUSE spectroscopy of blue horizontal branch stars in the core of $\omega$ Centauri and NGC\,6752\thanks{ Based on observations collected at the European Organisation for Astronomical Research in the Southern Hemisphere, Chile (Program IDs 094.D-0142(B), 095.D-0629(A), 096.D-0175(A), 097.D-0295(A), 098.D-0148(A), 099.D-0019(A), 0100.D-0161(A), 0101.D-0268(A), 0102.D-0270(A), 0103.D-0204(A), 0104.D-0257(B), and 105.20CR.002)}$^,$\thanks{Tables B.1 to B.6 are only available at the CDS via anonymous ftp to cdsarc.u-strasbg.fr (XXXX) or via http://cdsarc.u-strasbg.fr/viz-bin/cat/J/A+A/XXX/zzz}}

\author{M. Latour\inst{1}, S. H\"ammerich\inst{2}, M. Dorsch\inst{3,2}, U. Heber\inst{2},  T.-O. Husser\inst{1}, S. Kamann\inst{4}, S. Dreizler\inst{1}, and J. Brinchmann\inst{5,6} }

\institute{Institut für Astrophysik und Geophysik, Georg-August-Universität Göttingen, Friedrich-Hund-Platz 1, 37077 Göttingen, Germany
\\ \email{marilyn.latour@uni-goettingen.de}
         \and  Dr. Karl Remeis-Observatory and Erlangen Centre for Astroparticle Physics, Friedrich-Alexander-Universit\"at Erlangen-N\"urnberg, Sternwartstr. 7, 96049 Bamberg, Germany
         \and 
         Institut f\"ur Physik und Astronomie, Universit\"at Potsdam, Haus 28, Karl-Liebknecht-Str. 24/25, 14476, Potsdam-Golm, Germany
         \and
         Astrophysics Research Institute, Liverpool John Moores University, 
         IC2 Liverpool Science Park, 146 Brownlow Hill, Liverpool, L3 5RF, United Kingdom
         \and
         Leiden Observatory, Leiden University, PO Box 9513, 2300 RA Leiden, The Netherlands
         \and
         Instituto de Astrof{\'\i}sica e Ci{\^e}ncias do Espaço, Universidade do Porto, CAUP, Rua das Estrelas, PT4150-762 Porto, Portugal
         }

\date{Received 05/04/2023 / Accepted 23/06/2023}

  \abstract
    {}
   {We want to study the population of blue horizontal branch (HB) stars  in the centers of globular clusters (GC) for the first time by exploiting the unique combination of MUSE spectroscopy and HST photometry. In this work, we characterize their properties in the GCs \Cen\ and NGC\,6752. 
         }
   { We use dedicated model atmospheres and synthetic spectra grids computed using a hybrid LTE/NLTE modeling approach to fit the MUSE spectra of HB stars hotter than 8\,000~K in both clusters. The spectral fits provide estimates of the effective temperature (\teff), surface gravity (log~$g$), and helium abundance of the stars. The model grids are further used to fit the HST magnitudes, meaning the spectral energy distributions (SED), of the stars. From the SED fits, we derive the average reddening, radius, luminosity, and mass of the stars in our sample.
         }
   { The atmospheric and stellar properties that we derive for the stars in our sample are in good agreement with the theoretical expectations. In particular, the stars cooler than $\sim$15\,000~K follow neatly the theoretical predictions for the radius, log~$g$, and luminosity for helium-normal ($Y$=0.25) models. In \Cen, we show that the majority of these cooler HB stars cannot originate from a helium-enriched population with $Y>$0.35. 
   The properties of the hotter stars (radii and luminosities) are still in reasonable agreement with theoretical expectations, but the individual measurements have a large scatter. For these hot stars, we have a mismatch between the effective temperatures indicated by the MUSE spectral fits and the photometric fits, with the latter returning \teff\ lower by $\sim$3\,000~K. We use three different diagnostics, namely the position of the G-jump and changes in metallicity and helium abundances to place the onset of diffusion in the stellar atmospheres at \teff\ between 11\,000~K and 11\,500~K. Our sample includes two stars known as photometric variables, we confirm one to be a bona fide extreme HB object but the other is a blue straggler star.
   Finally, unlike what has been reported in the literature, we do not find significant differences between the properties (e.g., log~$g$, radius, and luminosity) of the stars in both clusters.
         }
   {
        We showed that our analysis method combining MUSE spectra and HST photometry of HB stars in GC is a powerful tool to characterize their stellar properties. With the availability of MUSE and HST observations of additional GCs, we have a unique opportunity to combine homogeneous spectroscopic and photometric data to study and compare the properties of blue HB stars in different GCs.   }

\authorrunning{Latour et al.}
\titlerunning{SHOTGLAS II. blue horizontal branch stars in the core of $\omega$ Centauri and NGC\,6752}
\maketitle


\section{Introduction}\label{Introduction}

Globular clusters (GC) may be considered ideal laboratories to study stellar evolution. However, evidence is mounting now that they are not simple stellar populations. Multiple stellar populations have been discovered on the main sequence (MS) and/or red giant branch
(RGB), showing up as distinct sequences in the color-magnitude diagrams (CMD) of numerous Galactic GCs, most prominently in massive clusters such as $\omega$\,Cen and NGC\,2808.
The formation mechanisms for these multiple
populations, however, remain unclear (see \citealt{renzini2015,Bastian2018}).
Peculiarities have also been found when comparing the morphology of the horizontal branches (HB) of globular clusters in CMDs, for example, the “second
parameter” problem. It became obvious that the globular clusters' HB morphology is not determined by metallicity alone. Other parameters such as age, helium abundance, and many others have been suggested to explain the HB stars' distributions in the cluster CMDs \citep[see, e.g.][]{recio2006,moehler01,catelan09,Miocchi2007,dotter2010}.
Traditionally, the HB is divided into a red and a blue part separated by the RR Lyrae instability strip at $\sim$8\,000~K.
Discontinuities along the HBs are ubiquitous for clusters with extended blue HBs, such as the "Grundahl jump'' (G-jump)
 at $\sim$11\,500 K \citep{grundahl99}, the "Momany jump'' (M-jump) separating the bluest "extreme'' HB (EHB) at $\sim$20\,000 K from the blue HB (BHB\footnote{BHB stars are hotter than the RR Lyrae instability strip but cooler than the EHB.})
 \citep{momany02,newell73,newell76}. Finally, there is the gap between the EHB and "blue-hook'' stars that was first identified by \citet{2000ApJ...530..352D} in $\omega$ Cen and later found 
in the most massive clusters near $\sim$32\,000–36\,000~K \citep{2010ApJ...718.1332B,2010MmSAI..81..838M}.
Discrete main sequences and RGBs may be linked to the HB morphology and discontinuities \citep{2008ASPC..392....3Y}. 

HB stars burn helium in their cores, and those massive enough (M $\gtrsim$ 0.55~\msun) also sustain hydrogen-shell burning. They are the progeny of low-mass red giant branch stars \citep{hoyle55,faulkner66}.
Generally, the mass of the helium-burning core ($\sim 0.5~\text{M}_\odot$) is the same across the entire sequence.
The mass of the hydrogen envelope surrounding the core, however, is different, making the HB a sequence of hydrogen-envelope mass.
The envelope mass decreases with increasing effective temperature (\teff), therefore, the HB is also a temperature sequence.
Consequently, the atmospheric structure is fundamentally different along the HB \citep{dorman92,brown16}.
The structural and atmospheric changes along the HB manifest themselves  as the gaps and discontinuities mentioned above.

BHB stars have a wide spread of temperatures (from 8000~K to 20\,000~K) and are mainly found in two spectral classes.
The A-type BHB stars (A-BHB) cover temperatures between $8000\,\rm K$ and $11\,500\, \rm K$, meaning that they are cooler than the G-jump.
Helium abundances of A-BHBs are at the solar level \citep{adelman96,kinman00,behr03} and the metal abundances are consistent with what is observed within the respective cluster population.
This chemical homogeneity is maintained by atmospheric convection and multiple sub-surface convection zones driven by the ionization of hydrogen and helium \citep{caloi99,sweigart02,brown16}.
With increasing effective temperatures, these zones are pushed towards the surface and disappear at $\sim$11\,500~K, which also marks the transition to the B-type BHB stars (B-BHB).
Due to the lack of atmospheric convection zones, B-BHBs have radiative atmospheres, which also give rise to atomic diffusion \citep[radiative levitation versus gravitational settling, see e.g.][]{huibon2000,quievy2009,2011A&A...529A..60M}.
Due to radiative levitation, heavy metals (e.g. iron) are enriched and high metal abundances are found among B-BHBs \citep{behr03,pace2006}.
The helium abundance, however, steadily decreases with temperature, due to gravitational settling, reaching a minimum at $\sim$15\,000~K \citep{monibidin2012}. At even higher effective temperatures, the He abundance increases again but remains sub-solar.
At a temperature of about $20\,000\,\rm K$ the hydrogen-envelope mass has decreased to a level no longer supporting hydrogen-shell burning. 
The last convection zone, from \ion{He}{ii} encroaches upon to the surface near this temperature and ceases to exist in the hotter stars \citep{brown16}. \citet{brown17} found that the stars hotter than the M-jump in \Cen\ have lower Fe abundances than their colder counterparts. 
It is these changes, happening at about $18\,000-20\,000\,\rm K$, that are believed to be responsible for the M-jump in the CMD of GCs. These hotter stars on the blue side of the M-jump form the extreme (or extended) horizontal branch (EHB).
In the Galactic field, the EHB stars are also referred to as hot subdwarfs, with spectral types B and O (sdB and sdO, \citealt{2009ARA&A..47..211H,heber16}). 

While many hot subdwarf stars in the field are known to be close binaries with periods of hours to days \citep{maxted2001,copperwheat2011}, hardly any such binary has been found in globular clusters \citep{monibedin2008,monibidin2018} including NGC\,6752 \citep{monibidin2006}. A population of binary subdwarf explains the excess UV emission observed for elliptical galaxies \citep{han2007} and the UV colours of early type galaxies in the Virgo cluster \citep{lisker2008}. 
\citet{pelisoli2020} concluded that binary evolution is required to explain the origin of all types of hot subdwarfs amongst the field population. In this scenario the single subdwarfs result from mergers of helium white dwarfs \citep{han02,han2003}. The lack of binaries in globular clusters could result from a much larger merger fraction than in the field \citep{han08}. The mass distribution of hot subdwarfs resulting from mergers is predicted to be much wider than that of binary subdwarfs and to contain subdwarfs more massive than those in close binaries \citep{han2003}. Consequently, the mass distribution of globular cluster subdwarfs should be wide with masses from 0.3 up to 0.9 M$_\odot$. Hence, it is of great importance to determine, accurately enough, the masses of sufficiently large samples of EHB (subdwarf) and BHB stars in globular clusters. This can allow to test whether BHB and EHB might form differently (single vs. binary evolution).

\begin{figure*}
\resizebox{\hsize}{!}{
   \includegraphics{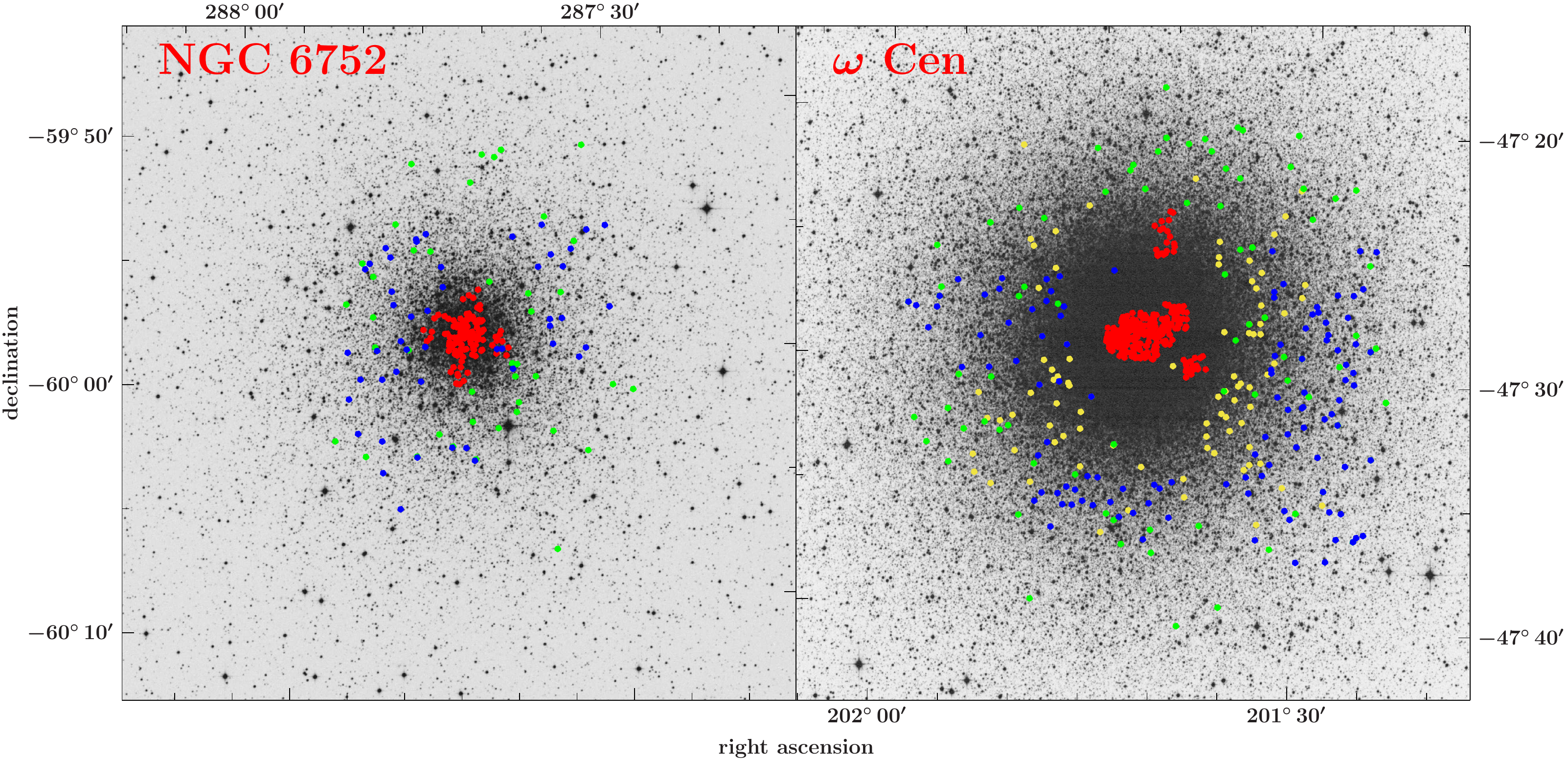}
   }
     \caption{Positions of the HB stars observed by MUSE in both GCs, left panel - \Cen, right panel - NGC\,6752, are shown in red. The positions of the stars analyzed in previous studies are also indicated. \cite{moehler2011} (green), \cite{monibidin2012} (blue), and \cite{latour2018} (yellow) are shown for \Cen\ and \cite{moehler1997} (green) and \cite{monibidin2007} (blue) are shown for \ngc. The images are from the Digitized Sky Survey\protect\footnotemark.}
     \label{pic:positions}
\end{figure*}

In this investigation, we analyze a spectral dataset of blue HB stars (bluer than the RR Lyrae gap) in two GCs: \Cen\ and \ngc. These two clusters have an extended and well-populated blue HB. Because the clusters are also nearby, their HB stars are relatively bright and they have been extensively studied in the past. Spectroscopic investigations include \citet{1986A&A...162..171H}, \citet{moehler1997,1999A&A...346L...1M,moehler00}, and \citet{monibidin2007} for \ngc, and \citet{moehler02, moehler2007,moehler2011}, \citet{monibidin2012}, and \citet{latour14,latour2018} for \Cen.
The EHB of \ngc\ hosts stars with \teff\ up to $\sim$30\,000~K that are depleted in helium, being the GC's counterparts to the field sdBs. In addition to the EHB, \Cen\ also harbors a blue hook population that extends at magnitudes fainter than the canonical EHB. The blue hook stars in \Cen\ are hotter than 30\,000~K and are also enriched in He and C compared to the EHB stars.

The previous ground-based spectroscopic investigations, such as the first paper of this series \citep{latour2018}, have targeted HB stars found in the outskirt of the clusters where crowding is not a severe issue. The previous investigations listed above used low to medium resolution (0.7$-$2.6 \AA) spectra obtained with various instruments (FLAMES, FORS, VIMOS) at the Very Large Telescope (VLT) to derive atmospheric parameters (\teff, log~$g$, He). In some cases, masses were also estimated, mostly using bolometric corrections as in \citet{moehler2011}.
In this work, we use spectra collected as part of the MUSE globular cluster survey (\citealt{kamann2018}, P.I.: S. Dreizler, S. Kamann) to gather a large and \textit{homogeneous} sample of HB stars located in the central regions of \Cen\ and \ngc\ (see Fig.~\ref{pic:positions}). 
Atmospheric parameters
are derived from the MUSE spectra using state-of-the-art model atmospheres.
The majority of the stars in our samples are located within the Hubble Space Telescope (HST) footprint and have magnitudes published as part of catalogs in several HST filters. We used our own model atmospheres to construct and fit the spectral energy distributions (SEDs) of the stars. Because the distances and reddening to the clusters are well constrained and the atmospheric parameters of the stars are known from the MUSE spectra, the SED fits allow us to derive stellar parameters, namely the radius, luminosity, and mass, to unprecedented precision. The parameters derived are then compared to evolutionary models.  

\footnotetext{copyright by Anglo-Australian Observatory/the Royal Observatory Edinburgh}

The paper is organized as follows. 
The MUSE observations and data processing are described in Sect.~\ref{sec:obs}. In Sect.~\ref{sec:method}, we present the model atmospheres and synthetic spectra used (Sect.~\ref{subsec:model_spec}) followed by
a detailed description of our analysis methods for the spectral fits (Sect.~\ref{subsec:spec_fit}-~\ref{subsec:metal}). 
The analysis of the SEDs based on the HST photometry is presented in Sect. \ref{subsec:sed_fit}. The final samples for both clusters are presented in Sect.~\ref{sec:sample}. 
Our resulting atmospheric parameters are discussed at length in Sect.~\ref{subsec:atm_par} where we present our samples in various parameter planes (e.g., \teff$-$log~$g$ and \teff$-$helium) and compare them with theoretical models and literature results. Section \ref{subsec:variable} presents our results for two variable stars (V16 and V17) and four hot blue straggler stars (BSSs) in \ngc. Our SED fit results, in terms of radius, luminosity, and mass are presented and discussed in Sect.~\ref{subsec:mrl}. Section~\ref{sec:comp_2clusters} presents a comparison between both clusters. Finally, we summarize our results and conclude
in Sects.~\ref{sec:summary} and~\ref{sec:conclusion}.

\section{MUSE Observations}
\label{sec:obs}

We used the Multi Unit Spectroscopic Explorer (MUSE, \citealt{bacon2010}) GTO observations of \Cen\ and NGC\,6752 obtained in wide field mode (1$\arcmin \times$ 1$\arcmin$) between April 2014 and May 2022. The observations from 2018 to 2022 benefited from the use of the adaptive optics system installed on UT4 of the VLT.
A total of ten and eight 1$\arcmin \times$ 1$\arcmin$ fields were observed in \Cen\ and NGC\,6752, respectively. These fields include the clusters' most central and crowded regions (see Fig~.\ref{pic:positions}), where the use of an integral field spectrograph is particularly efficient. A summary of the MUSE observations is presented in Table 
\ref{table_obs}. 

\begin{table}[h]
\scriptsize
\caption{MUSE observations of \Cen\ and NGC\,6752}
\label{table_obs}      
\centering                    
\begin{tabular}{c c c c c c}        
\toprule\toprule
\noalign{\vskip4bp}
Field &  RA & DEC & \multicolumn{2}{c}{\# Epochs} & Total exp. time \\
 & & & non-AO & AO & (s) \\
 \midrule
 \multicolumn{6}{l}{\Cen} \\
 1 & 13:26:45.0 & $-$47:29:09 & 8 & 7 & 2025 \\
 2 & 13:26:45.0 & $-$47:28:24 & 7 & 7 & 1890 \\
 3 & 13:26:49.5 & $-$47:29:09 & 7  & 10 & 2250 \\
 4 & 13:26:49.5 & $-$47:28:24 & 7  & 10  & 2295 \\
 5 & 13:26:40.6 & $-$47:28:31 & 7  & 7  & 3280 \\
 6 & 13:26:53.1 & $-$47:29:01 & 7  & 10  & 4080 \\
 7 & 13:26:36.8 & $-$47:27:54 & 6  &  9 & 4500 \\
 8 & 13:26:31.0 & $-$47:29:55 & 7  & 9  & 7200 \\
 11 & 13:26:40.3 & $-$47:25:00 &  6 & 8  & 12600 \\
 12 & 13:26:47.2 & $-$47:24:03 & 4  &  8 & 21600\\ 
 \midrule
\multicolumn{6}{l}{NGC\,6752} \\
 1 & 19:10:49.10 & $-$59:59:26.84 & 2 & 1 & 1080 \\
 2 & 19:10:49.12 & $-$59:58:41.84 & 2 & 1 & 1080 \\
 3 & 19:10:55.10 & $-$59:59:26.95 & 2  & 1 & 1080 \\
 4 & 19:10:55.12 & $-$59:58:41.95 & 2  & 1  & 1080 \\
 11 & 19:11:04.29 & $-$59:58:41.19 & 1 & 2 & 3000 \\
 12 & 19:10:39.13 & $-$59:59:29.03 & 0 & 9 & 17400 \\
 13 & 19:10:48.62 & $-$59:57:37.58 & 0  & 2 & 2000 \\
 14 & 19:10:55.35 & $-$60:00:38.19 & 0  & 2  & 2000 \\
\bottomrule
\end{tabular}
\end{table}

   \begin{figure*}
   \centering
   \includegraphics[width=0.3\textwidth]{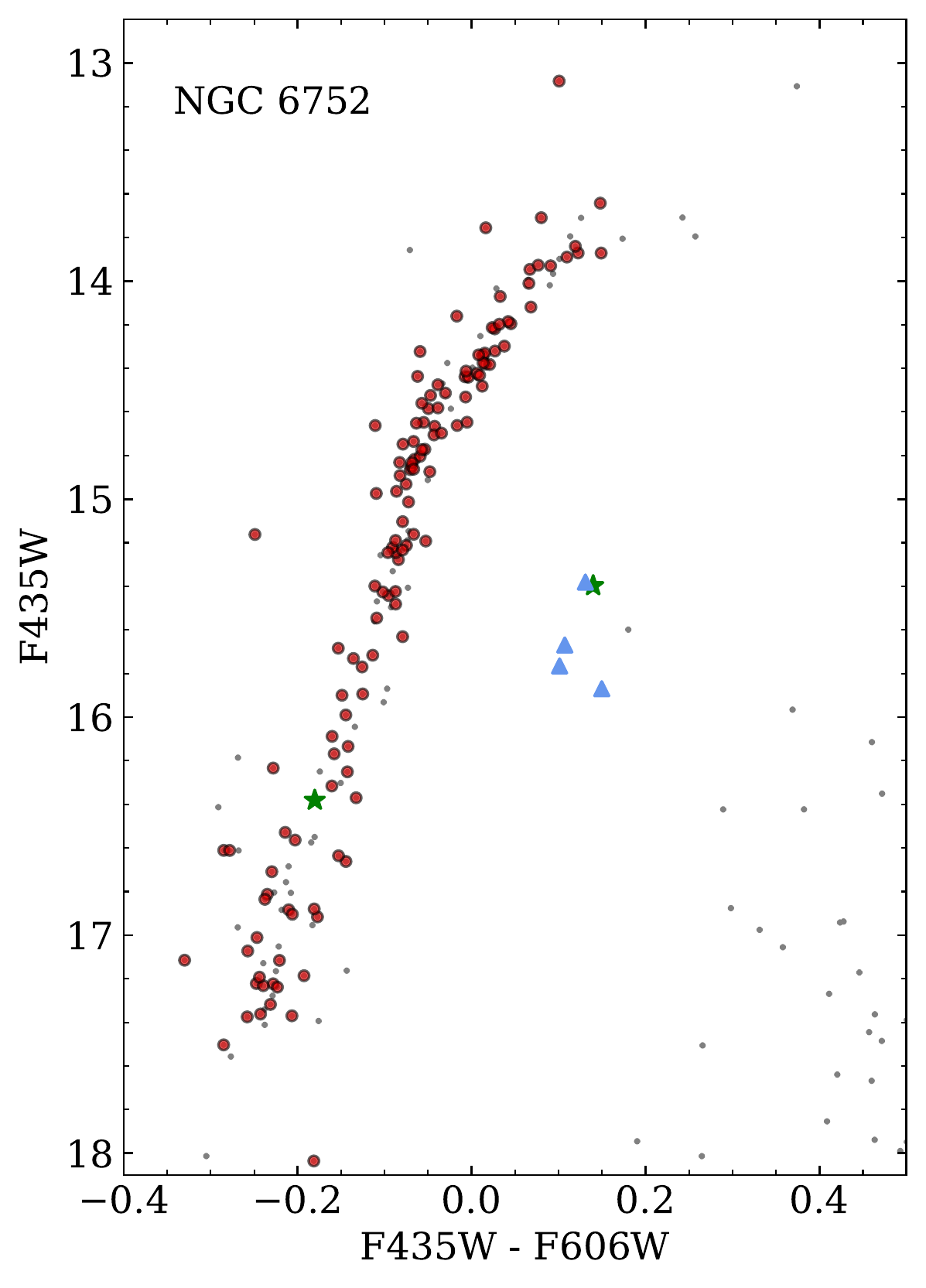}
\includegraphics[width=0.3\textwidth]{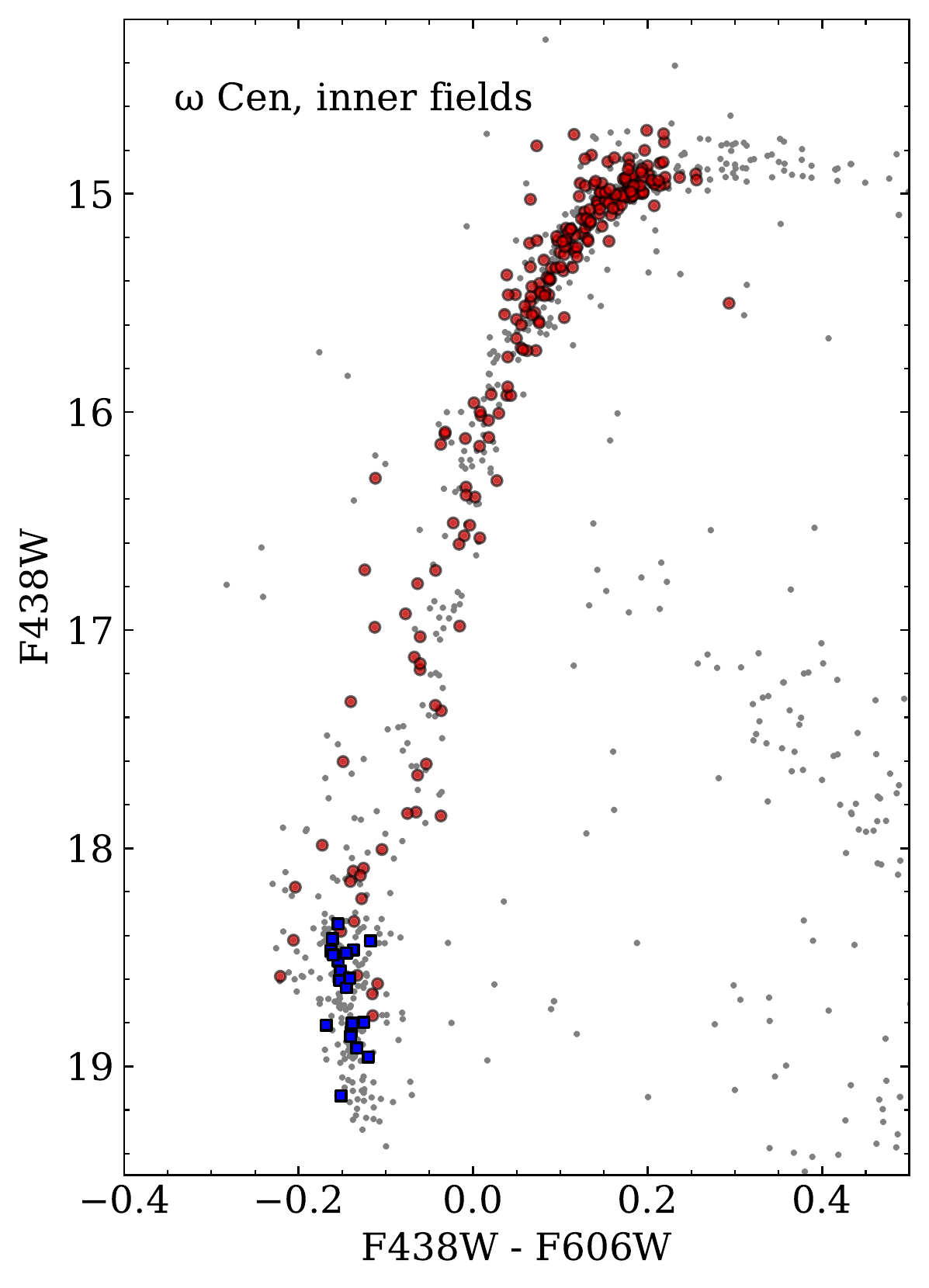}
\includegraphics[width=0.3\textwidth]{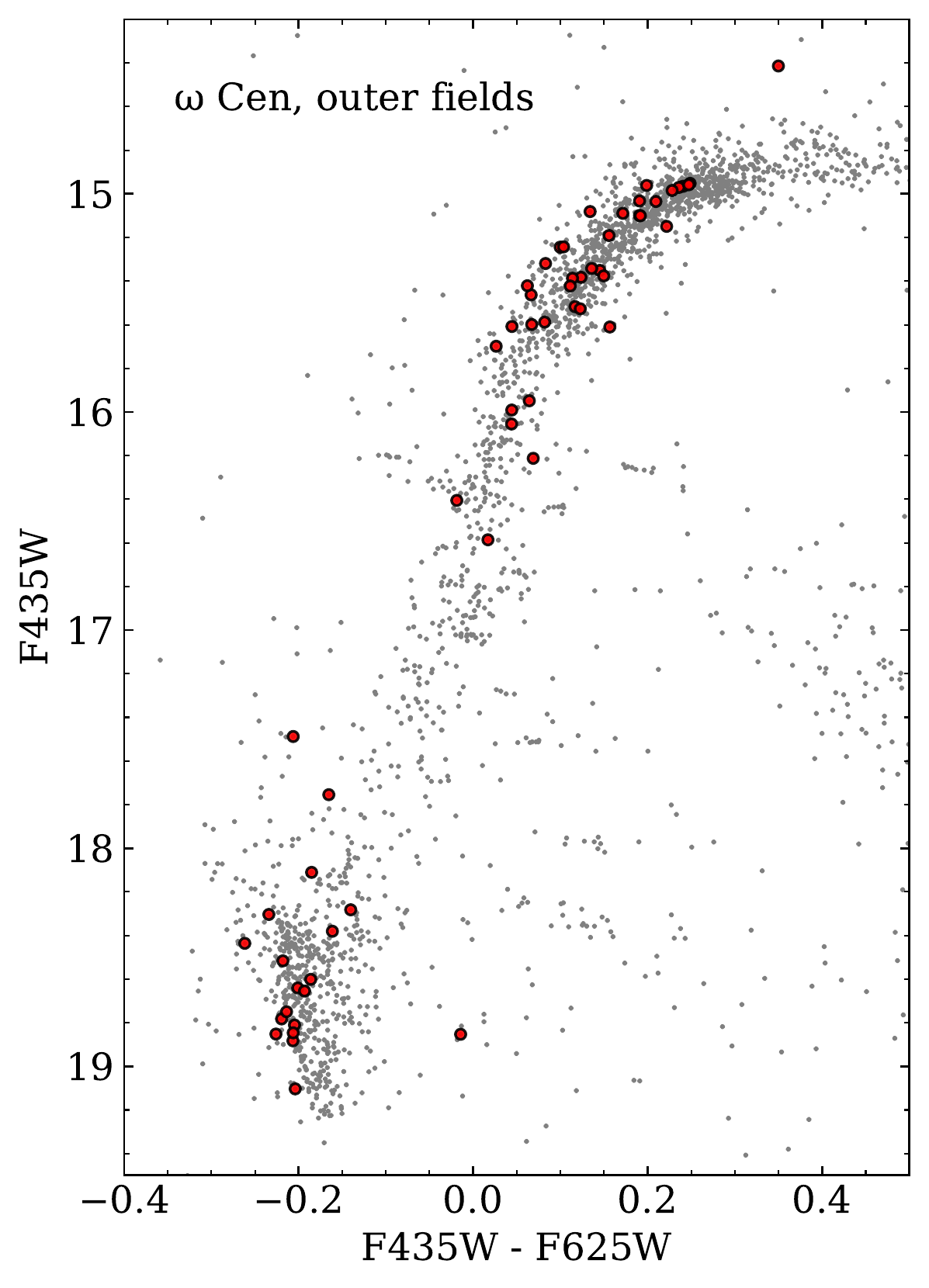}    
      \caption{Optical CMDs of NGC\,6752 and \Cen. The stars included in the spectroscopic samples are identified with large red dots. NGC\,6752: Also shown are the variable stars V16 and V17 (stars) and the four bright blue stragglers (triangles). The magnitudes are from the HUGS survey \citep{nardiello2018}. \Cen: The blue hook stars in \Cen\ are marked with blue squares. 
      The magnitudes for the central fields of view (1 to 6 in Table~\ref{table_obs}) are from the catalog of \citet{bellini2017a} and the magnitudes for stars in the outer fields (7, 8, 11, and 12) are from \citet{anderson10}. } 
         \label{pic:CMDs}
    \end{figure*}
   
\begin{figure*}
\sidecaption
   \includegraphics[width=12cm]{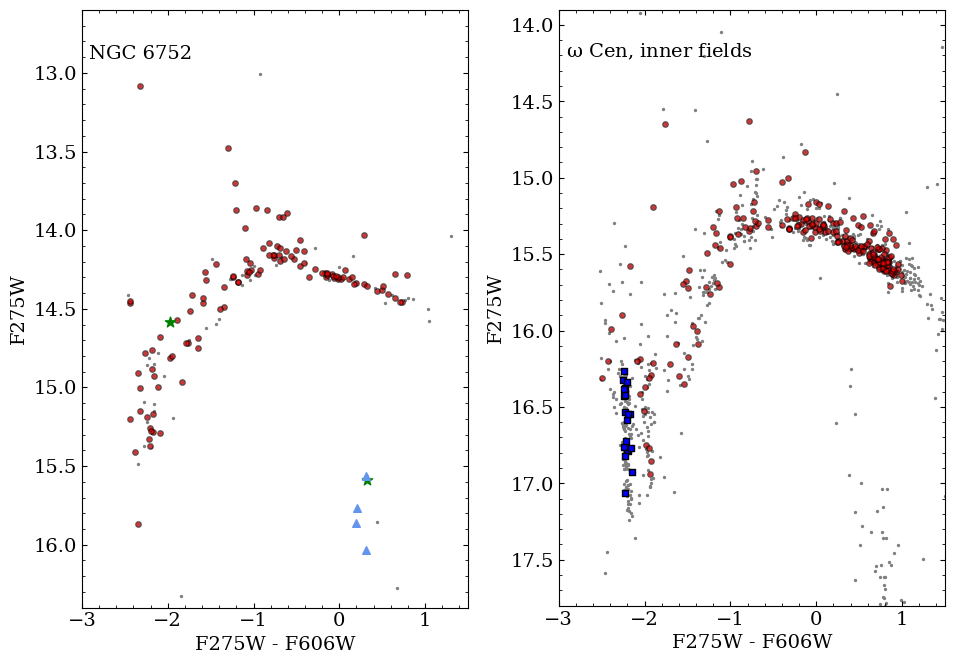}   
     \caption{ F275W$-$F606W CMD of \ngc\ and \Cen. Symbols are the same as in Fig.~\ref{pic:CMDs}.
     }
     \label{pic:app_CMD}
\end{figure*}

The spectra cover the 4750$-$9350 \AA\ range with an average spectral resolution of $\sim$2.5 \AA\ \textbf{($R \sim3000$)}, although this varies slightly across the wavelength range \citep{husser2016}. This range is redder than what is normally used to study HB stars; it only includes the two Balmer lines H$_\alpha$ and H$_\beta$. However, the Paschen lines from H$_{3-9}$ to the Paschen jump are covered. 
The data reduction is done with the standard MUSE pipeline \citep{weilbacher2020} and a general description of the different steps is presented in \citet{kamann2018}. 
 The stellar spectra are extracted with the \textsc{pampelmuse} software \citep{kamann2013,pampelmuse2018} that relies on the existence of a photometric catalog. We used HST catalogs to identify the sources present in the field of view and deblend the individual spectra \citep{sarajedini2007,anderson2008}.  
We use the photometry of \citet{anderson10} for the spectral extraction of stars in the external fields (6 to 11 in Table~\ref{table_obs}) of \Cen\ because these regions are not fully covered by the \citet{anderson2008} catalog.

Each field of view was observed at multiple epochs, thus the individual spectra were co-added to obtain the final, high signal-to-noise (S/N) spectrum. Each individual spectrum is fitted with synthetic spectra from the G\"{o}ttingen spectral library of PHOENIX models \citep{husser2013}. This grid only covers effective temperatures up to 15~kK, but the goal is to achieve a fit that is good enough to provide a radial velocity ($v_r$) and to reproduce the telluric lines. This allows the individual spectra to be shifted to restframe velocity, to have the telluric absorption removed, and to be co-added. More details on this procedure are presented in \citet{husser2016} with the main difference that for HB stars, the surface gravity is fitted along with \teff. 
Their work also presents the case of the sdO star ROB\,162 in NGC\,6397 \citep{1986A&A...169..244H} showing that the hydrogen lines are reproduced surprisingly well with a colder model, in this case that of an F-type star. This example showed that the $v_r$ and telluric lines can be corrected for in stars hotter than 15~kK, even if the best-fit solution is not realistic\footnote{In the future, we plan to further improve our data reduction method by using our HB synthetic spectral grid described in section \ref{subsec:model_spec} for individual spectra, to remove the telluric absorption before combination, instead of the Phoenix models.}. 

The final, co-added, telluric-free spectra are then fitted with proper model atmospheres as described in Sect.~\ref{sec:method}. In general, the $S/N$ decreases with increasing magnitudes but is also strongly dependent on the number of individual spectra that were collected. This depends on the position of the star in the cluster, that is in which field of view it is located, and whether it is found in an overlapping region between two fields.
We initially selected stars based on their positions in the CMD.
Our final samples of stars in both clusters, described in more details in Sect.~\ref{sec:sample}, are shown in their respective CMDs in Fig.~\ref{pic:CMDs} and \ref{pic:app_CMD}.

\section{Analysis method}\label{sec:method}
\subsection{Model atmospheres and synthetic spectra}\label{subsec:model_spec}

The model atmospheres and synthetic spectra used in this work were computed using the so-called ADS approach.
ADS is a hybrid LTE/NLTE method (local thermodynamic equilibrium and non-local thermodynamic equilibrium), which was first described by \citet{przybilla2006} and \citet{nieva2007}, and has since been improved by various authors \citep{przybilla2011,irrgang2014,irrgang2018}. This approach is consistent with results achieved by the means of full NLTE methods for hot stars (\teff\ $\lesssim 35$~kK, \citealt{przybilla2011}).
The calculation is done using the procedure described in \cite{irrgang2018} and \cite{kreuzer20}.
The final synthetic spectra are obtained by subsequently running three different codes.
At first, an LTE line-blanketed, plane-parallel, homogeneous, and hydrostatic model atmosphere is calculated using ATLAS12 \citep{Kurucz96}. The resulting LTE atmospheric structure is then used by DETAIL \citep{giddings81,butler85} to calculate the population numbers of hydrogen \citep{przybilla04} and helium \citep{przybilla05} assuming NLTE and using appropriate model atoms.
Other chemical elements are considered assuming a scaled solar abundance pattern \citep{asplund09} in ATLAS12 and DETAIL as background opacities.
The NLTE population numbers of H and He are then used in ATLAS12 to obtain a refined atmospheric structure \citep{2018A&A...620A..48I}. 
The process of passing the NLTE population numbers between the two codes is repeated until convergence is reached.
The final model atmosphere is then used by SURFACE \citep{giddings81,butler85} to compute a synthetic spectrum including lines of hydrogen and helium.
In this process, we use the occupation probability formalism \citep{hummer88} for H and He by \cite{hubeny94} and the line-broadening data of \cite{tremblay09} for hydrogen. For the most recent improvements of the code see \citet{2021A&A...650A.102I,2022NatAs...6.1414I}. 

We note that our models are calculated without microturbulence.
We computed five overlapping grids of model atmospheres and synthetic spectra
in order to cover the whole parameter range of the blue HB stars in terms of \teff, log\,$g$, and helium abundance.
The helium abundance is given as the logarithm of the fractional particle number with respect to all particles, which we denote as log $N$(He)/$N$(tot). 
The surface gravity is varied in steps of 0.2\,dex and the helium abundance in steps of 0.25\,dex.
We used solar-scaled chemical mixtures to produce model atmospheres at metallicities [$M/H$]\footnote{[$M/H$]=log($M/H$)-log($M/H$)$_\odot$} between $-2.0$ and $0.5$ in steps of 0.5\,dex.
The steps in \teff\ are not uniform across all grids.
The coverage of the individual grids is listed in Table~\ref{table_grids}.

\begin{table}[h]
\footnotesize
\caption{Properties of the model grids.}
\label{table_grids}      
\centering                    
\begin{tabular}{c c c c c}        
\toprule\toprule
\noalign{\vskip4bp}
Grid \# &  \teff\  & \teff\ step & log\,$g$  & log $N$(He)/$N$(tot)  \\
 & (K) & (K) & (cm s$^{-1}$) &  \\
 \midrule
 \midrule
 1 & 8\,000 - 12\,500 & 250 & 2.4 - 4.4 & $-$5.0 - $-$0.50 \\
 2 & 11\,000 - 17\,000 & 250 & 2.8 - 6.0 & $-$5.0 - $-$0.25 \\
 3 & 15\,000 - 26\,000 & 1\,000 & 3.0 - 6.4 & $-$5.0 - $-$0.25 \\
 4 & 22\,000 - 40\,000 & 1\,000 & 4.0 - 6.6 & $-$5.0 - $-$0.25 \\
 5 & 38\,000 - 55\,000 & 1\,000 & 4.6 - 6.6 & $-$5.0 - $-$0.25 \\
\bottomrule
\end{tabular}
\end{table}

\begin{figure*}
\centering
   \includegraphics[width=0.98\textwidth]{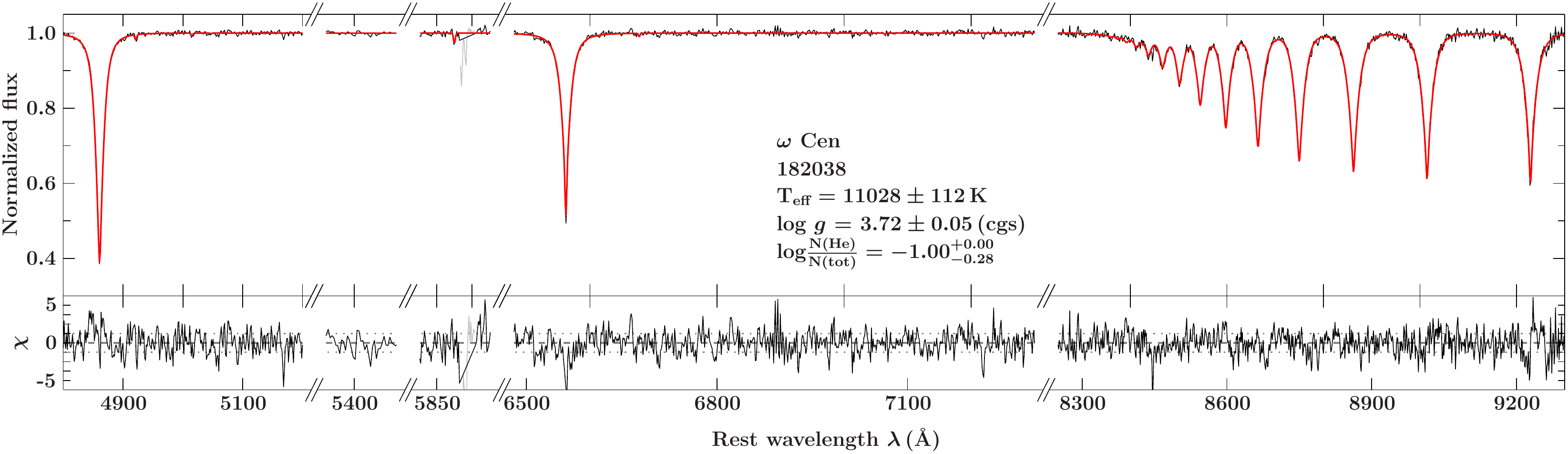}
   \includegraphics[width=0.98\textwidth]{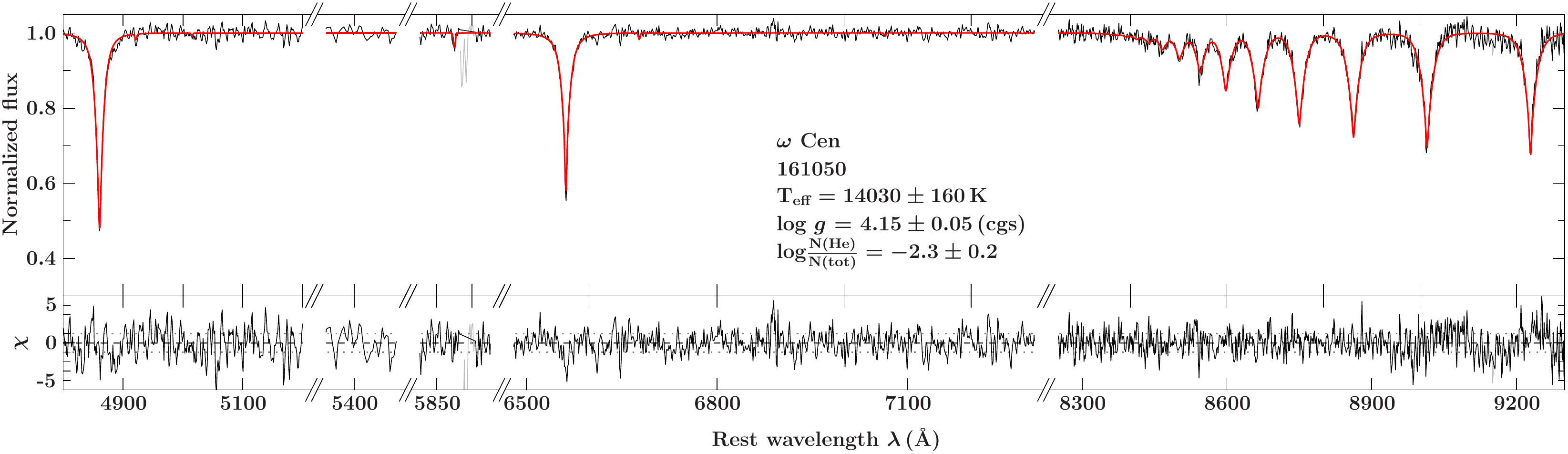}
\includegraphics[width=0.98\textwidth]{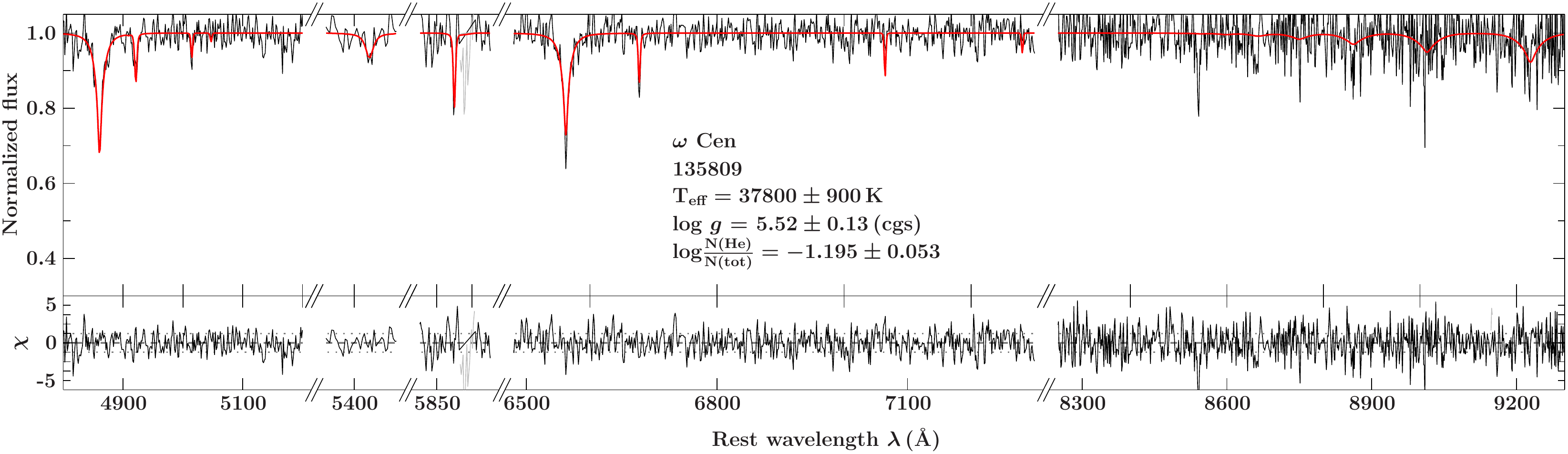}
   \includegraphics[width=0.98\textwidth]{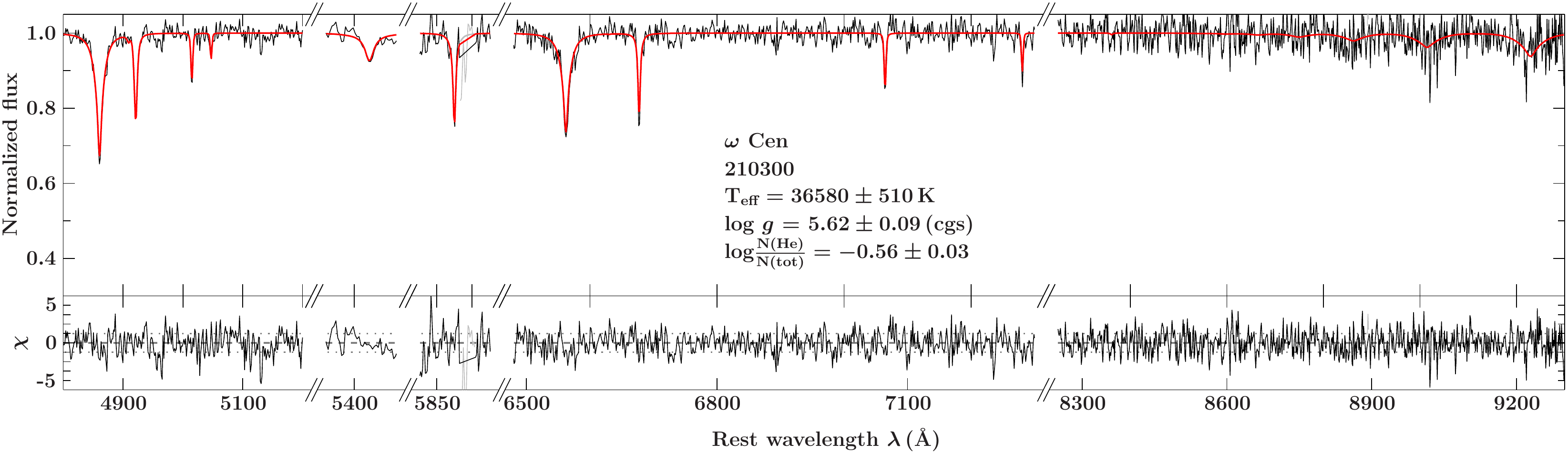}
     \caption{Examples of spectroscopic fits for stars in \Cen. Best fit (red) to the normalized spectrum (black) of, from top to bottom, an A-BHB, B-BHB, EHB, and blue hook star. 
     The residuals are shown below each fit. Only the regions with spectral lines of hydrogen and helium are plotted. 
     The cluster name, star identification number, and resulting spectral parameters are indicated for each fit.}
     \label{pic:spectral_fits}
\end{figure*}

  \begin{figure*}
  \centering
     \includegraphics[width=0.98\textwidth]{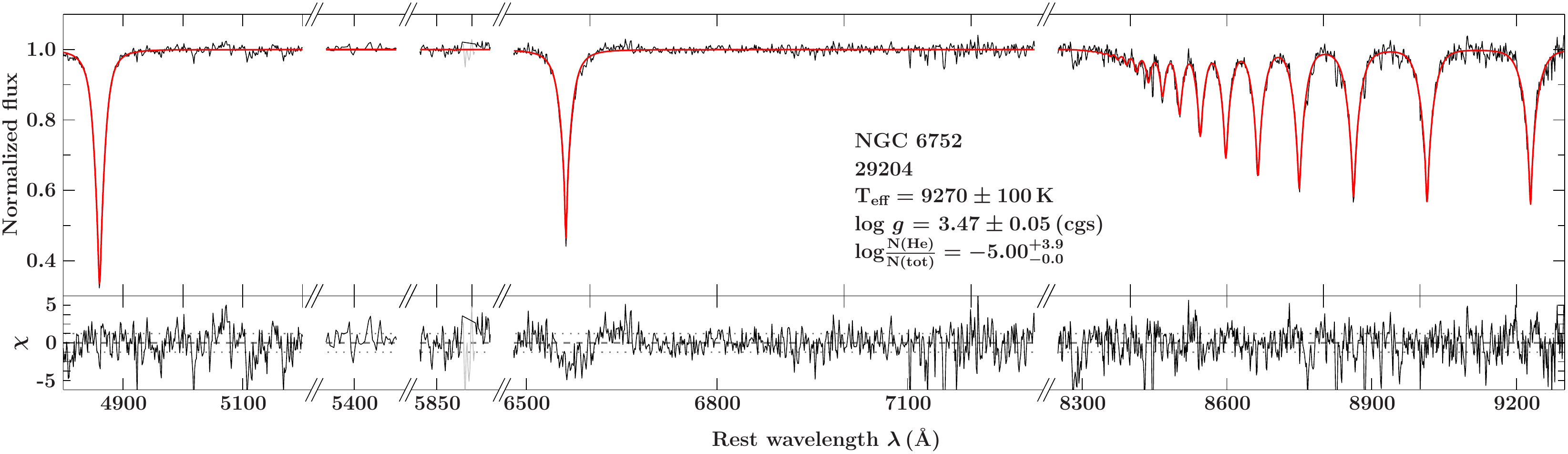}
     \includegraphics[width=0.98\textwidth]{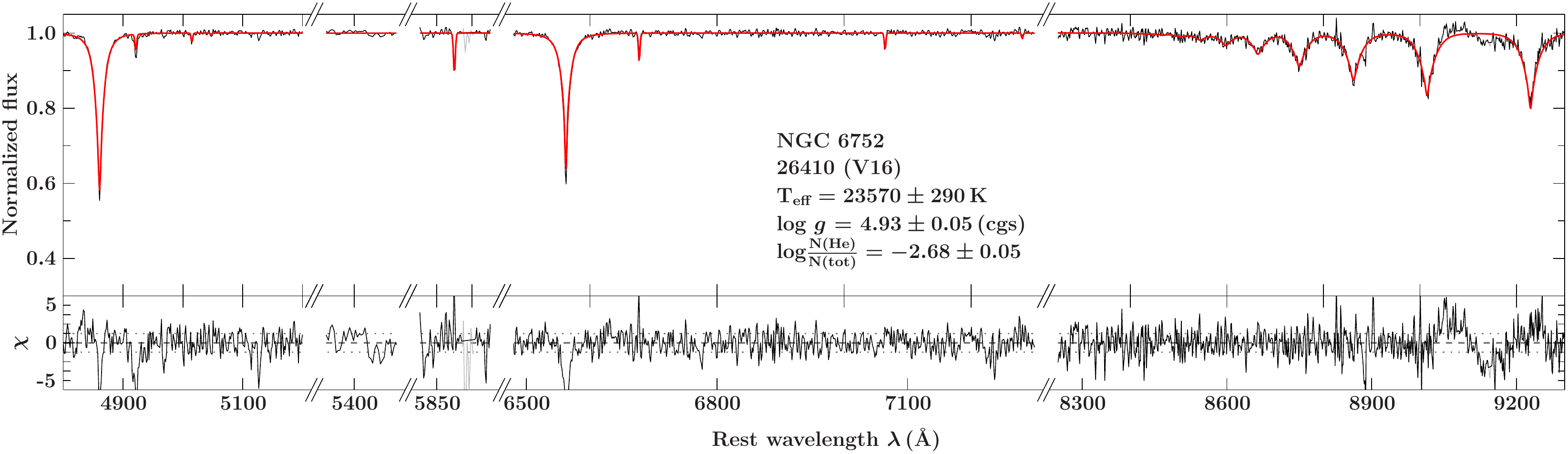}
  \includegraphics[width=0.98\textwidth]{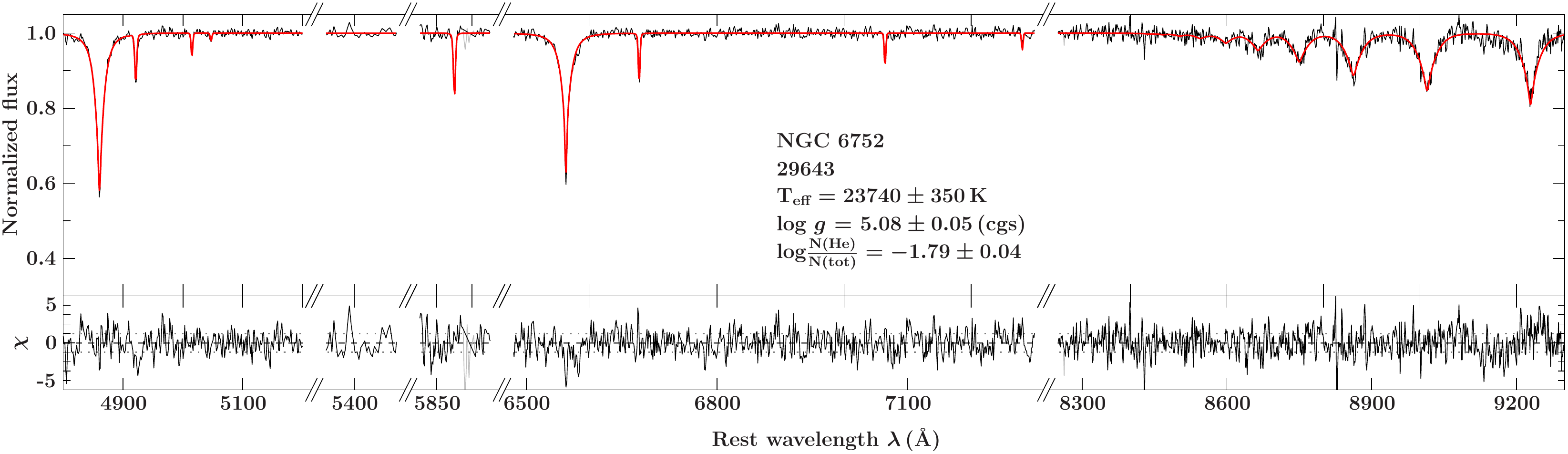}
     \includegraphics[width=0.98\textwidth]{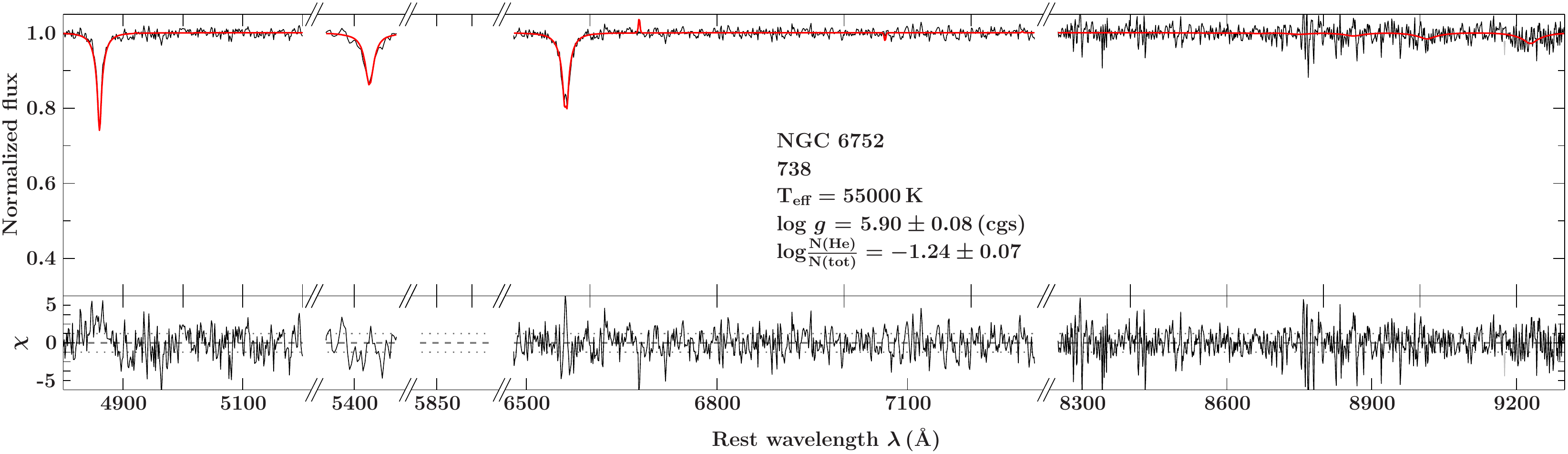}
      \caption{Same as Fig.~\ref{pic:spectral_fits} but for stars in \ngc. From top to bottom, an A-BHB, V16, an EHB with similar \teff\ as V16, the hot post-EHB (H-sdO) star. 
Star 738 was only observed with AO, thus the gap in the Na D lines region.}
       \label{pic:spectral_fits2}
  \end{figure*}

\subsection{Spectral fitting procedure}\label{subsec:spec_fit}
Spectral fitting is carried out using the \textit{Interactive Spectral Interpretation System} (\textit{ISIS,} \citealt{houck00}) with a modified, version of the $\chi^2$-minimization method presented by \cite{irrgang2014}.
To normalize the observed spectra, the continuum is modeled using a spline with anchor points placed every $100\,\rm\AA$ while avoiding the hydrogen and helium lines.
The spectral region containing the interstellar NaD lines ($5882.0 - 5901.0\,\rm\AA$) is excluded.
During the fitting process, the resolving power ($R$) of MUSE is considered to be a linear function, where the resolution increases with the observed wavelength.
The equation for $R$ was obtained from fitting the value of $R$ at different wavelengths from the fit of the hydrogen lines.
The macroturbulence and projected rotational velocity are set to $0\,\rm km\,s^{-1}$.
A low projected rotational velocity is fully consistent with what is expected for most of the HB stars \citep{geier12,stevenmsc}. For A-BHB stars, $v$sin$i$ are expected to be below 40 km\,s$^{-1}$ (see, e.g., \citealt{behr03}). Given the low spectral resolution of MUSE, and considering that we are fitting broad hydrogen lines, such a $v$sin$i$ has no measurable effect on the resulting atmospheric parameters.
The model spectra are convolved with a Gaussian following the relationship between $R$ and $\lambda$, and linear interpolation within the grid is used to determine the best fit atmospheric parameters.

Every fit is first carried out using the lowest temperature grid (\#1).
If the resulting \teff\ is within 4\% of the grid's upper limit, the fitting procedure is performed again using the following grid with a higher temperature.
Once the best solution within a grid is found, wavelength regions with metal lines and artifacts are identified using $3\sigma$ outliers in the $\chi^2$ and ignored as well (regions around the hydrogen and helium lines, however, are protected in this procedure), in order to assure an appropriate fit. Afterwards, the fit is repeated not taking the ignored regions into account.
Based on $\chi^2$ statistics, statistical errors are calculated by computing the confidence limits (68\%) as presented in \citet{irrgang2014}.
Systematic errors are considered as well, assuming an uncertainty of 1\% in \teff\ and $0.04$ dex for the surface gravity.
The final uncertainties are given as the quadratic sum of the systematic and statistical uncertainties.
The radial velocity $\mathrm{v}_\mathrm{rad}$ is left as a free parameter, in order to account for possible deviations from $0\,\rm km\,s^{-1}$, to which the spectra were shifted to in the combination process.

Example fits of various stars along the HB in both clusters are shown in Fig.~\ref{pic:spectral_fits} and \ref{pic:spectral_fits2}. 
Some things are important to keep in mind after the examination of these figures. The $S/N$ generally decreases with increasing temperature because the hotter stars are fainter. The number and strength of the hydrogen lines also decrease with increasing temperature because hydrogen is ionized. This means that with the MUSE spectra, we expect the atmospheric parameters to be less precise for the hot stars. As extreme examples, we show in Fig.~\ref{pic:spectral_fits} the fits of a low $S/N$ spectrum (135809, $S/N \approx 20$) and in Fig.~\ref{pic:spectral_fits2} the fit of the hottest star (738) in \ngc. In the latter case, the fit reached the border of the model grid at 55~kK. We have a few such hot stars in our samples, and we are aware that the \teff\ of these objects is a rough estimate, but we can nevertheless state that they are likely to have \teff\ larger than 50~kK and assign them an H-sdO spectral type (see \citealt{latour2018} for the different spectral types of EHB stars). The Paschen lines are prominent in the A-BHB and B-BHB stars and these lines provide a good constraint on the surface gravity. However, the strength and the number of Paschen lines diminish as hydrogen is ionized in the hot objects. In stars hotter than   $\sim$30~kK they often vanish in the noise. 

\subsection{Treatment of helium}\label{subsec:helium}

In the cooler stars (\teff~$\lesssim$ 11~kK) the helium lines are very weak and usually not visible in the MUSE spectra. Some studies have shown that helium abundances in these A-BHB stars are consistent with the solar value \citep{adelman96,kinman00,behr03,2009A&A...499..755V,2012ApJ...748...62V,2014MNRAS.437.1609M}. This is also in line with the fact that the stars cooler than the G-jump have convective atmospheres.
This is why, in previous studies, the helium abundance has been fixed to the solar value for A-BHB stars (see, e.g. \citealt{monibidin2007}). 
To assess the impact of fixing the helium abundance on the fits and resulting atmospheric parameters, we used sets of synthetic spectra 
with solar helium abundance and fitted them in the MUSE spectral range while keeping the helium abundance fixed to different values (from log $N$(He)/$N$(tot) = $-3.0$ to $-0.5$).
These tests revealed that the helium abundance assumed influences the resulting atmospheric parameters significantly, with differences up to $400\,\rm K$ in \teff\ and $0.4\,\rm dex$ in log\,$g$.
This indicates that the helium abundance, even when helium lines are weak or not visible, has an impact on the hydrogen lines that are used as temperature and surface gravity indicators. Therefore, we performed our fits with the helium abundance as a free parameter for the whole temperature range with the only constraint that the maximum helium value is set to the solar abundance in the cool stars (\teff~$<$~11.5~kK).
However, the fact that the helium lines are weak in these stars is reflected in the large uncertainties obtained for their He abundance.

\subsection{Treatment of metallicity}\label{subsec:metal}

\begin{figure*}
\resizebox{\hsize}{!}{
   \includegraphics[width=0.33\textwidth]{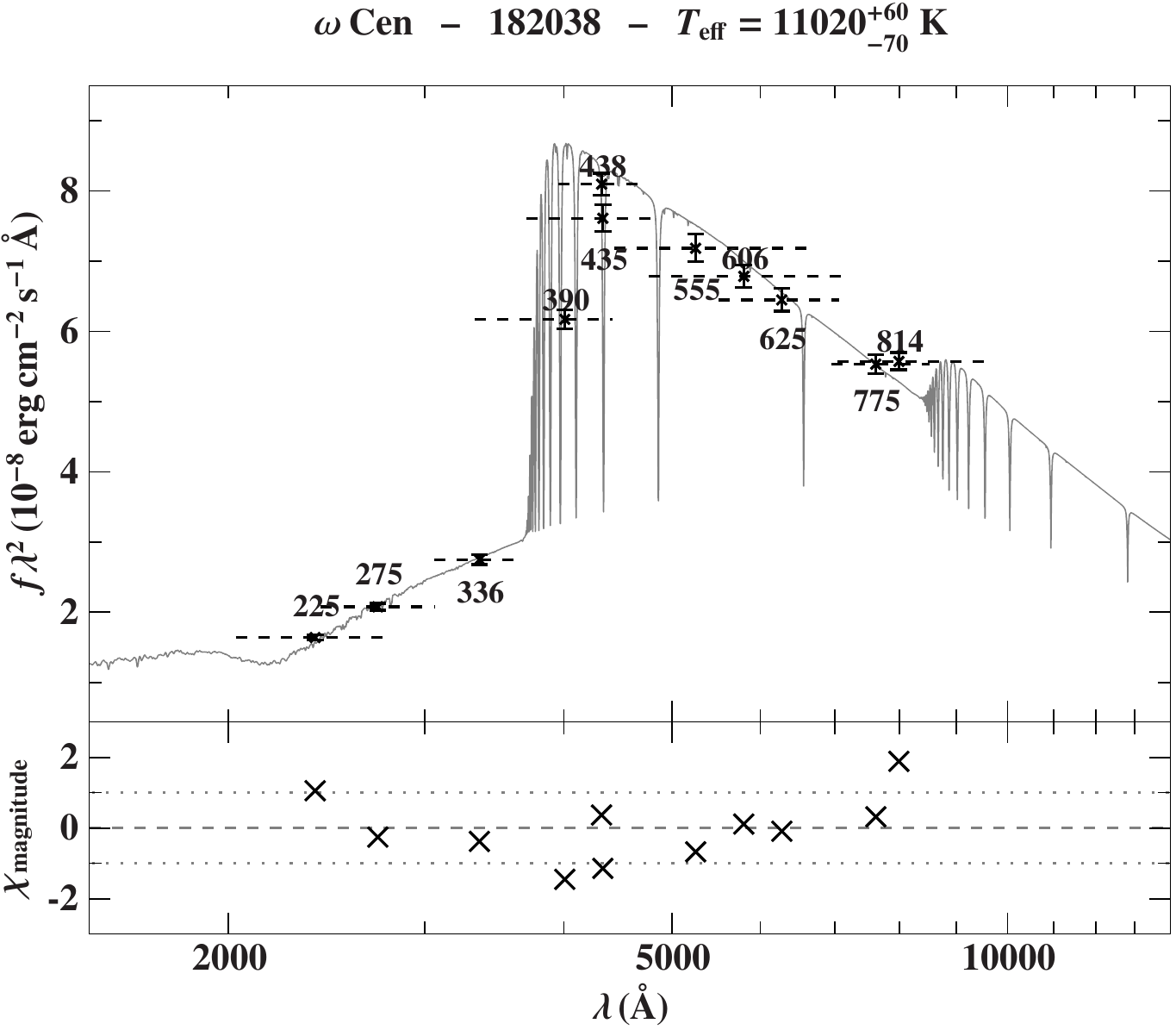}
   \includegraphics[width=0.33\textwidth]{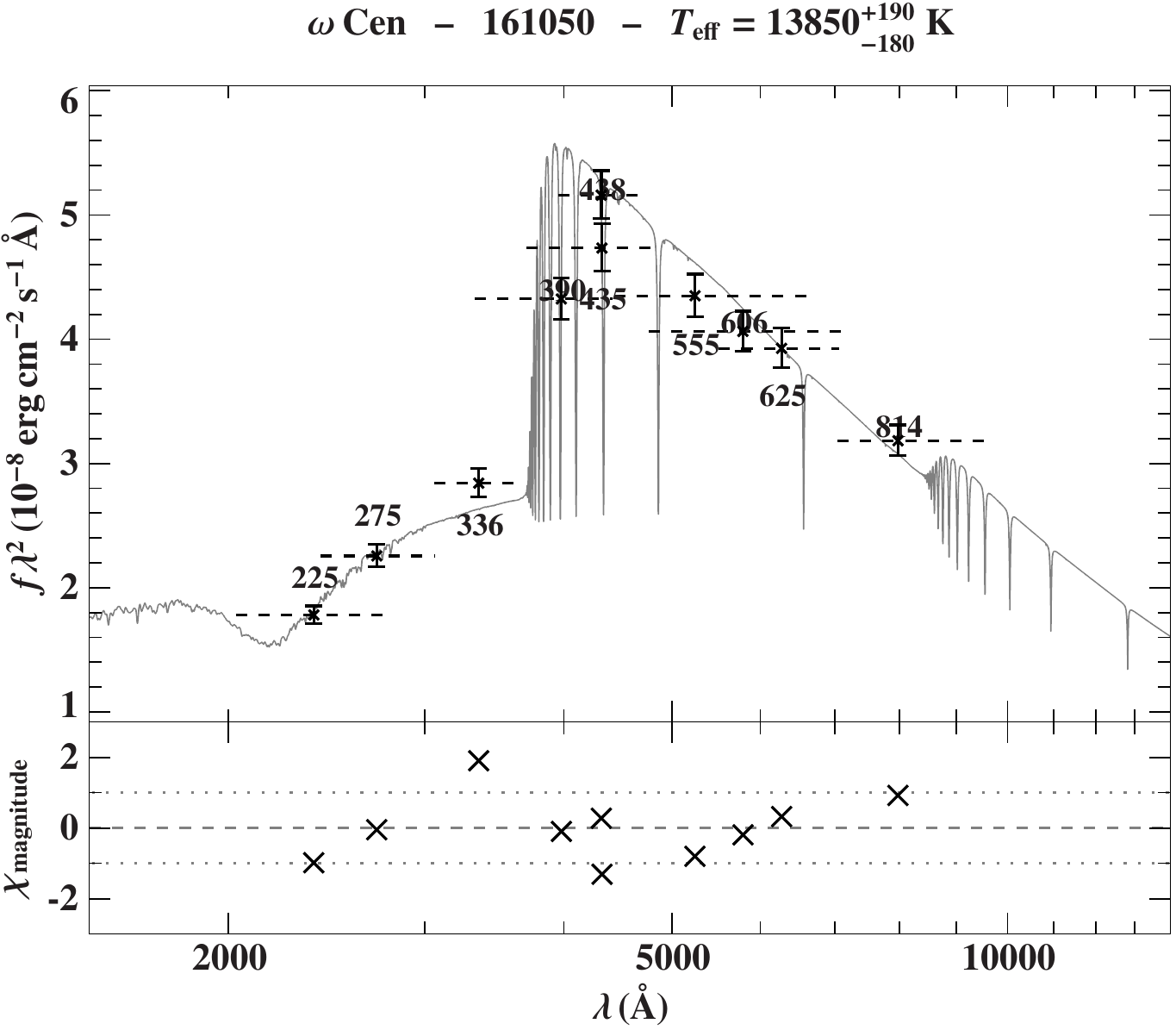}
  \includegraphics[width=0.33\textwidth]{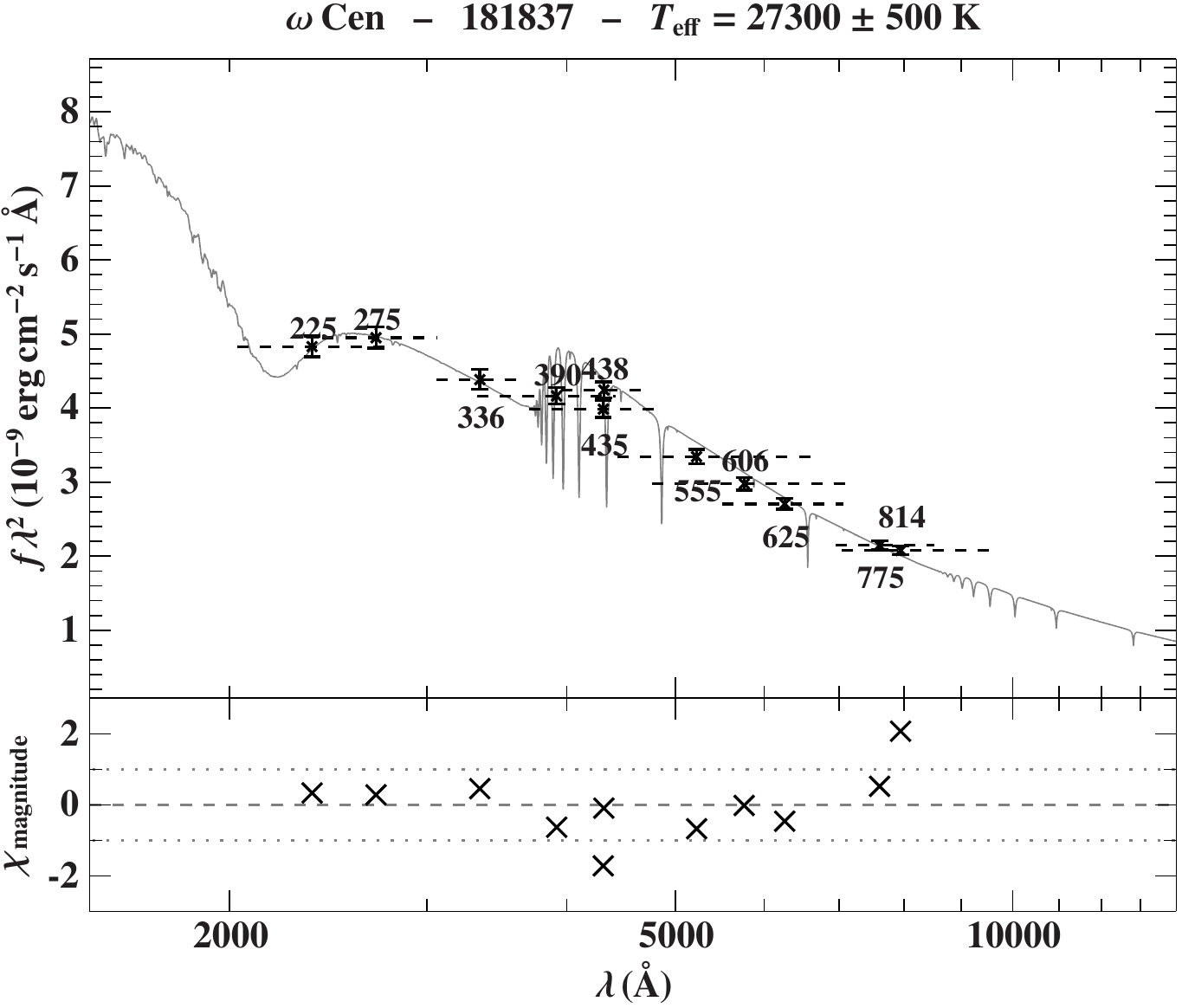}
 }\vspace{3pt}
   \resizebox{\hsize}{!}{
\includegraphics{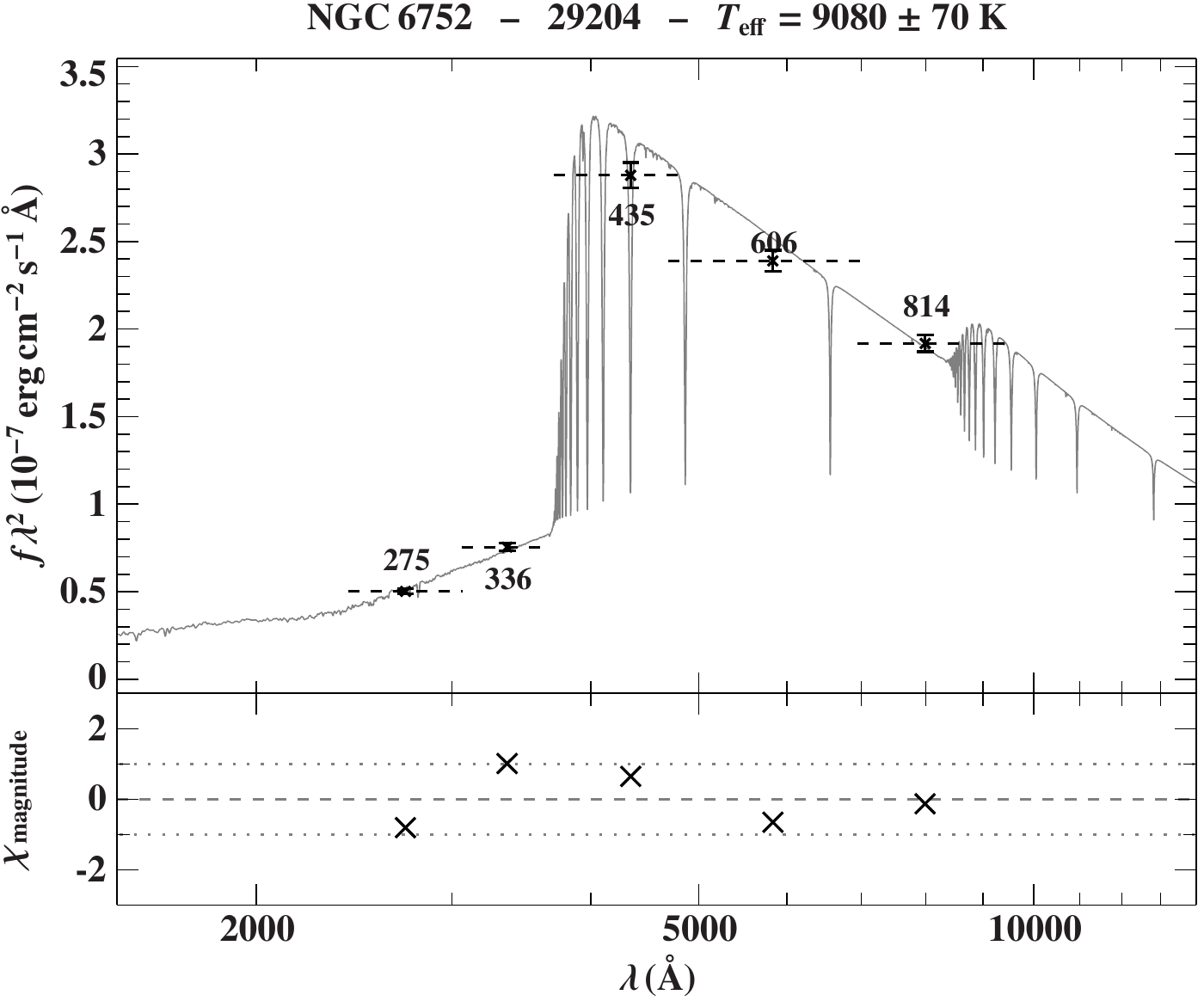}
   \includegraphics{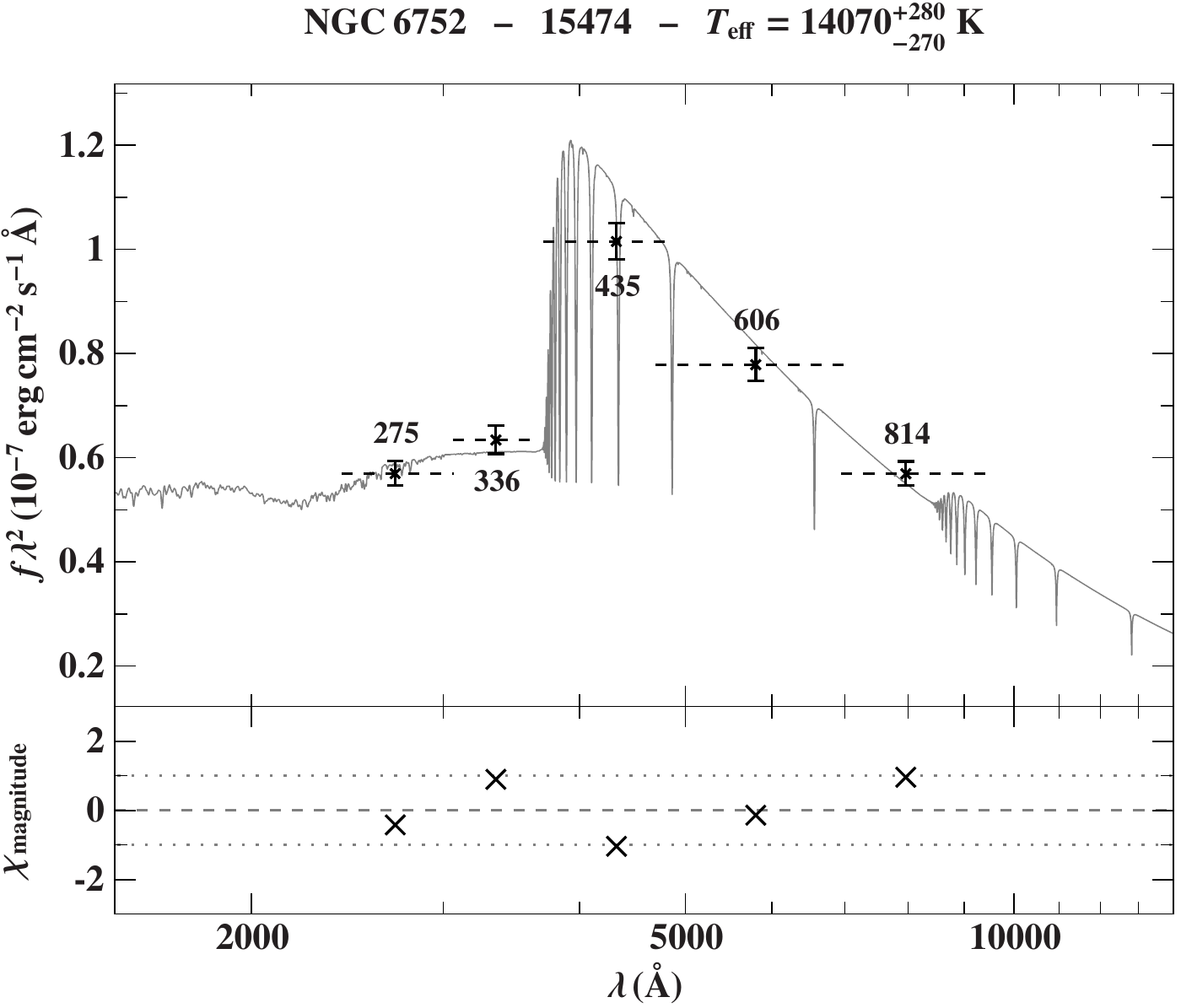}
  \includegraphics{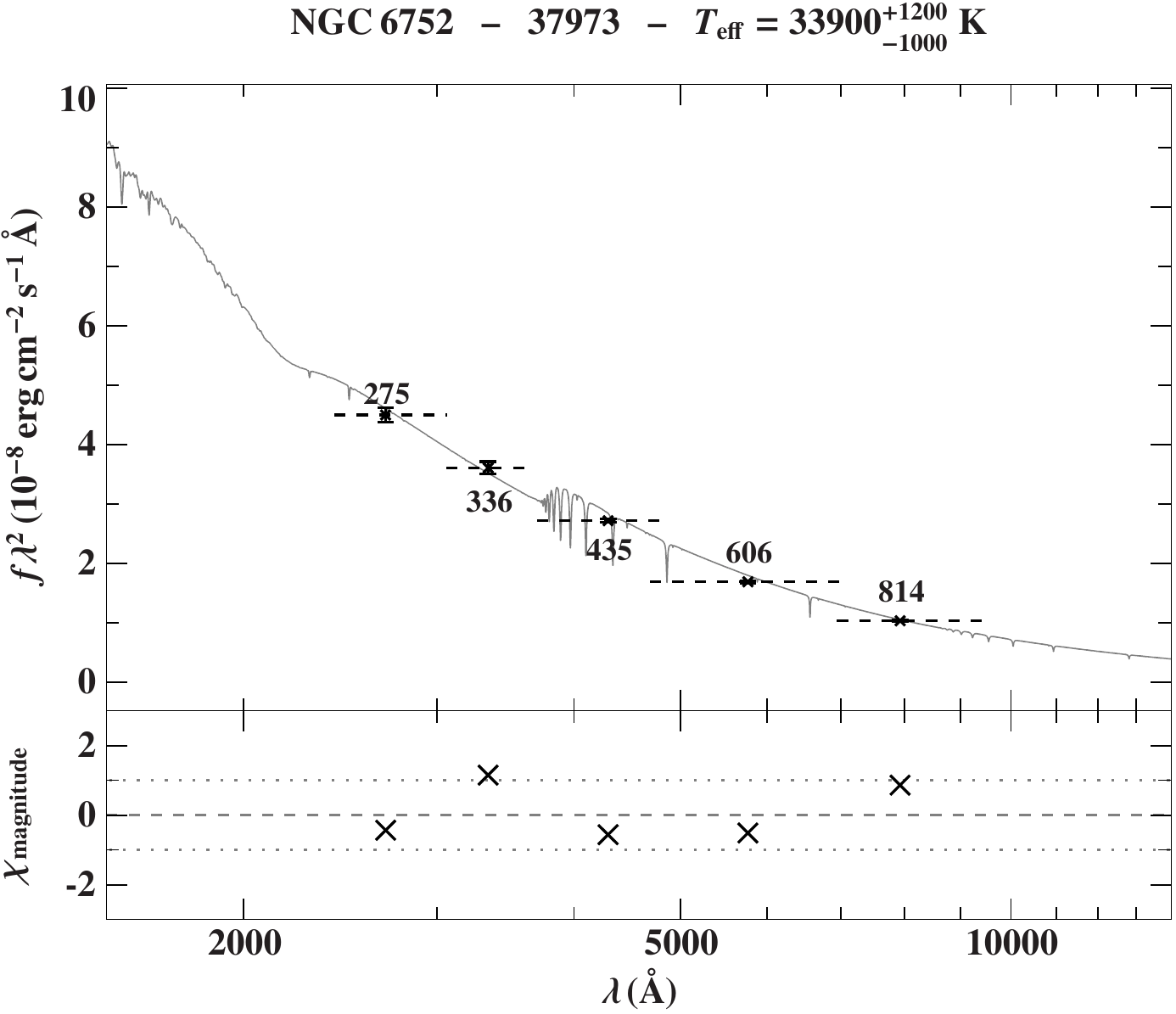}
  }\vspace{3pt}
   \resizebox{\hsize}{!}{
\includegraphics{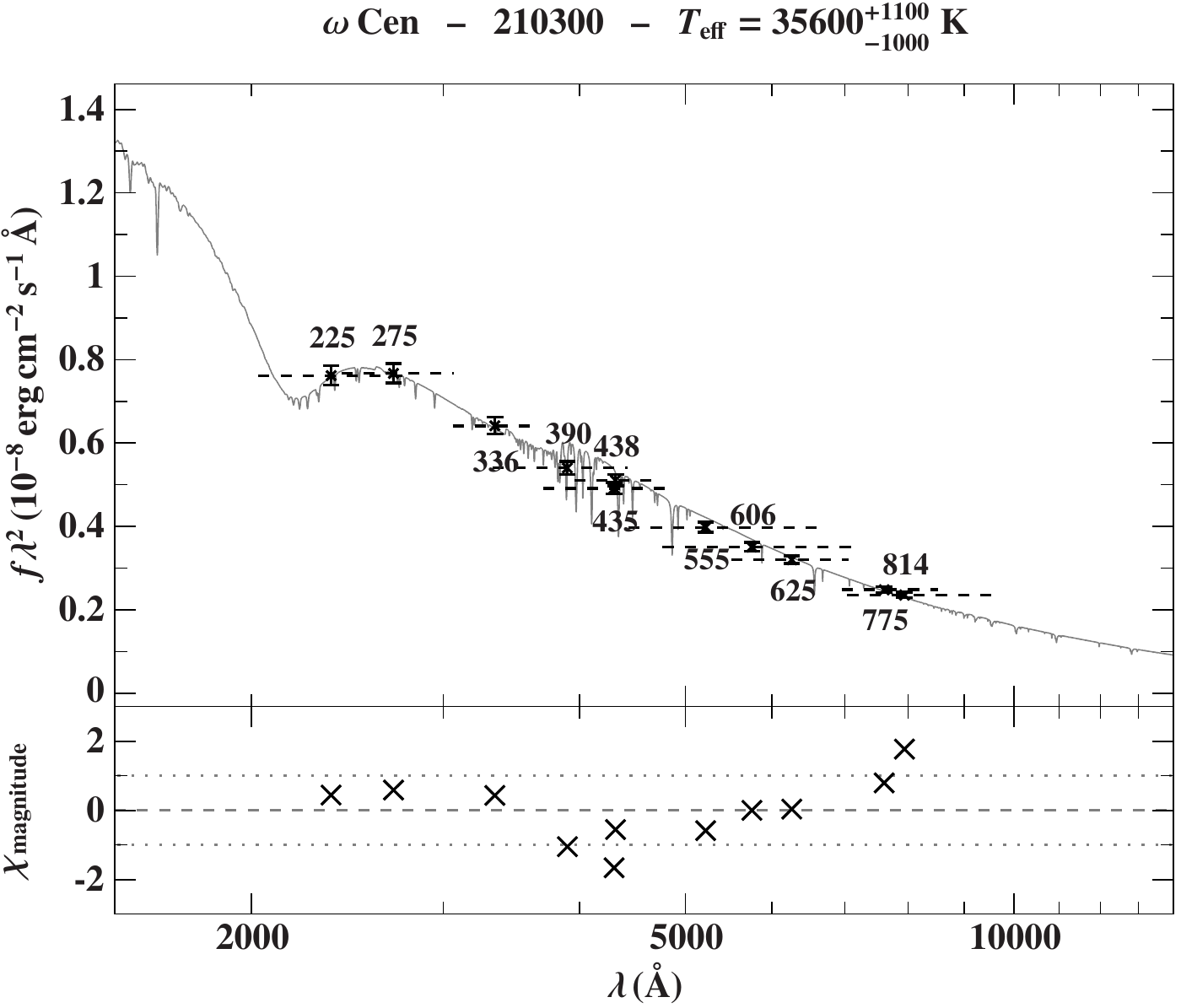}
   \includegraphics{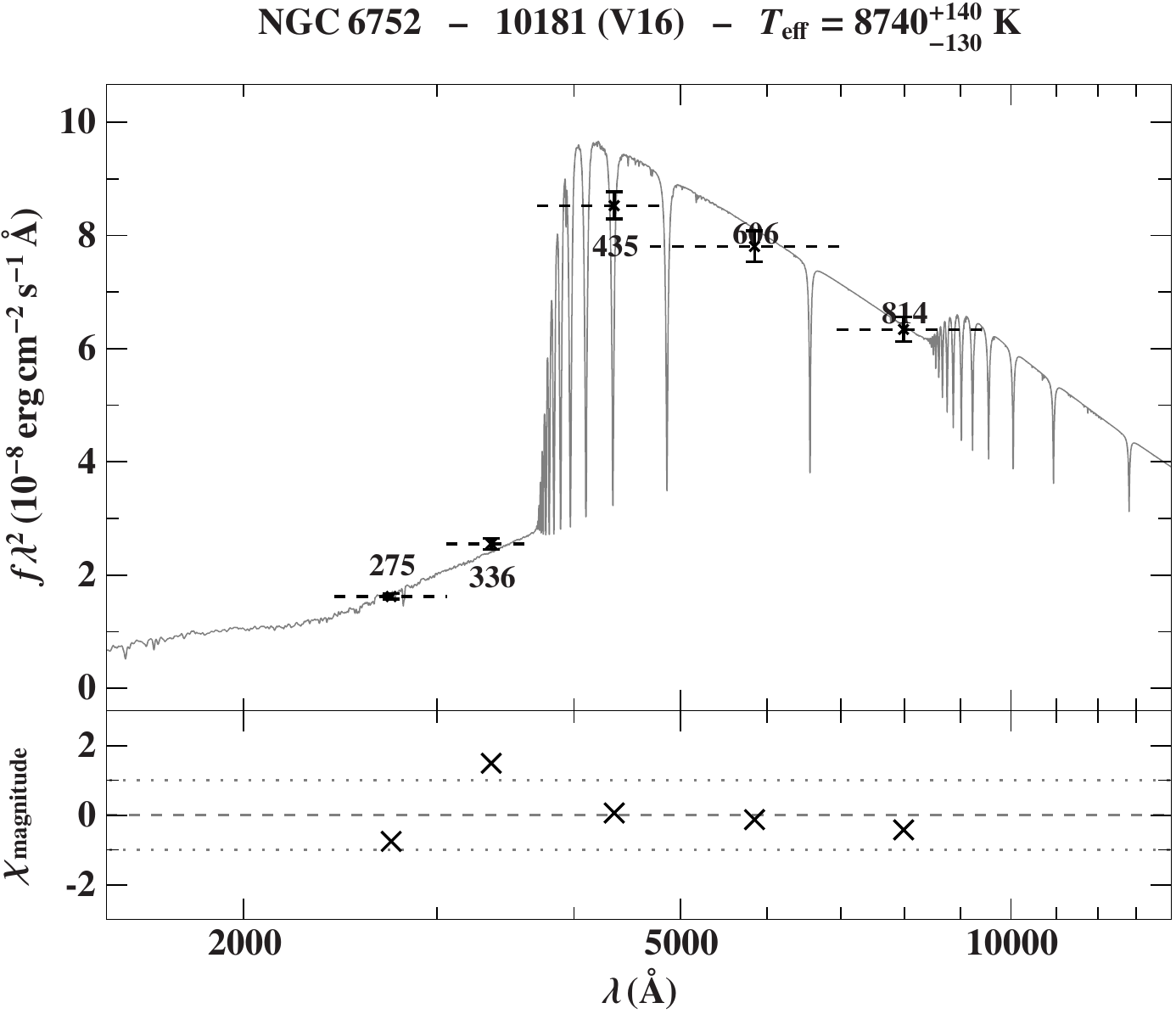}
    \includegraphics{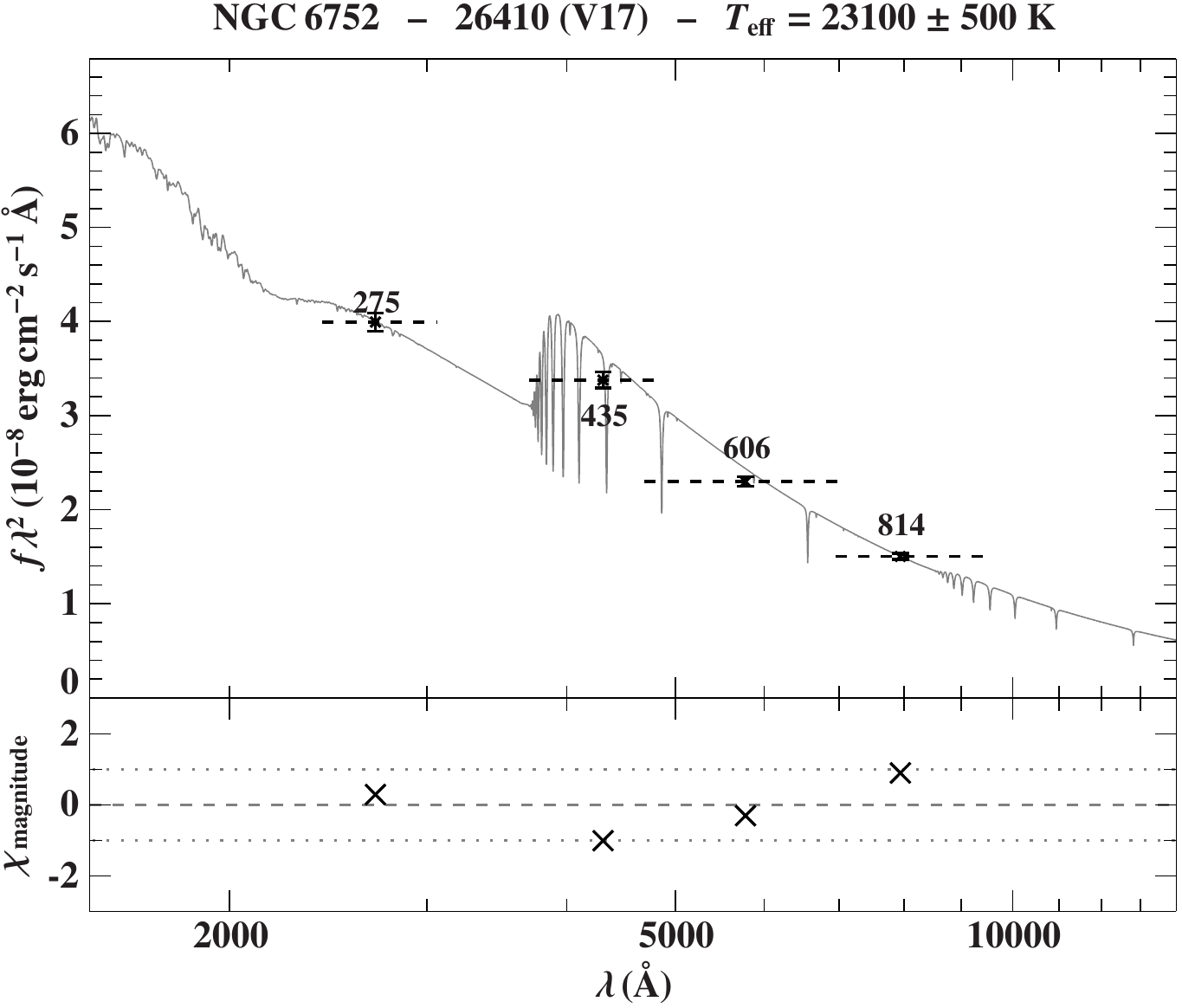}
  }
     \caption{Examples of photometric (SED) fits with free parameters $\Theta$ and \teff. On the y-axis, we plot the flux $f_\lambda$ multiplied by $\lambda^2$. The best-fit model is plotted in grey while the flux corresponding to the observed magnitude in each filter is indicated along with the central wavelength of the filter. The horizontal dashed lines show the wavelength coverage of each filter.
The top panels show the best fits for an A-BHB, B-BHB, and EHB type star in \Cen. The middle panels show fits for counterpart stars in \ngc. The bottom panels show fits for a blue hook star in \Cen, the BSS V16, and the EHB star V17 in \ngc. 
The uncertainty-weighted residuals ($\chi$ = (mag$_{\rm model}$-mag$_{\rm observed}$) / uncertainty) are plotted at the bottom of each fit. The cluster name, star identification number, and resulting \teff$^{\rm SED}$ are indicated for each fit.}
     \label{pic:sed_fits}
\end{figure*}

As mentioned in the introduction, the stars along the HB do not have the same atmospheric composition due to the onset of diffusion at $\sim$11.5~kK. In \Cen, the intrinsic metallicity and abundance spreads within the cluster are also likely to affect the atmospheric composition of the HB stars. 
We fitted all stars with the $[M/H]=-1.5$ grid and with the $[M/H]=0$ (solar metallicity) grid. For stars colder than the G-jump, meaning those with a convective atmosphere, we keep the results obtained with the $[M/H]=-1.5$ models, this metallicity is in agreement with the mean metallicity of both clusters. For the hotter stars, where diffusion changes the atmospheric composition, we keep the results obtained with the solar metallicity grid. The use of solar metallicity is a crude, but yet reasonable estimate. The element-to-element abundances resulting from diffusion are more complex but most of the atomic species become more abundant under the effect of radiative levitation \citep{behr03,brown17,2011A&A...529A..60M}. Our analysis of the spectral energy distribution of the stars in \Cen\ further supports this decision (see Sect.~\ref{subsec:metallicity_var} and Fig.~\ref{pic:sed_Z}).
The use of an appropriate metallicity, at least as much as possible, in the model atmospheres adopted for the spectral fits is important because, as for the helium abundance, $[M/H]$ influences the resulting parameters. To quantify this, we compared the \teff\ and log $g$ obtained from our fits with the $[M/H]=0$ and $-1.5$ models. The differences in log $g$ are at most $\pm$0.1 dex. In terms of \teff , for the A-BHB stars the difference is up to 200~K, while it reaches 1000~K in the EHB stars at $\sim$20\,000~K. 

\subsection{Spectral energy distribution and stellar parameters}\label{subsec:sed_fit}

To complete our analysis we derive mass, radius, and luminosity for the stars in our samples. This is done by fitting the spectral energy distribution (SED) of the stars defined by their magnitude at different wavelengths and making use of the known distances of the clusters. We construct grids of synthetic fluxes in the various HST filters from our ATLAS12 model atmosphere grids. A general description of the SED fitting method, used for field sdBs, is presented in \citet{heber2018}. 

For \ngc, we used the magnitudes provided in the five filters available from the HST UV Globular Cluster Survey (HUGS): ACS/WFC F435W, F606W, F814, and WFC3/UVIS F275W, F336W \citep{piotto2015,nardiello2018}\footnote{\url{https://archive.stsci.edu/prepds/hugs/}}.  
For \Cen, we use eight WFC3/UVIS magnitudes from the catalog of \citet{bellini2017a}\footnote{We note that for both the HUGS and \citet{bellini2017a} data we used the Method 1 catalogs.} (F225W, F275W, F336W, F390W, F438W, F555W, F606W, F775W, F814W) and the ACS/WFC F435W and F625W magnitudes of \citet{anderson10}.
The error on the magnitudes is computed by adding in quadrature the error provided in the catalogs (the RMS of individual measurements) when available, and a systematic uncertainty, typically 0.01$-$0.02 mag, related to the photometric calibration zero points. The uncertainties for \teff\ and log~$g$ come from the spectroscopic fits, but we add in quadrature 0.08 dex to the log~$g$ error. This stems from our previous experiences in fitting spectra of hot subdwarf and BHB stars.

We adopt a distance of $D=4.125 \pm 0.04$ kpc 
and $D= 5.43 \pm 0.05$ kpc 
for \ngc, and \Cen\ respectively \citep{Baumgardt2021}.
With the distance to the globular clusters fixed, the parameters that influence the shape of the SED are the angular diameter $\Theta$ ($=2\,R/D$), the interstellar reddening $E$(44$-$55)\footnote{$E$(44$-$55) is analogous to $E$($B$$-$$V$), but with the monochromatic measures of the extinction at 4400 and
5500 \AA\ substituting for measurements with the $B$ and $V$ filters. 
Conversion factors to the $UBV$ systems are given in table 4 of \citet{2019ApJ...886..108F}. They are close to 1 for hot stars.}
$E$(44$-$55)\, teff, log $g$, metallicity, and, to a lesser extent, the helium content of the model atmospheres.
The surface gravity and helium abundance cannot be well constrained by photometry, so we used the spectroscopic values obtained from the MUSE spectra. The metallicity is fixed in the same way as for the spectroscopic analysis (see the previous subsection). 

We account for interstellar extinction using the functions of  \citet{2019ApJ...886..108F} and adopt a ratio of total-to-selective extinction of $R(55)$=3.02, 
which is the Milky Way average.
Because we realized that the results are sensitive to the adopted reddening $E$(44$-$55), the SED fits were performed in two iterations. 
In the first step, we leave the reddening and $\Theta$ as free parameters while \teff\ is fixed to its spectroscopic value. Then we used the reddening obtained for the stars colder than 13~kK, for which the spectroscopic and photometric \teffs\ are in good agreement, to derive an average reddening for each cluster (see Appendix A for additional details). That way, we obtained $E$(44$-$55) = 0.041\,mag for \ngc\ and 0.119\,mag for \Cen. We note that these values are in excellent agreement with the literature (\citealt{harris1996}, 2010 edition). 
The second step is to perform the final fit with $E$(44$-$55) fixed to the values mentioned above and to have \teff\ and $\Theta$ left as free parameters.
From $\Theta$, we directly obtain the radius of the star because the clusters' distances $D$ are well known. We compute the luminosity and the mass via the formulae
\begin{equation*}
\centering
L=4\pi R^2 \sigma T^4 _{\rm eff}\qquad \text{and} \quad M=\frac{gR^2}{G}.
\end{equation*}
Because the luminosity and mass have an additional dependence on the effective temperature and surface gravity, respectively, these two parameters bear larger uncertainties than the radius. 
All uncertainties were propagated using the Monte Carlo method; the resulting best-fit values and their uncertainties are stated as the median with 68\% uncertainties throughout this work.

\begin{figure*}
\sidecaption
   \includegraphics[width=12cm]{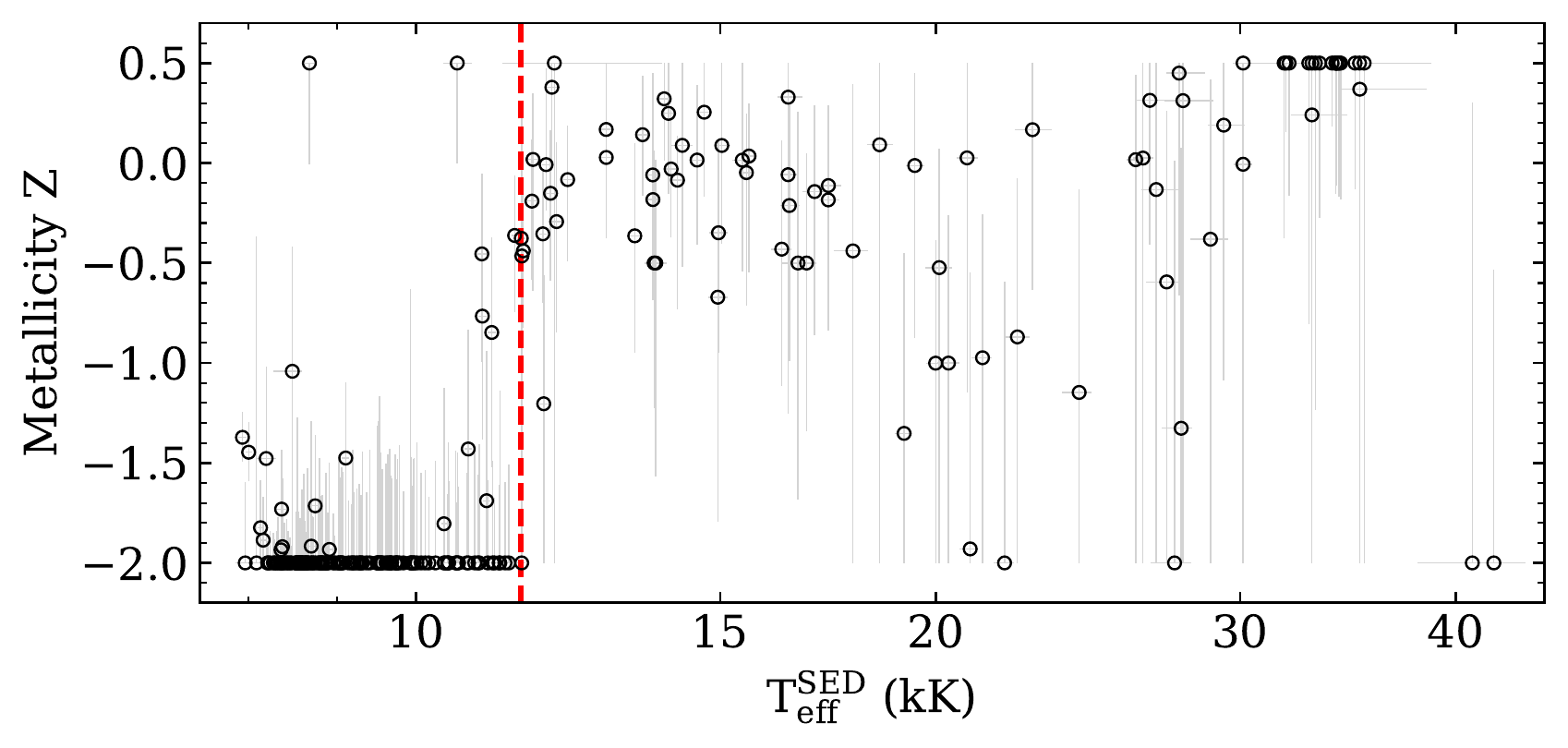}
     \caption{Metallicity versus effective temperatures obtained from the photometric fits of the stars in \Cen. The vertical line at 11.5~kK indicates the position of the G-jump.  }
     \label{pic:sed_Z}
\end{figure*}

We show in Fig.~\ref{pic:sed_fits} some examples of SED fits for stars at various temperatures in both clusters.
The main feature of the SED for the A-BHB and B-BHB stars is the Balmer jump. This feature is a good indicator of the stellar effective temperature in BHB stars\footnote{With magnitudes on both sides of the Balmer jump, the surface gravity can also be estimated from the SED.}. The hotter EHB stars have a less prominent Balmer jump and are characterized by an increasing flux at short wavelengths (keeping in mind that the $y$-axis is the flux multiplied by $\lambda^2$). Above 30~kK, the \teff\ obtained from the SED fits are generally less precise. The dip in the UV flux of the models is due to the 2200~\AA\ bump present in the interstellar extinction curve. This feature is stronger in \Cen\ than in \ngc\ because the reddening of the former is higher.

\paragraph{Metallicities from SED fits}\label{subsec:metallicity_var}

The surface metallicity of BHB stars is difficult to determine from MUSE spectroscopy, due to a lack of iron-group spectral lines and the low resolution.
However, the forest of spectral lines from heavy elements in the near-UV and UV regions of the spectra can block a significant amount of flux. Generally speaking, a higher metallicity increases the strength of the metal lines in the UV and consequently suppresses the UV flux, thus affecting 
the WFC3/UVIS filter at the shortest wavelength (F225W and F275W). The flux that is blocked is then emitted at longer wavelengths for a fix \teff. 
As a test, we performed SED fits of the stars in \Cen\ where $\theta$, \teff, and the metallicity $Z$ were left free to vary,  while $E$(44$-$55) was fixed to 0.12 mag. We used the data in \Cen\ to perform this test because the \citet{bellini2017a} catalog includes magnitudes in the F225W filter. 
Figure~\ref{pic:sed_Z} shows the resulting metallicity as a function of \teff. Although $Z$ is not well constrained, the sudden increase in atmospheric metallicity due to the transition from a convective to a radiative atmosphere is clearly visible and happens, as expected, around 11.5~kK. For the hotter stars, the metallicities scatter around the solar value ($Z$=0). This supports our decision to use a solar metallicity in the spectral analysis of the stars hotter than 11.5~kK. This exercise demonstrates that SED fits can be a powerful investigation tool, more so when UV magnitudes are available. In \Cen, we have ideal conditions to probe the atmospheric metallicity: well-calibrated near-UV and UV magnitudes combined with well-constrained distance and reddening for the stars.

\begin{figure*}
\resizebox{\hsize}{!}{
   \includegraphics{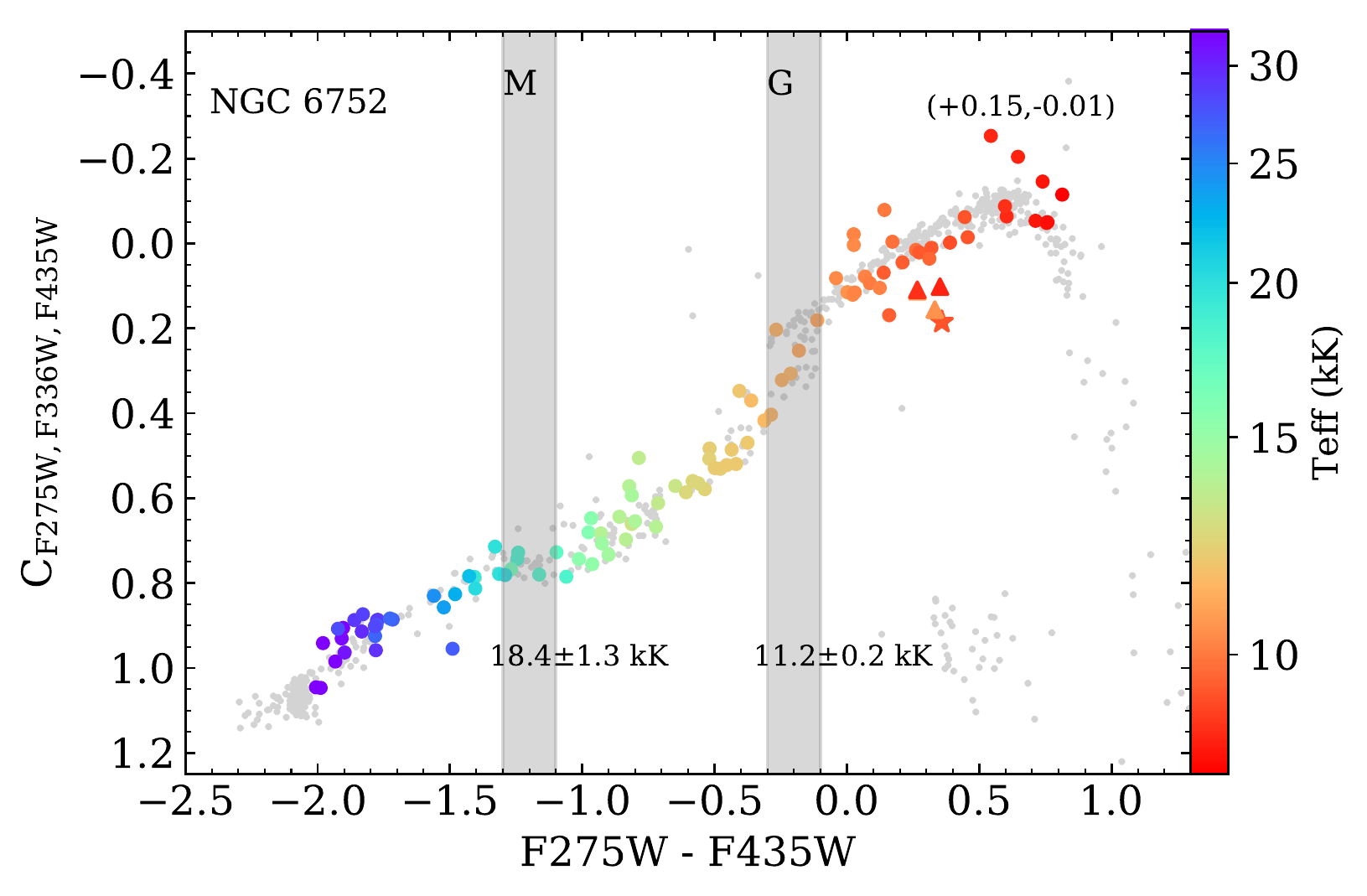}
   \includegraphics{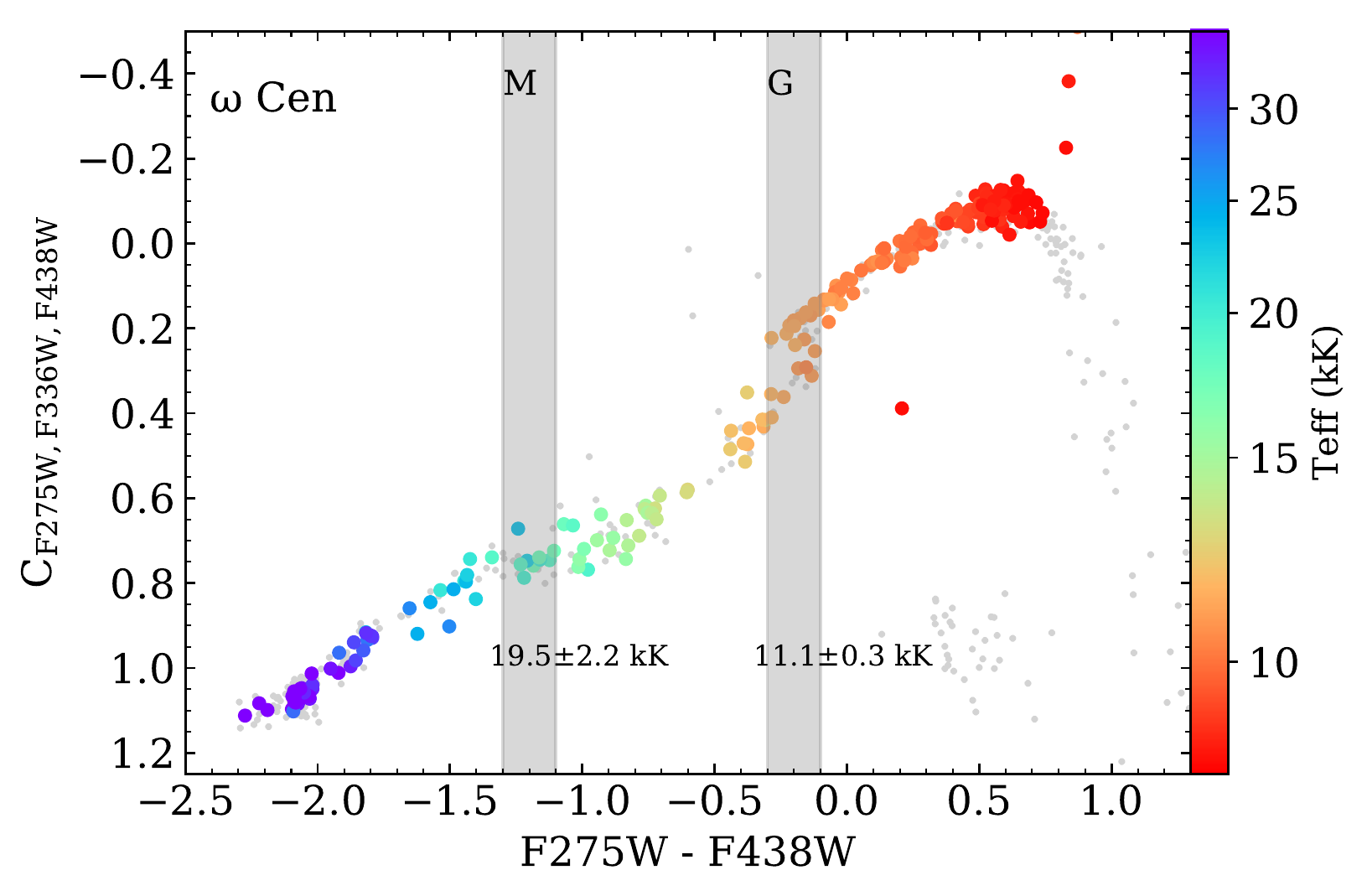}
   }
     \caption{Color-color plot of the HB stars in our samples. The \teffs obtained from the spectral fits are color coded. The position of the G- and M-jump is indicated by shaded areas, as well as the average temperature and standard deviation of the stars that are included in the area of the jumps. Three hot BSSs (triangle) and V17 (star) in \ngc\ are also included.
     We note that some stars are missing in this plot because they do not have a magnitude in all three filters. For both clusters, we plotted the position of the stars from the photometric catalog of \Cen\ as small grey dots. For \ngc, we also indicate the shift applied to align the stars with those of \Cen.  }.
     \label{pic:Cindex}
\end{figure*}

\section{The final samples}\label{sec:sample}
We used spectra with a S/N $\geq$ 20 in our analysis. 
The resulting fit for each star was visually inspected and those with poor fits, in terms of reduced $\chi^2$ and residuals were removed from the sample. Stars that were outliers in some of the parameters derived (e.g., \teff, radius, mass) were also individually inspected. A few additional stars were excluded from the sample after these checks. In most cases, the MUSE spectra were contaminated by the light of a very close-by companion. 
Even though \textsc{pampelmuse} is efficient at ``deblending" the spectra, it is limited, especially for exposures with poor atmospheric conditions and very faint stars. We also found issues when one or more spectra from individual exposures were of especially poor quality, most often due to the star being very close to the edge of the field of view on these particular exposures.
With these criteria, we ensure that our final sample contains, to the best of our knowledge, stars for which we have good atmospheric parameters. We note that the radial velocity of all stars in our final sample is consistent with cluster membership. For \Cen, we also verify their membership via the proper motions of \citet{bellini2017a}.
The final samples contain 302 and 130 HB stars in \Cen\ and \ngc, respectively. In Sect.~\ref{subsec:variable}, we discuss the analysis of five hot BSSs in \ngc.

\section{Atmospheric parameters}\label{subsec:atm_par}

We present here the atmospheric parameters derived from the fits of the MUSE spectra. The tabulated results are only available online as Tables B.1 and B.2 (see Appendix B). 
In Sect.~\ref{subsec:ccp} we use a color-color plane to identify the location of the HB jumps in terms of their effective temperatures. In Sect.~\ref{subsubsec:tef_logg} we construct the Kiel diagram (\loggt), and in Sect. \ref{subsubsec:helium} we discuss the variation in helium abundances along the HB. Finally, we compare our results with previous surveys from the literature in Sect.~\ref{sec:comp_litt}.

\subsection{Color-color plane and the location of the HB jumps}\label{subsec:ccp}

\citet{brown16} used a particular combination of HST magnitudes from optical and near-UV filters to study the properties of the HB in 53 galactic globular clusters, including \Cen. We use the same combination of colors in Fig.~\ref{pic:Cindex} to display the stars in our samples. 
On the y-axis, we use the color index C$_{\rm{F275W,F336W,F438W}}$ = ($m_{\rm{F275W}}-m_{\rm{F336W}}$) - ($m_{\rm{F336W}}-m_{\rm{F438W}}$). The effective temperatures obtained from the spectral fits are color-coded and illustrate very well the temperature progression along the HB.
In this particular color-color plane, the G- and M-jump are  visible at $m_{\rm{F275W}}$-$m_{\rm{F438W}}$ $\approx$$-$0.3 and $-$1.3 respectively. For \Cen, we define the position of the jumps, with shaded areas in Fig.~\ref{pic:Cindex}, as in \citet{brown17}. 
For \ngc, we used the same method as in \citet{brown17} and we align the stars in our sample with those of \Cen\ by applying a shift of +0.15 and -0.01 in the x and y direction respectively.
There is a larger scatter of the stars in the color-color plane of \ngc\ and the jumps are not as clearly visible as in \Cen. This is due to the use of the significantly wider ACS/WFC F435W filter instead of the WFC3/UVIS F438W filter, which is why this cluster was not included in the HUGS photometric sample at the time of \citet{brown16}. Most importantly, the F435W filter covers the Balmer jump while the F438W filter does not.

Finally, we computed the mean \teff\ of the stars found in the shaded area\footnote{A few more stars than seen in the figure are contributing to the mean \teff\ at the position of the jumps. This is because the jumps are defined in terms of F275W$-$F435(438)W. Some stars have magnitudes in these two filters but not in F336W, thus they do not appear on the plots.}
and indicate the resulting values and standard deviations on Fig.~\ref{pic:Cindex}.
The mean \teff\ of the stars at the G-jump in both clusters 
is in good agreement with the expected value of 11.5~kK. As for the M-jump, we found mean \teffs of 18.4$\pm$1.3~kK and 19.5$\pm$2.2~kK for \ngc\ and \Cen, respectively. This is still in reasonable agreement with the theoretical expectations that are around 20~kK.

\begin{figure*}
\resizebox{\hsize}{!}{
   \includegraphics{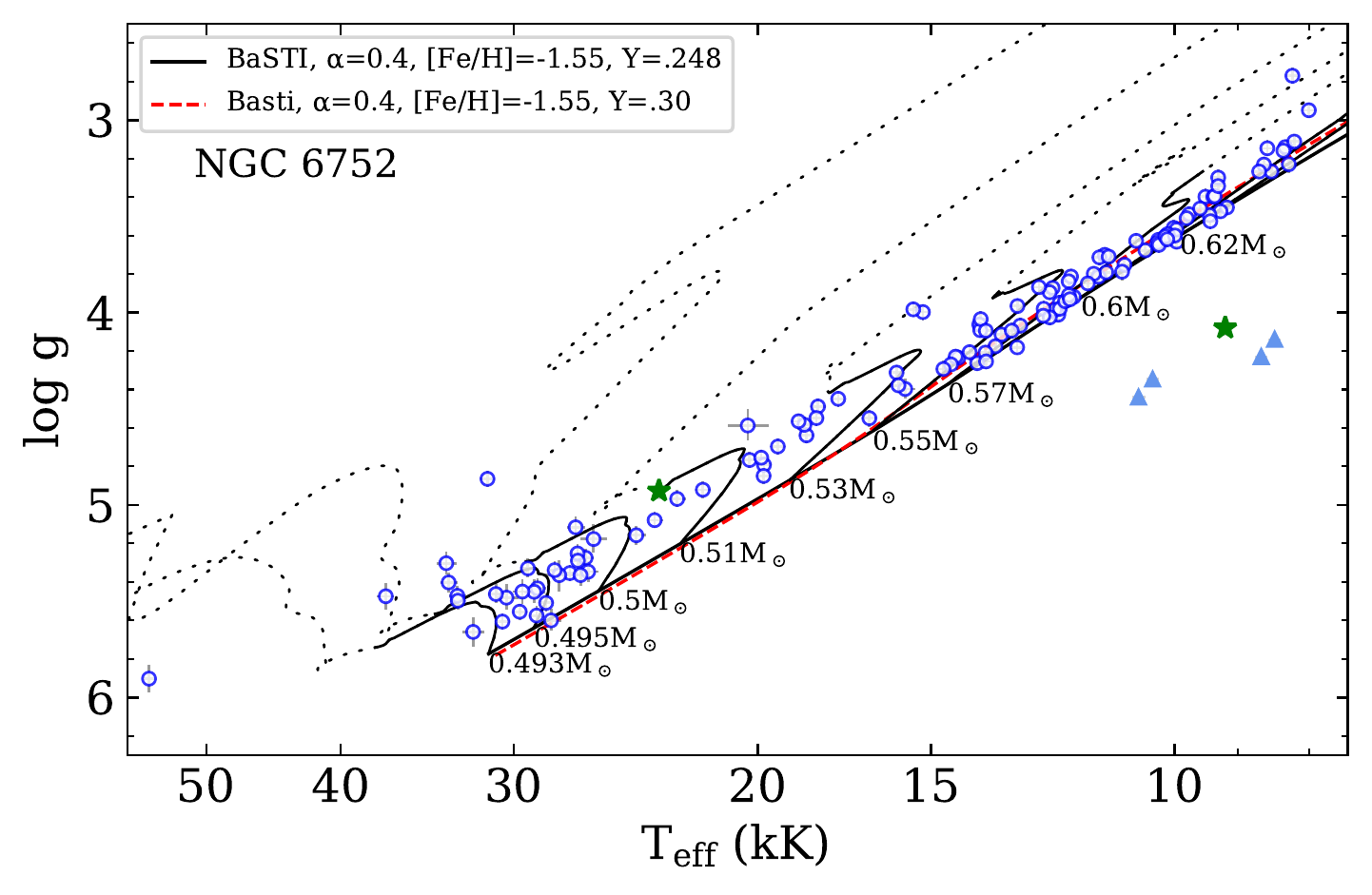}
   \includegraphics{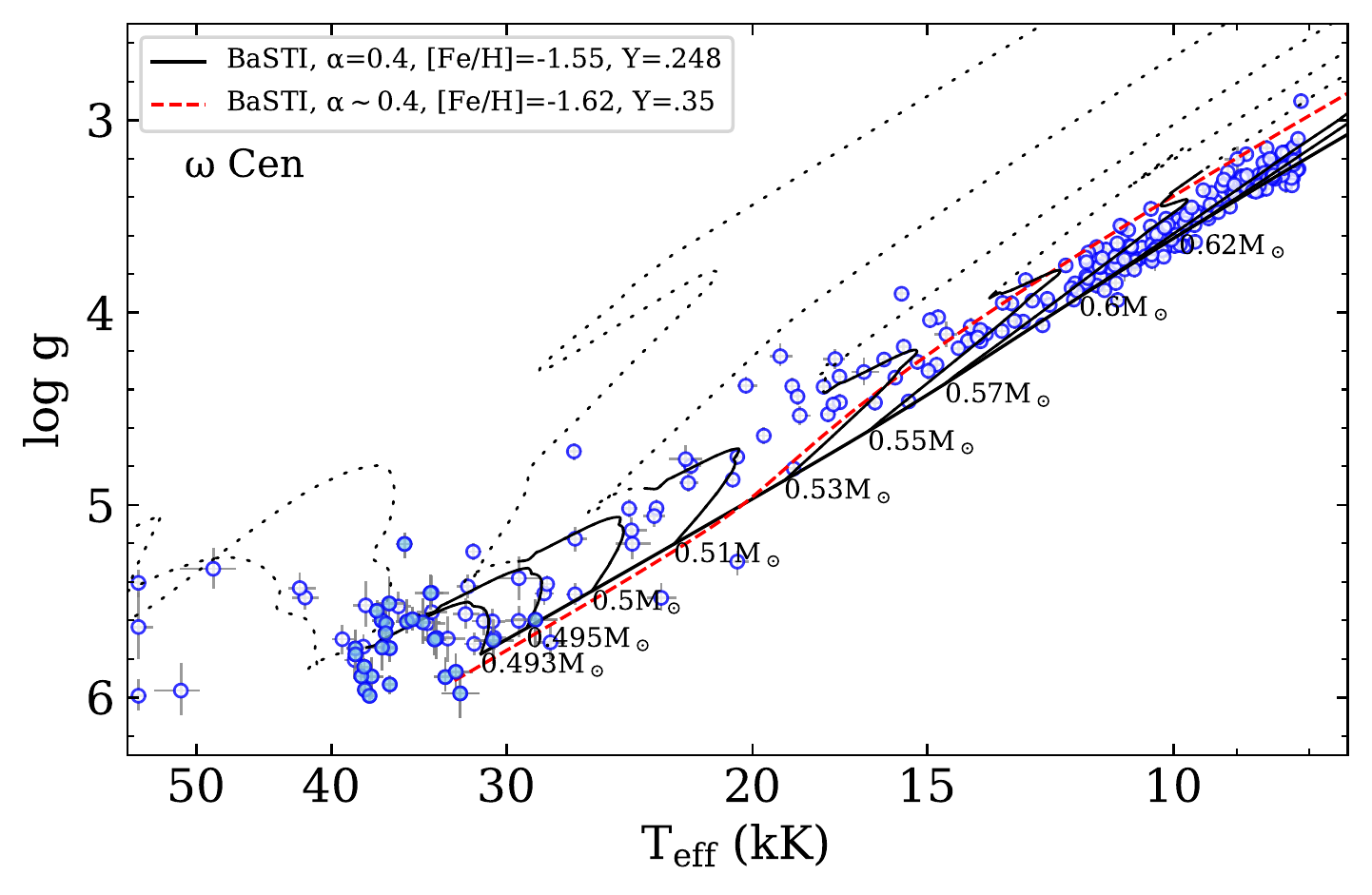}
   }
     \caption{Kiel diagrams for \ngc\ (left) and \Cen\ (right). We show the theoretical ZAHBs from BaSTI models at a metallicity representative of the clusters and with normal (solid line) and enhanced (red dashed line) helium ($Y$) abundances. Evolutionary tracks from BaSTI for different masses are also shown with solid lines for the He-core burning phase (i.e, the HB phase per se) and with dotted lines for the He-shell burning phase (post-HB). The He-rich (blue hook) objects in \Cen\ are indicated with filled symbols. Also shown in \ngc\ are the two variable stars V16 and V17 (green star symbol), and the four hot BSSs (blue triangles). 
     }
     \label{pic:Kiel_dia}
\end{figure*}

\subsection{Effective temperature and surface gravity}\label{subsubsec:tef_logg}

In Fig.~\ref{pic:Kiel_dia}, we present the HB stars of our samples in the \teff$-$log~$g$ diagram (hereafter Kiel diagram) for each cluster. 
We also include the position of the theoretical zero-age horizontal branches (ZAHB) taken from the BaSTI database \citep{basti2021}\footnote{\url{http://basti-iac.oa-teramo.inaf.it/}}.
Additional (post-)HB evolutionary tracks, from BaSTI, are also shown for different masses along the HB. The part of the tracks shown with solid lines represents the central He-burning phase, the HB phase per se. The subsequent, and faster by a factor of ten, post-HB evolution with He-shell burning is shown with dotted lines. 
For both clusters, we selected theoretical HB models with parameters matching the properties of each cluster. We used the new BaSTI $\alpha$-enhanced ($\alpha$=0.4) models \citep{basti2021} with [Fe/H]$= -1.55$ ($Z$=0.000886) and normal helium ($Y$=0.248).
For both clusters, we also show additional theoretical ZAHBs for helium-enhanced models ($Y$=0.30 for \ngc\ and $Y$=0.35 for \Cen\footnote{We note here that the updated $\alpha$-enhanced BaSTI models only include, so far, $Y$ up to 0.30. So the $Y$=0.35 model is from \citet{basti2006}.}).

\begin{figure*}
\resizebox{\hsize}{!}{
   \includegraphics{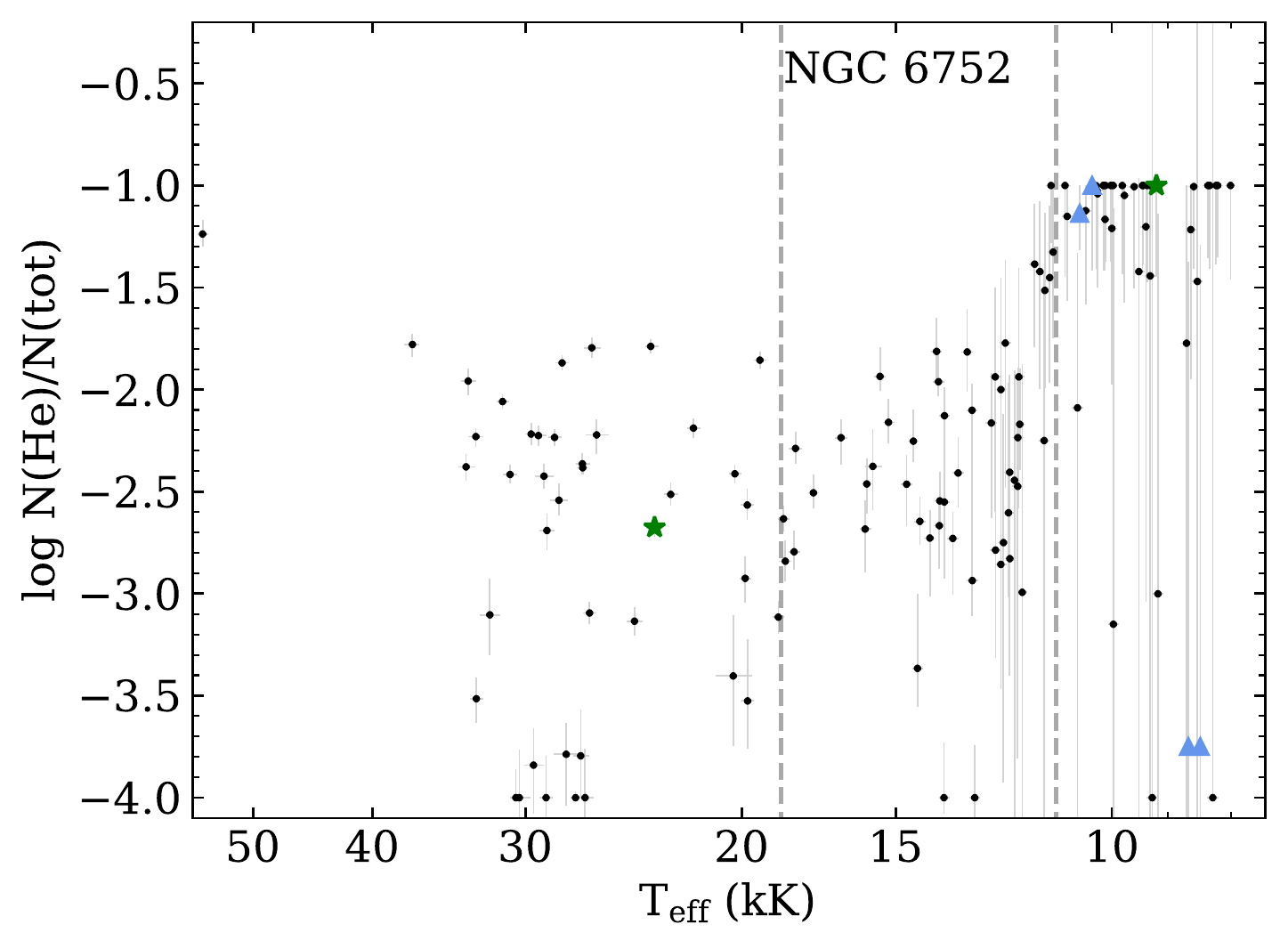}   \includegraphics{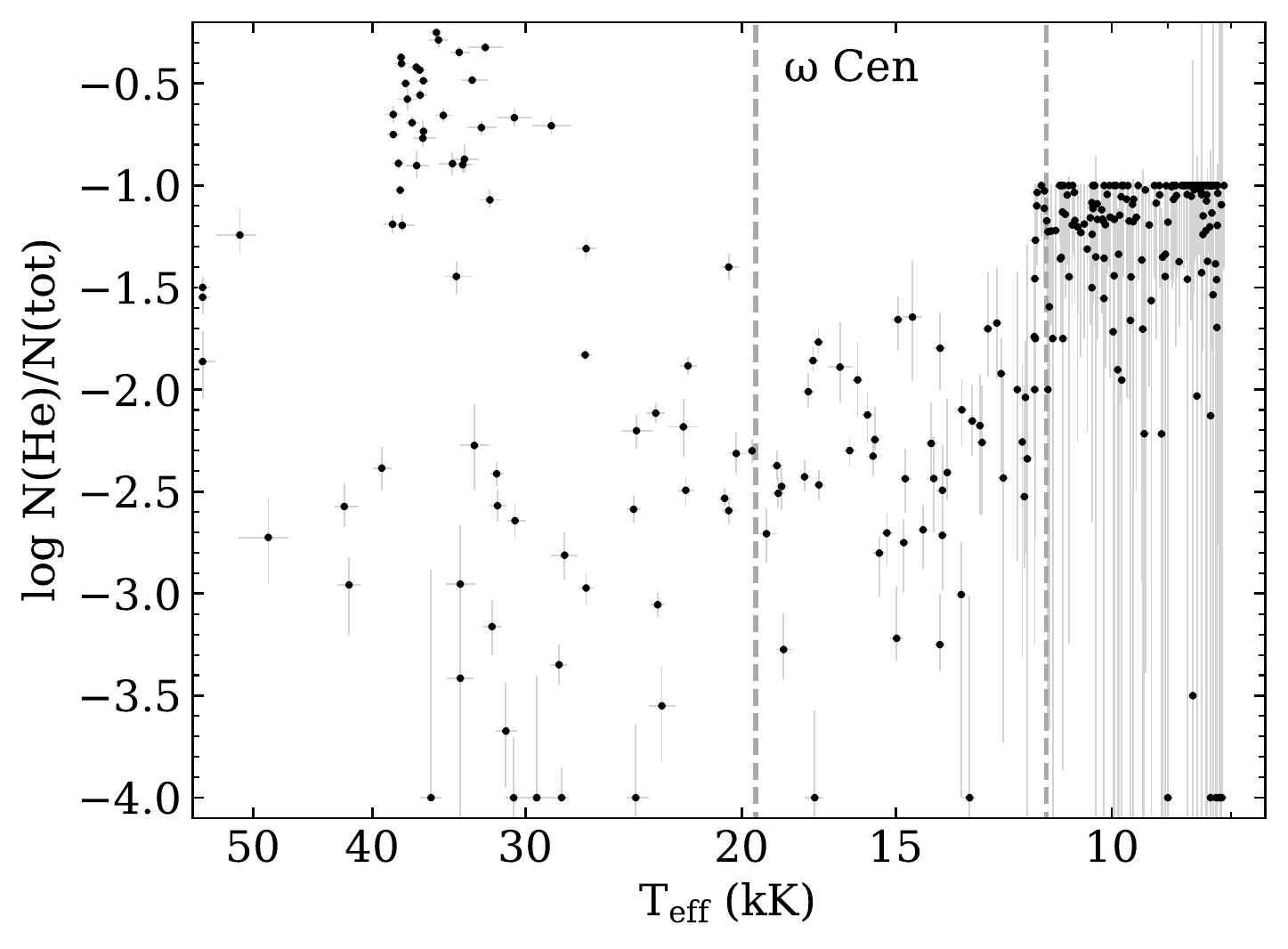}
   }
     \caption{Helium abundance as a function of \teff. The two vertical lines indicate the positions of the G- and M- jumps at the \teff\ determined in Fig.~\ref{pic:Cindex}.}
     \label{pic:HE_plot}
\end{figure*}

In both clusters, the cool HB stars (\teff\ $\lesssim$ 14-15 kK) sit on the ZAHB as predicted from the helium normal models. 
In \ngc, the distribution of the cooler HB stars is tightly clustering on the ZAHB. This is expected since the evolutionary tracks of stars with $M \gtrsim 0.57$\msun\ evolve towards the asymptotic giant branch essentially parallel to the ZAHB. 
This clustering of the cold stars on the ZAHB is also present in \Cen, although with a larger scatter. Interestingly, while the A-BHB stars in \ngc\ form a very narrow sequence in the F275W-F606W CMD (see Fig.~\ref{pic:app_CMD}), the equivalent stars in \Cen\ have a larger scatter, reminiscent of what is seen in the Kiel diagram. Thus, this feature is unlikely to be an artifact coming from our spectral analysis. The scatter could be due to the spread in metallicity among the stars of \Cen, as the metallicity affects the position of the theoretical ZAHB in the Kiel diagram. In addition, a metallicity spread among the A-BHB stars could also produce some variations in the \teff\ derived (see Sect.~\ref{subsec:metal}).
It is clear from Fig.~\ref{pic:Kiel_dia} that the cool HB stars in \Cen\ do not originate from the helium-enriched population, these models predict the stars to have lower surface gravity (i.e. higher luminosity) than what we measure. This is consistent with the findings of \citet{Joo2013} and \citet{tailo2016} who used population synthesis of the main sequence, sub-giant and horizontal branches to reproduce the main features of \Cen's CMD. That way, they populated most of the cool part of the BHB with metal-poor and helium-normal objects. 

For both clusters, we see in Fig.~\ref{pic:Kiel_dia} that most of the stars hotter than $\sim$15~kK lie above the ZAHB, but still below the terminal-age HB (TAHB, i.e. the end of the He-core burning phase).
We do not know the reason behind this shift for the hotter stars, but \citet{leblanc2010} showed that elemental stratification in the radiative atmosphere of B-BHB has an effect on the hydrogen line profiles. The authors showed that this could result in an underestimation of the log~$g$ derived with homogeneous model atmospheres like the ones we use for this work.

The few objects found above the TAHB correspond to the evolved post-HB phase where the nuclear burning occurs in a shell. These stars are more luminous and have larger radii than those on the HB. They are also brighter than the bulk of HB stars and are often referred to as UV-bright objects (see e.g., \citealt{moehler2019}). 
In NGC~6752, the low-gravity star (id 15070, \teff = 31.3~kK, log~$g$~=~4.86) is much brighter ($F275W=13.1$) than stars of similar color in the CMD shown in Fig.~\ref{pic:app_CMD}. This object is also known as UIT-1 (from the Ultraviolet Imaging Telescope, \citealt{landsman1996}). An independent study found \teff = 32$\pm$2~kK and log $g$ = 4.9$\pm$0.2 dex from an optical HST spectrum (P. Chayer, priv. communication). 
In fact, we found that most of the stars with a post-HB position in the Kiel diagram of both clusters correspond to brighter objects than the bulk of HB stars in the NUV-optical CMD (Fig.~\ref{pic:app_CMD}). In the optical CMD (Fig.~\ref{pic:CMDs}) they appear shifted to the left compared to the other stars.
In \ngc, we have identified, for the first time, a hot hydrogen-rich sdO (H-sdO, id 738) with \teff$\sim$55~kK that
correspond to the bluest object in the optical CMD of \ngc\ (Fig.~\ref{pic:CMDs}). 
Finally, the five hot BSSs in \ngc, including V17 (see Sect.~\ref{subsec:variable}), are significantly below the ZAHB as expected.

\subsection{Helium abundances}\label{subsubsec:helium}

The fact that the B-BHB stars, essentially those between the two jumps (Fig.~\ref{pic:Cindex}), are shifted downward in the color-color plane is explained by an increase in atmospheric metallicity and a decrease in helium abundance, both being the result of diffusion processes as the surface convection vanishes \citep{brown16}.
Figure~\ref{pic:HE_plot} shows our results in the \teff$-$He plane.

As mentioned previously, the He lines in stars colder than the G-jump are very weak and not necessarily visible in the MUSE spectra. This is reflected in the very large error bars on the helium abundance of the coldest stars, meaning that it is poorly constrained. Nevertheless, the spectral fit for the majority of stars colder than the G-jump is consistent with a solar helium abundance.  
Figure~\ref{pic:HE_plot} shows the expected decrease in He abundance in the stars hotter than the G-jump (indicated with a dashed line) until $\sim$15~kK. In stars hotter than 15~kK, the He values scatter mostly between $-2$\,dex to $-$3\,dex. In \Cen, Fig~\ref{pic:HE_plot} clearly shows that our sample includes a handful of He-rich stars with \teff\ between 30 and 40~kK. As expected, we did not find any such objects in \ngc. This is one of the differences between the HB morphology of both clusters.
These He-rich stars form the blue hook population of \Cen\ that is located at the faint end of the HB in the optical CMD of Fig.~\ref{pic:CMDs}. In the F275W-F606W CMD of \Cen\ (Fig.~\ref{pic:app_CMD}), the end of the HB appears to be split into two narrow vertical strips. The He-rich, blue hook stars in our sample are all found to lie on the bluest strip.
With the MUSE spectra, we did not aim, nor pretend, to derive accurate individual He abundances. However, taken globally, our results are in good agreement with the previous spectroscopic analyses in \ngc\ \citep{moehler00,monibidin2007} and \Cen\ \citep{moehler2011,monibidin2012,latour2018}.

\begin{figure*}
\resizebox{\hsize}{!}{
   \includegraphics{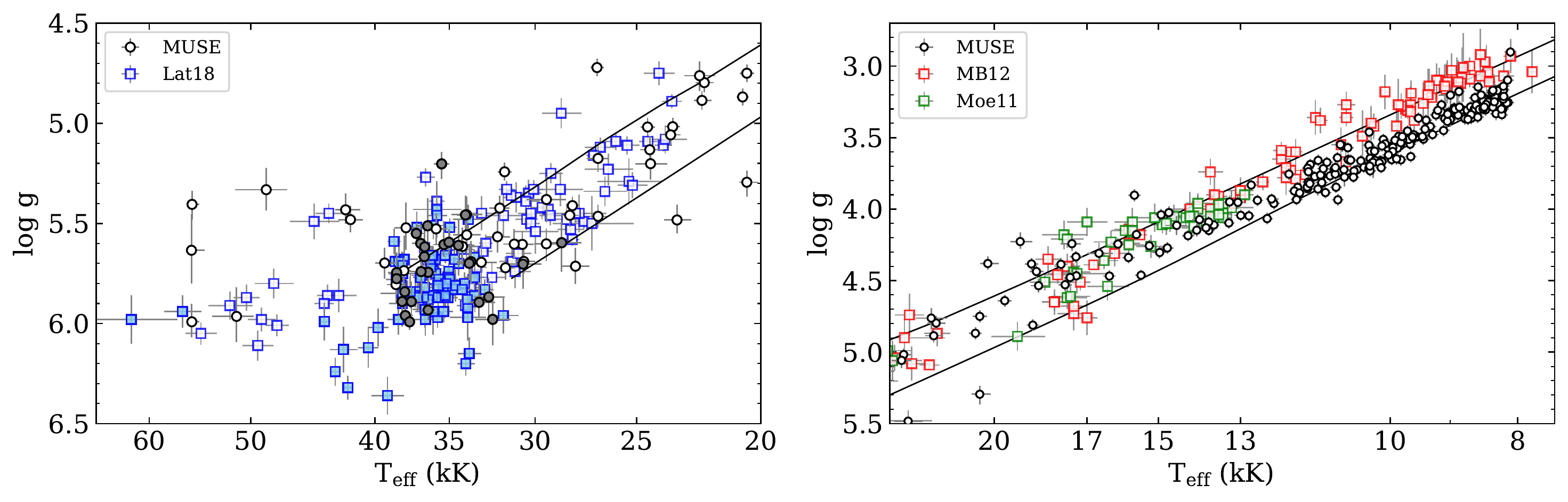}
   }
     \caption{
     Kiel diagrams showing the position of the stars in \Cen\ for our MUSE sample (black circles) and for samples taken from the literature. Left panel for the EHB stars (blue squares, \citealt{latour2018}) and right panel for the BHB stars (red squares, \citealt{monibidin2012}, green squares \citealt{moehler2011}). Filled symbols on the left panel indicate the He-rich, blue hook stars.
     The theoretical HB band for the BaSTI He-normal models is also shown.}
     \label{pic:kiel_5139_litt}
\end{figure*}

\begin{figure}
\resizebox{\hsize}{!}{
   \includegraphics{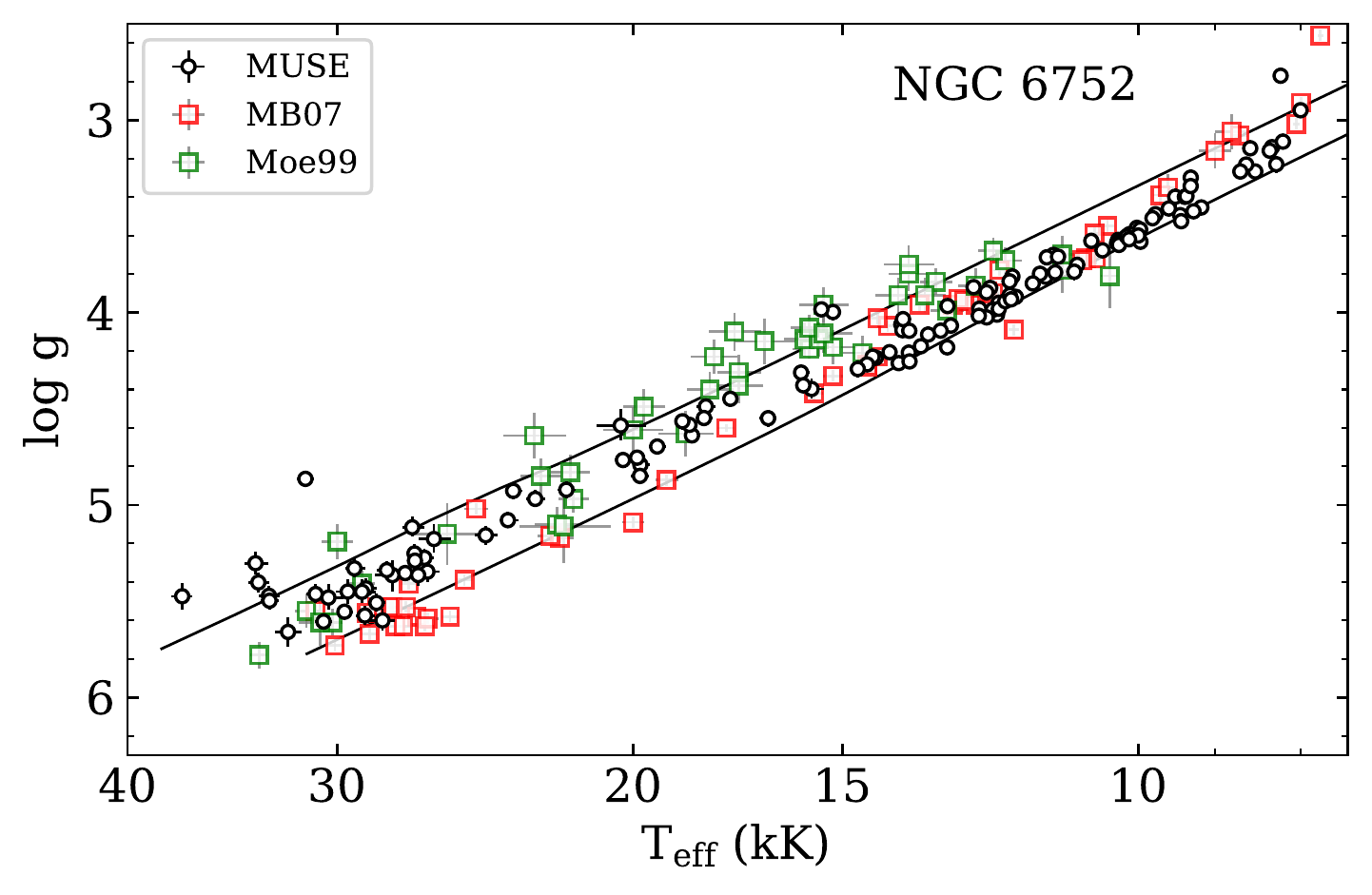}
   }
     \caption{Kiel diagram showing the position of the stars in \ngc\ for our MUSE sample (black circle) and for samples taken from the literature (\citealt{monibidin2007}, red squares and \citealt{1999A&A...346L...1M}, green squares).
     }
     \label{pic:kiel_6752_litt}
\end{figure}

\subsection{Comparison with literature results} 
\label{sec:comp_litt}
In terms of the stellar parameters derived from optical spectra, \citet{monibidin2012} found that the surface gravity of the HB stars colder than $\sim$18~kK in \Cen\ was systematically lower than the canonical ZAHB and lower than the log $g$ of their counterparts in three other clusters studied by the same group. This is shown in the right panel of Fig.~\ref{pic:kiel_5139_litt} where the \citet{monibidin2012} stars lie on the TAHB while our MUSE stars are on the ZAHB. 
Interestingly, the B-BHB stars with \teff$\gtrsim$13~kK lie above the ZAHB in our analysis as well as in the literature samples \citep{monibidin2012,moehler2011}. This makes us wonder whether it could be a real feature caused by stratification in the atmosphere \citep{leblanc2010} given that the literature analyses were also performed with chemically homogeneous model atmospheres. 
For the hot stars in \Cen\ we compare our results with the EHB sample of \citet{latour2018}. We notice that the stars scatter similarly across the EHB in both studies. The He-rich, blue hook, stars are more tightly grouped at the end of the EHB, at \teff\ close to 35~kK, in the \citet{latour2018} sample than in our MUSE sample. Taking into consideration that the typical $S/N$ of our blue hook spectra (see, e.g., star 210300 in Fig.~\ref{pic:spectral_fits}) is rather low, we find our results to be very reasonable. The sample of \citet{latour2018} also includes some very He-enriched stars at higher temperatures and surface gravities than the bulk of the blue hook objects. Our sample does not include such objects. This does not mean that they are not present in the core of \Cen, but is most likely due to the fact that we are not sampling the faintest part of the blue hook (see Fig.~\ref{pic:CMDs}) due to the $S/N$ limit. 

Figure~\ref{pic:kiel_6752_litt} shows an equivalent comparison as Fig.~\ref{pic:kiel_5139_litt} but for \ngc, including the two literature studies available for this cluster \citep{1999A&A...346L...1M,monibidin2007}. This time, the sample of \citet{monibidin2007} closely follows the ZAHB on almost the full \teff\ range. The slightly lower log~$g$ for stars with \teff $\gtrsim$~15~kK discussed in Sect.~\ref{subsubsec:tef_logg} is similarly seen in the sample of \citet{1999A&A...346L...1M}. 
When looking at the comparisons in Figs.~\ref{pic:kiel_5139_litt} and \ref{pic:kiel_6752_litt}, we understand why Moni~Bidin et al. were puzzled by their results in \Cen. 
We believe that there were some issues with their analysis of the (cold) stars in \Cen, but it is unclear what causes the difference.

\section{Variable stars and hot blue stragglers in NGC\,6752}\label{subsec:variable}

We looked for variable HB stars that could have been observed by MUSE in our two clusters. We found only two such variable stars, both in \ngc. The two stars are among the HB variables identified by \citet{momany2020} who refer to them as vEHB-1 and vEHB-3. However, these two stars were already known as V16 and V17 in the variable star catalog of \citet{kaluzny2009}\footnote{In the following, we adopt the nomenclature of \citet{kaluzny2009}}. According to the literature, the periods of V16 and V17 are $\sim$ 19.5 and 3.2 days, respectively. We indicated the position of the two variables with green star markers in the CMD (Fig.~\ref{pic:CMDs}) and atmospheric parameters diagrams (Figs.~\ref{pic:Kiel_dia}-\ref{pic:HE_plot}). 
V16 is located directly on the EHB in the CMD and has \teff\ = 23.6~kK and log $g$= 4.9. Its spectrum and best fit are shown in Fig.~\ref{pic:spectral_fits2}. With these parameters, the star blends in with the other EHB stars in the log $g-$\teff\ and He$-$\teff\ planes. Its spectrum does not show any conspicuous features compared to that of other stars with similar parameters (see also Fig.~\ref{pic:spectral_fits2}).  

\begin{figure}
\resizebox{\hsize}{!}{
   \includegraphics[width=12cm]{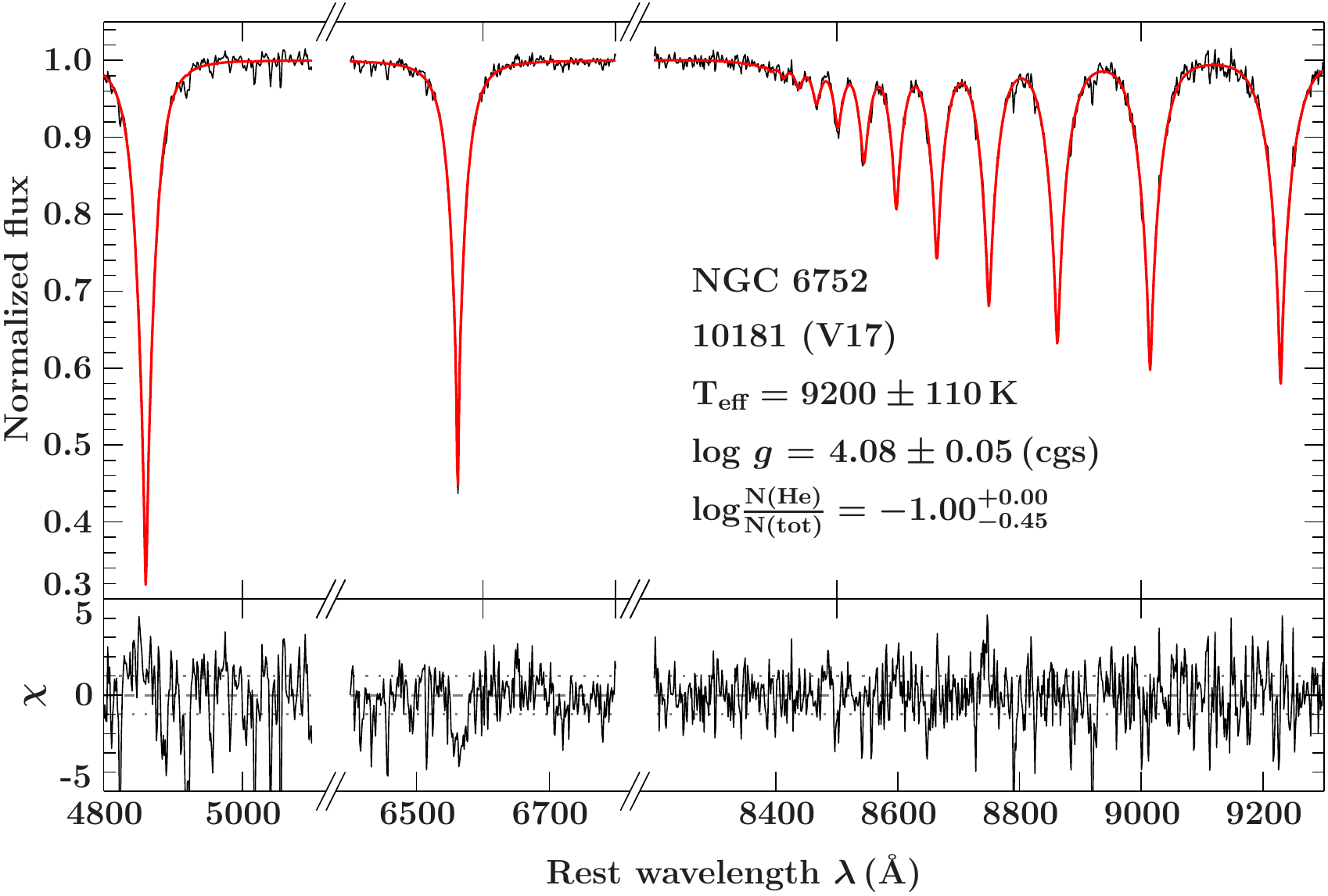}
   }
     \caption{Best fit of the spectrum of the variable BSS V17 in NGC\,6752. The top panel shows the best fit (red) to the normalized spectrum (black). The lower panel displays the residuals. Only the regions with spectral lines of hydrogen are shown.}
     \label{pic:fit_v17}
\end{figure}

V17 sits among a conspicuous small group of six objects located in between the blue straggler region and the HB (see Fig. \ref{pic:CMDs}). The atmospheric parameters obtained for V17 (\teff\ = 9200 K, log $g$ = 4.08) confirm that the star does not belong to the HB; it lies below the ZAHB (see Fig.~\ref{pic:Kiel_dia}). Our best fit of V17 is shown in Fig.~\ref{pic:fit_v17}.
This star is also among the sample of BSSs studied with high-resolution spectroscopy by \citet{lovisi2013}. The authors derived \teff = 9016~K and log $g$ = 4.1 from comparison with isochrones\footnote{They identify it as BSS10.}. In their sample, they have two additional BSSs located in the same region of the CMD as V17. They found that these three objects, being the three hottest ones of their sample, have iron abundances larger than the cluster value of [Fe/H]=$-$1.5. The authors interpreted this in terms of radiative levitation affecting the chemistry of the hot BSSs. From the BSSs studied in a few other GGCs by the authors, the onset of radiative levitation, indicated by an increase in Fe abundance but also a decrease in oxygen, starts at \teff\ $\sim$7800~K \citep{lovisi2013}. Among the small group of hot and bright BSSs in the CMD of NGC\,6752, four additional stars were observed by MUSE (triangles in  Fig.~\ref{pic:CMDs}). We retrieved their spectra and fitted them just like the other stars of our sample. As seen in Fig.~\ref{pic:Kiel_dia}, all of them lie below the HB and we obtained \teffs between 8400~K and 10\,600~K. Because they are affected by diffusion, we fitted the stars with the $Z=0.0$ model grid.

\citet{momany2020} argue that the variability they observed (typically over periods of 2$-$10 days and with $\Delta U$ $\sim$0.05-0.2 mag) in a subset of EHB stars in three GCs, including V16 and V17, is caused by magnetic spots present at the surface of the stars. The (weak) magnetic fields would be generated by the \ion{He}{ii} convective zone as it reaches the surface in the atmosphere of EHB stars with \teff\ close to that of the M-jump. While V16 is a genuine EHB star with \teff=23~kK, V17 is a different type of object. It is a hot BSS that is significantly colder than the other EHB variables. It is not clear whether the same mechanism can also produce such variability in a BSS. We note that while some BSSs are known to be in binary systems \citep{giesers2019}, V17 does not show RV variations and its variability cannot be explained by the presence of a close companion \citep{momany2020}.

\section{Mass, radius, and luminosity}\label{subsec:mrl}

\begin{figure*}
\resizebox{\hsize}{!}{
  \includegraphics{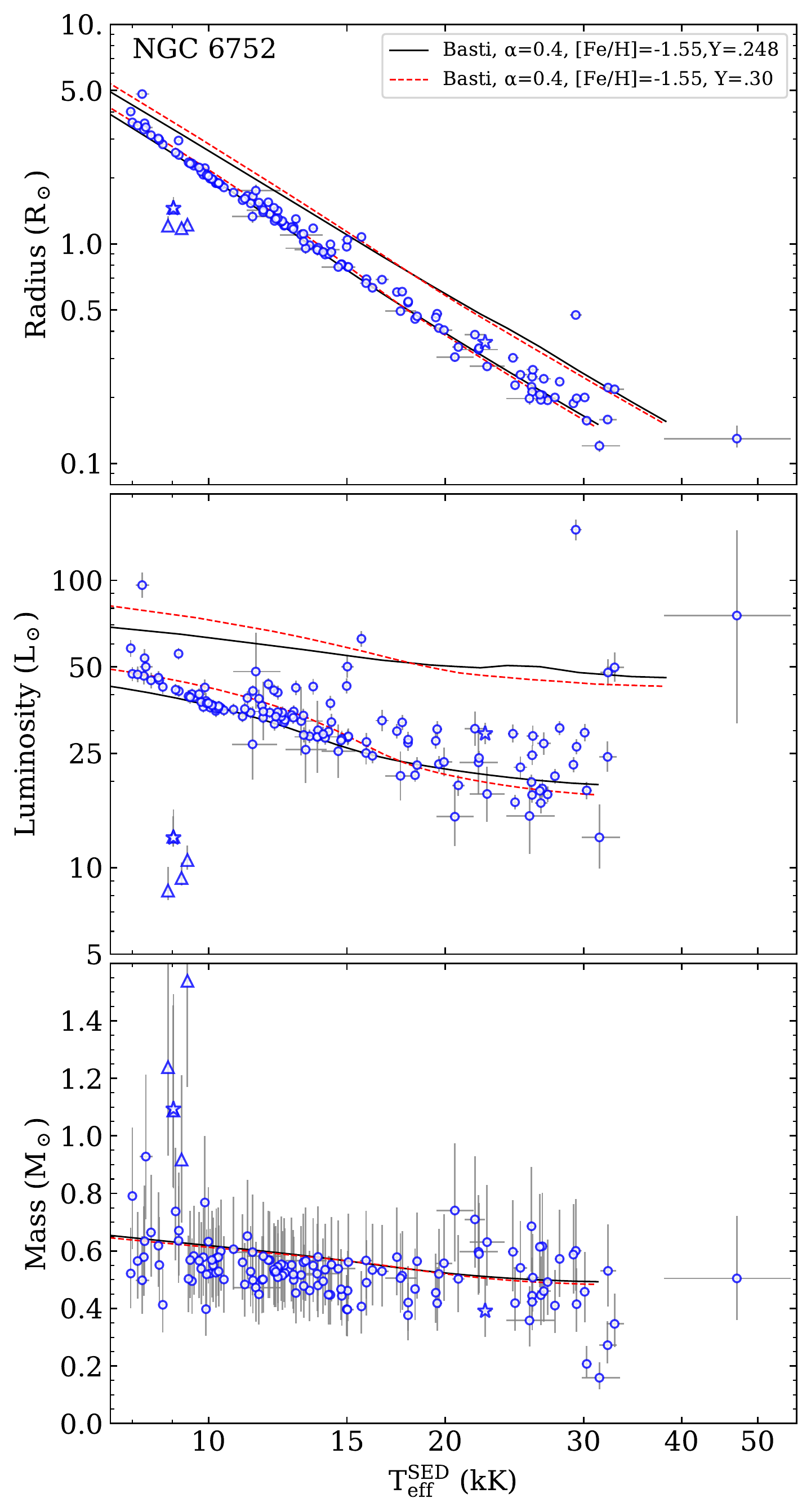}
  \includegraphics{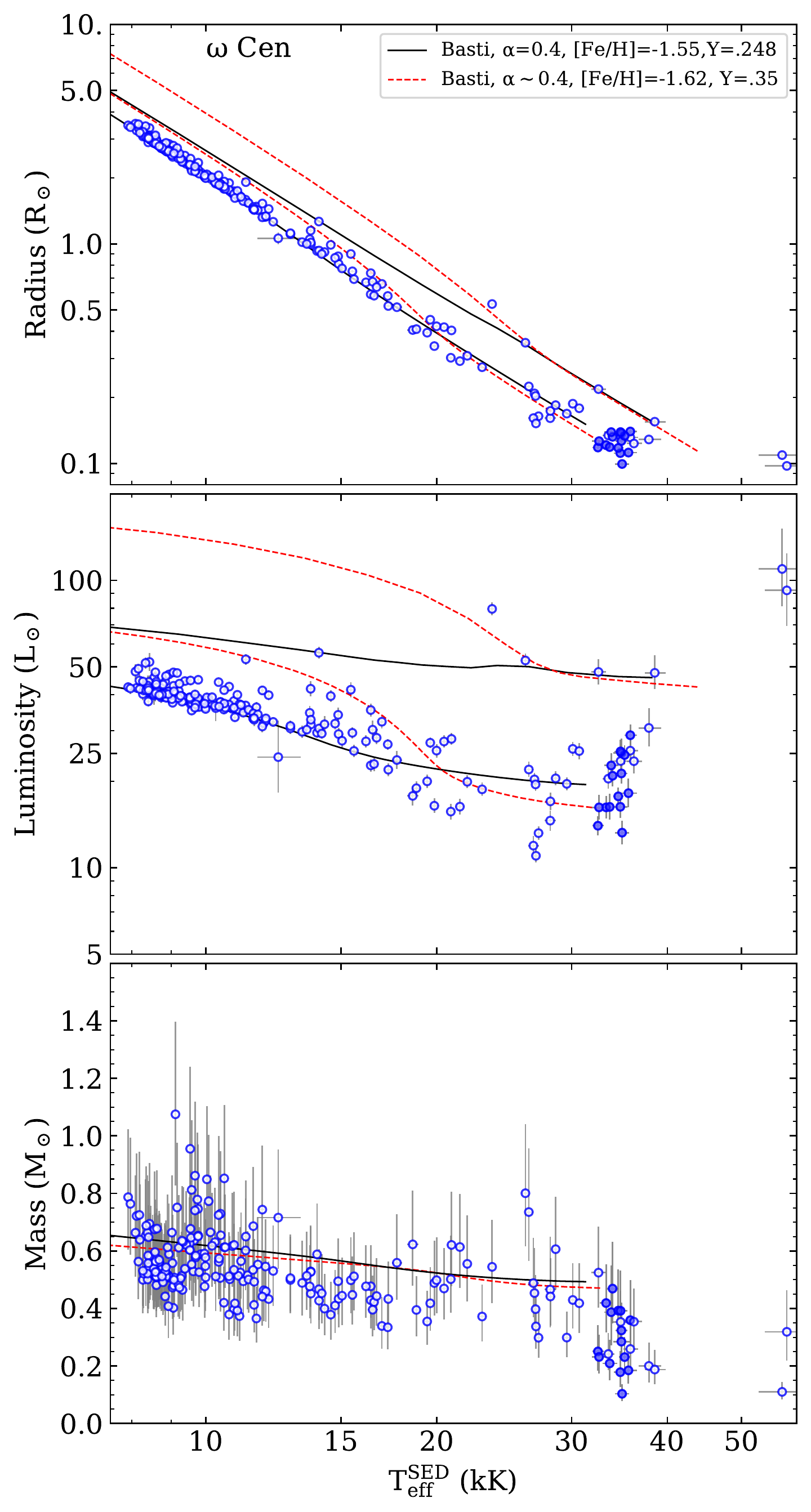}
   }
     \caption{Radius, luminosity, and mass versus \teff\ for the stars in \ngc\ (left panels) and \Cen\ (right panels). \teff\ here is obtained from the SED fit. The two variables (asterisks) and three BSSs (triangles) in \ngc\ are marked. The filled symbols in \Cen\ indicate the He-rich (blue hook) stars. We show the theoretical ZAHB and TAHB for He-normal (solid) and He-enriched (dashed) models.  }
     \label{pic:mass_radius}
\end{figure*}

The mass of an HB star is tightly correlated with its effective temperature. The mass of the hottest EHB stars, as well as the blue hook objects, is essentially that of their He-core because their very light hydrogen envelope (M$\lesssim$0.01~\msun) has a negligible contribution. 
There is only a small range of possible masses for the He-core ($\sim$0.45$-$0.50~\msun) dictated by the mass required for the He-flash \citep{dorman93}. Thus, the HB forms a sequence of increasing stellar mass as the hydrogen envelope becomes "thicker" with decreasing effective temperature.

For EHB stars in the Galactic field, mass determinations using various methods, including asteroseismology and eclipsing binaries, are in line with theoretical expectations (see, e.g. \citealt{fontaine12}, \citealt{schaff2022}, \citealt{Schneider2022}). 
The situation is different in globular clusters, especially for the two clusters involved in our study. In \ngc, \citet{monibidin2007} reported groups of stars with anomalously low or high masses along the HB, with the peculiarity that all BHB stars colder than 10 kK had too low masses. A similar issue was also reported in \Cen; \citet{monibidin2011} derived masses lower than the canonical values for the BHB and EHB stars. Similarly, low masses were reported by \citet{moehler2011} for stars colder than 20 kK and by \citet{latour2018} for the EHB and blue hook stars. A mass distribution for BHB or EHB stars peaking around 0.35~\msun\ is difficult to reconcile with evolutionary models. Although various methods were used to derive stellar masses in these previous studies, the results were never fully consistent with theoretical predictions. 
Here, we make an attempt at reconciling spectroscopic masses with evolutionary prescriptions by combining the latest state-of-the-art data and methods, as described in Sect.~\ref{subsec:sed_fit}, to ultimately derive masses.

In Fig.~\ref{pic:mass_radius} we show the radii, luminosities, and masses of the stars in both clusters as a function of the effective temperatures obtained from the photometric fits (\teffsed). The results are available in the online Tables B.3 and B.4 (see also Appendix B).
The theoretical predictions from the ZAHB and TAHB models with normal and enhanced helium abundances are indicated as well. 

In both clusters, the stars follow very well the theoretical \teff$-$radius relation of the ZAHB. The agreement between our derived luminosity and the predictions is also very good in both clusters for stars up to $\sim$18~kK. Beyond that temperature, the scatter increases for the stars in \ngc, but the position of the stars remains consistent with the theoretical HB. In \Cen, there is a significant fraction of "underluminous" stars in this hot regime. However, we note that the underluminous objects are also the stars having radii smaller than the theoretical ZAHB prediction. 
In \ngc, the stellar masses follow the expected decreasing trend with \teff, but are systematically lower (by $\sim$0.05~\msun) than the theoretical prediction up to $\sim$18~kK where the masses are then in better agreement with the models. Typical uncertainties on the masses are of $\pm$0.15~\msun, thus considerably larger than the systematic difference observed.
The behavior is similar in \Cen, but with a large scatter in the masses of the cool stars compared to \ngc. This is related to the larger scatter also seen in log~$g$ in the Kiel diagram.
As for the hottest stars (\teff\ $\gtrsim$ 30~kK), they have masses that are too low. This is especially obvious in the case of the He-rich (blue hook) stars in \Cen.

 We also fitted the SEDs of the stars with \teff\ fixed to the spectroscopic values (see Fig.~\ref{pic:app_rlm}). This has little effect on the results for the cooler stars where \teffsed\ and \teff$^{\rm spectro}$ are in good agreement. For the hotter stars, it does not lead to a better agreement between observations and theoretical models. Especially, the masses remain lower than expected, and that for the whole temperature range. We discuss these results in more detail in Appendix A.

For both clusters, in addition to the HB tracks for He-normal ($Y=0.248$) models, we also show the tracks for He-enriched models. While the He-enriched population in \Cen\ is believed to have a helium content as high as $Y=0.35-0.4$ \citep{Norris2004,king2012}, that of \ngc\ would have, at most, $Y=0.3$ \citep{milone2018,2021A&A...650A.162M}. In the latter case, the small difference in helium content does not significantly affect the position of the HB tracks. 
However, the models with $Y=0.35$ (shown for \Cen) predict larger and more luminous stars than what we derived for the stars with \teff\ below $\sim$20~kK. At higher \teff, the He-normal and He-enriched tracks predict similar properties.
As concluded from the Kiel diagram, we can exclude that the cold HB stars in \Cen\ originate from a He-rich subpopulation with $Y\gtrsim0.35$. This is consistent with the conclusions from the population synthesis analysis of \citet{tailo2016}, who populated the colder part of the HB with helium-normal stars. However, they populate everything hotter than $\sim$13~kK with stars having increasingly more helium, from $Y=0.30$ to $Y=0.37$ and the blue hook with model stars having $Y=0.37$. Unfortunately, it is in the hottest part of the HB that the He-enriched models show the smallest difference from the helium normal models.

In the course of our SED fits in \Cen, we realized that two stars (id 240693, 142667) appear to be hotter than estimated from the spectroscopic fits (\teff\ between 50$-$55~kK). We fitted the photometry of these objects with a grid of DA white dwarf models \citep{reindl2016} and estimated \teff\ $\sim$ 110 kK and 75 kK, respectively for these two hot stars\footnote{They are outside the range plotted in Fig.~\ref{pic:mass_radius}}. With such high temperatures, these stars are possibly in a post-AGB phase. 

We also fitted the SEDs of the five BSSs in \ngc.
As expected, these stars are smaller and less luminous than the He-core burning HB. We note that the range of \teff\ obtained from the SED fits of the BSSs stars is restricted to 8800-9400 K. 
We find for these five objects masses between 0.9 and 1.5 \msun, which is within the expected values for BSSs in old GCs \citep{2005ApJ...632..894D}.

\begin{figure}
\resizebox{\hsize}{!}{
   \includegraphics{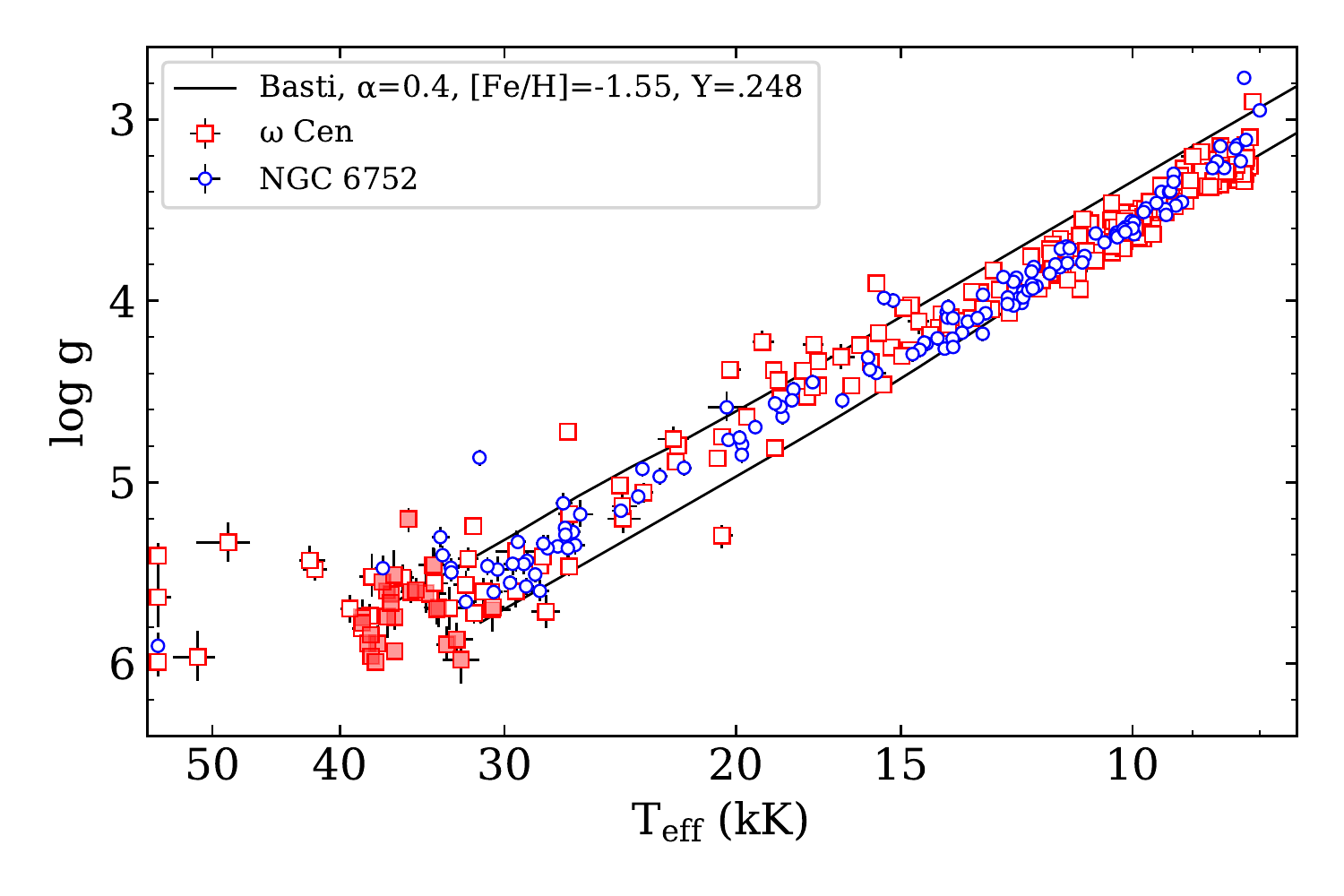}
   }
     \caption{Kiel diagram showing the position of the stars in \ngc\ (circle) and \Cen\ (square). The He-rich (blue hook) objects are indicated with filled symbols.}
     \label{pic:kiel_2clusters}
\end{figure}

\section{Comparison between both clusters}\label{sec:comp_2clusters}

 Although they have a similar HB morphology, \Cen\ and \ngc\ are fundamentally different as globular clusters. The former is the most complex cluster in the Milky Way and is believed to be the nuclear star cluster of a dwarf galaxy accreted by the Milky Way or the result of the merger of two or more clusters (see, e.g., \citealt{bekki2003,ibata2019,pfeffer2021}). Its stars have a metallicity spread of more than one order of magnitude ($ -2.2 \lesssim [Fe/H] \lesssim -0.6 $, \citealt{johnson10}) and a significant spread in helium abundance ($\delta Y \approx0.1-0.15$) as well \citep{Norris2004,king2012}. 
 On the other hand, \ngc\ is a relatively simple globular cluster, essentially mono-metallic \citep{carretta2009} with a modest spread in helium abundance of $\delta Y\lesssim$\,0.04 \citep{milone2018} and showing a Na-O anticorrelation typical of Milky Way GCs \citep{carretta09_nao}. 

\begin{figure}
\resizebox{\hsize}{!}{
   \includegraphics{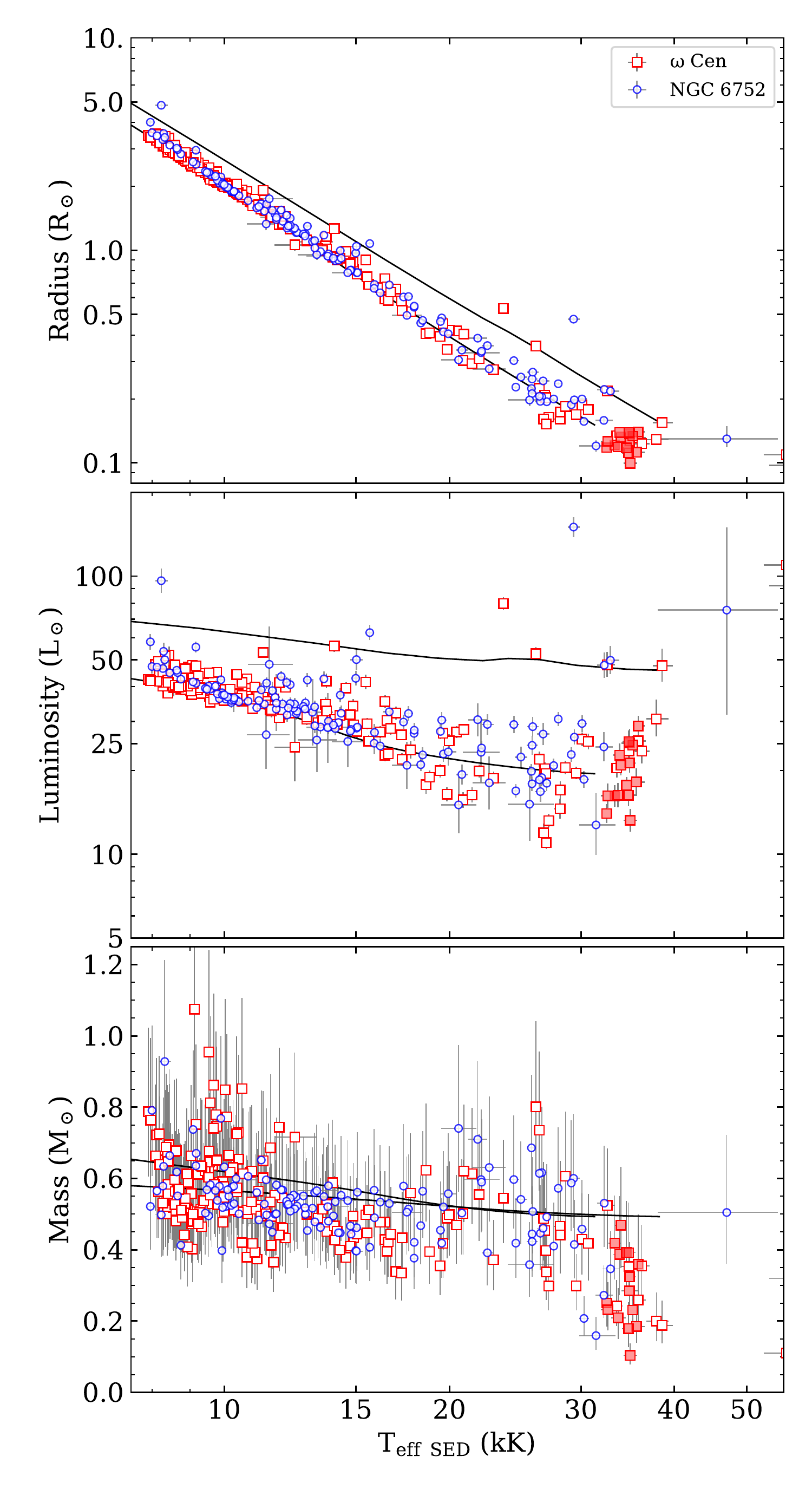}
   }
     \caption{Same as Fig.~\ref{pic:mass_radius} but comparing the results for \ngc\ (circle) and \Cen\ (square).}
     \label{pic:RLM_2clusters}
\end{figure}

 Given these fundamental differences between the two GCs, we thought it worthwhile to compare the properties of their HB stars in the Kiel diagram (Fig.~\ref{pic:kiel_2clusters}) and in terms of radius, luminosity, and mass (Fig.~\ref{pic:RLM_2clusters}). 
 We do not see any significant difference in the position of the stars in the Kiel diagram between NGC\,6752 and \Cen, besides the absence of He-rich (blue-hook) stars in NGC\,6752.  
 This is different from the conclusion of \citet{monibidin2011,monibidin2012} who found differences in log~$g$ and mass between the HB stars of \Cen\ and those of three other clusters (namely NGC\,6752, M80, and NGC\,5986). 
\citet{prabhu2022} also reported differences in magnitudes (in the UV-optical CMD), radius, and luminosity between the HB stars in \Cen\ and those in M13. They found the hot HB stars in \Cen\ to be fainter than model predictions by $\sim$0.5 mag in the FUV. According to their study, these same stars also appear to have smaller radii and lower luminosity than their counterparts in M13.  
The comparison between the radius, luminosity, and mass (see Fig.~\ref{pic:RLM_2clusters}) obtained for the stars in our two samples does not suggest any fundamental difference in these parameters between the HB stars in \Cen\ and \ngc.
We mentioned in Sect.~\ref{subsec:atm_par} that the cold stars in \Cen\ scatter more than those of \ngc\ in the Kiel diagram. This is also the case in the radius, luminosity, and mass plots. As mentioned previously, this behavior might have to do with the metallicity spread in \Cen.

 The remaining major difference between the HB morphology of both clusters is the He-rich, blue hook stars that are present in \Cen\ but absent in \ngc. Only a few Galactic GCs have a sizable population of blue hook stars. Their presence appears to be related to the cluster's mass, meaning that blue hook populations are only found in massive clusters \citep{dieball2009,2010ApJ...718.1332B,johnson2017}. However, the presence of a stellar population with large He-enhancement ($\delta Y$ $\gtrsim$0.09, \citealt{milone2018}) also seems to favor the formation of blue hook stars\footnote{M54, the nuclear star cluster of the Sagittarius Dwarf Galaxy, is somewhat an exception as it does not show a particularly high He-enhancement \citep{milone2018} but has a large population of blue hook stars \citep{brown16}.}. The recent characterization of NGC~6402 supports this idea \citep{dantona2022}. As for NGC~6752, it is not massive enough and does not have a sufficiently He-enriched population, to produce blue hook stars.

\section{Summary}\label{sec:summary} 
 
We analyzed the MUSE spectra of more than 400 HB stars hotter than 8\,000~K found in the central regions of the GCs \Cen\ and \ngc. The MUSE spectra cover the 4750$-$9350 \AA\ spectral range and include H$_\alpha$, H$_\beta$, the Paschen series, \ion{He}{ii} $\lambda$5412, and a handful of \ion{He}{i} lines depending on the spectral type. We fitted these spectral features with dedicated grids of hybrid LTE/NLTE model atmospheres in order to derive \teff, log~$g$, and helium abundance. We also used our model atmospheres to fit the HST photometry, in up to five (for \ngc) and eleven (for \Cen) filters, of the stars in our samples. We used the SED of the stars colder than 13~kK to estimate the average reddening of the clusters. We obtained values in perfect agreement with the literature. From the SED fits of the stars in our sample, we derive radii, luminosities, and masses by making use of the known distances of the clusters. 

When plotted in the \loggt\ diagram, the position of the stars colder than $\sim$15~kK, in both clusters, are in excellent agreement with theoretical $\alpha$-enhanced BaSTI ZAHB models having normal helium abundance ($Y$=0.247) and a metallicity representative of the mean metallicity of the clusters (i.e. [Fe/H] = -1.55). In \Cen, the position of these colder stars (\teff\ $\lesssim$ 15~kK) in the Kiel diagram excludes the possibility that they come from the He-rich ($Y$=0.35) subpopulation of the cluster. Their luminosities and radii 
support that conclusion as well. A milder He-enrichment (e.g. $Y$=0.3) does not produce measurable differences in terms of log~$g$ and radius, and only a slightly higher luminosity in stars colder than 15~kK. Thus we cannot exclude the possibility that some of the BHB stars have a modest He-enrichment.

We detected the onset of atmospheric diffusion that separates the A-BHB from the B-BHB stars 
using three different indicators: via the \teff\ of the stars at the position of the G-jump in the color-color plane (Fig.~\ref{pic:Cindex}), via the sharp drop in helium abundance measured from the MUSE spectra (Fig.~\ref{pic:HE_plot}), and via the increase in atmospheric metallicity measured from the photometric fits of the stars in \Cen\ (Fig.~\ref{pic:sed_Z}). All of these diagnostics indicate that the transition happens between 11~kK and 11.5~kK in both clusters.

We estimated the effective temperatures of the stars also as part of the SED fits. The spectroscopic and photometric \teff\ are in good agreement for stars colder than 15~kK (see Fig.~\ref{pic:delta_teff}). For the stars hotter than 15~kK, the spectral fits generally return a higher temperature than the photometric fits. This discrepancy between \teff$^{\rm spectro}$ and \teff$^{\rm SED}$ also affects the reddening that is estimated from the SED when fixing \teff\ to its spectroscopic value: the hot stars require a larger reddening in order to reproduce the photometric measurements (see Fig.~\ref{pic:reddening}). This is because lowering \teff\ or increasing E(B-V) has a similar effect on the shape of the SED.
For now, it remains unclear where this discrepancy in \teff\ (and reddening) for the hot stars comes from and if it can be solved. 
Concerning the masses of the HB stars, we found them to be systematically lower than theoretical expectations by about 0.05~\msun\ for the HB stars colder than $\sim$18~kK. However, this difference is well within the typical uncertainty of $\pm$0.15~\msun\ on the individual masses. As for the masses of the EHB stars (\teff\ $>$20~kK), they are significantly influenced by the adopted temperature, either \teffsed\ or \teff$^{\rm spectro}$. The former leads to a better agreement with the theory. However, in both cases, the blue hook stars in \Cen\ have masses significantly lower than expected.

We analyzed two of the periodic variables in \ngc, namely V16 and V17 \citep{kaluzny2009}. These stars were also presented among the sample of 22 periodic EHB stars discovered in \ngc, \Cen, and NGC\,2808 by \citet{momany2020}. 
We showed that while V16 is a genuine EHB star with \teff=23~kK, V17 is a hot blue straggler star with a \teff\ $\sim$9000~K, thus significantly colder than the other EHB variables discovered by \citet{momany2020}. It is not clear whether the mechanism invoked by \citet{momany2020}, magnetic fields produced by the presence of surface/sub-surface convective layers, can also produce such variability in a BSS. The hot BSSs are clearly too cold for the \ion{He}{ii} convection zone to reach the surface but could have instead the \ion{He}{i} convection zone close to their surface. If a similar mechanism is indeed responsible for the periodic variations in EHB and in V17, other hot BSSs in GCs might show similar luminosity variations.

\section{Conclusion and outlook}\label{sec:conclusion}

In conclusion, we carried out a pioneering spectroscopic and photometric investigation of the population of HB stars in the core of \ngc\ and \Cen. Thanks to MUSE, we obtained a first glimpse into the BHB and EHB stars in the center of these two clusters. We found a rich variety of spectral types, resembling closely those studied in the outskirts of the clusters. Our spectroscopic analysis demonstrates
that the MUSE spectra, in spite of their "red'' wavelength coverage, are well-suited for the study of blue HB stars in GCs. These spectra, combined with the HST photometry of the stars, allowed us to derive the usual spectroscopic parameters (\teff, log~$g$, helium abundance) but also stellar parameters (radius, luminosity, and mass) that were compared with theoretical evolutionary models. The analysis method provided us with precise measurements following well the theoretical predictions for radii, luminosities, and position in the Kiel diagram for stars with \teff\ $\lesssim$ 15~kK. Although for hotter stars some discrepancies between theoretical expectations and observations arise, our results are comparable to those of previous studies.

The numerous observations taken as part of the MUSE globular cluster Survey provided an unprecedentedly large number of homogeneous spectra of HB stars, not only in \Cen\ and \ngc, but also in other GCs with a blue HB such as NGC\,2808, NGC\,1851, NGC\,5904, NGC\,6656, NGC\,6093, and NGC\,7078.
This opens up the avenue to future detailed studies of several other GCs. The only limitation is the faintness of the EHB stars, but the A-BHB and B-BHB stars are bright enough to have good $S/N$ spectra. Because the MUSE observations are targeting the core regions of GCs, the stars observed also have HST photometry in the five filters included in the HUGS survey \citep{nardiello2018,piotto2015}, thus providing a reliable dataset to perform SED fits. We
also want to include additional FUV and NUV magnitudes from
UVIT/AstroSat \citep{Sahu2022} and STIS/HST (e.g. for NGC 2808, \citealt{brown2001}). This will hopefully provide further constraints for the photometric fits, especially for the hot EHB stars.
In future papers, we want to analyze the spectra of HB stars in the GCs listed above, but also from additional MUSE observations of \Cen\ \citep{nitschai2023}. These new observations fill most of the spatial gap between the data presented here and the surveys from the literature (see Fig.\ref{pic:positions}). They include a few stars in common with the previous studies, allowing us to directly compare the spectroscopic parameters derived from MUSE and other spectra with a bluer spectral range.

In the future, Blue-MUSE \citep{2019arXiv190601657R}, with a planned spectral coverage of 3500-6000 \AA, a resolution R$\sim$4000 and 2 arcmin$^2$ field of view, will be a perfect instrument to study the HB and especially EHB stars in GCs. The wavelength range and resolution will provide spectra similar to those of the FORS2 instrument, which has been used for most literature studies of EHB stars in GCs, but with all the advantages of an IFU.

\begin{acknowledgements}
We are most grateful to Andreas Irrgang for the development of the spectrum and SED- fitting tools, his contributions to the model atmosphere grids and his advice to use the isis scripts.
 We also thank Annalisa Calamida for providing us information about the calibration of HST filters.
 We acknowledge funding from the Deutsche Forschungsgemeinschaft (grant LA 4383/4-1 and DR 281/35-1) and from the German Ministry for Education and Science (BMBF Verbundforschung) through grants 05A14MGA, 05A17MGA, 05A14BAC, 05A17BAA, and 05A20MGA. 
 SK gratefully acknowledges funding from UKRI in the form of a Future Leaders Fellowship (grant no. MR/T022868/1). JB acknowledges financial support from the Fundação para a Ciência e a
Tecnologia (FCT) through research grants UIDB/04434/2020 and UIDP/04434/2020, national funds PTDC/FIS-AST/4862/2020 and work contract 2020.03379.CEECIND.
 This research has made use of NASA’s Astrophysics Data System Bibliographic Services and the \textsc{python} packages pandas \citep{reback2020pandas,mckinney-proc-scipy-2010} and \textsc{matplotlib} \citep{Hunter:2007}.

\end{acknowledgements}

\bibliographystyle{aa} 

\clearpage
\newpage
\begin{appendix} 

\section{Additional material on SED fits}

\subsection{Reddening estimate}\label{subsec:reddening}

\begin{figure}
\resizebox{\hsize}{!}{
   \includegraphics{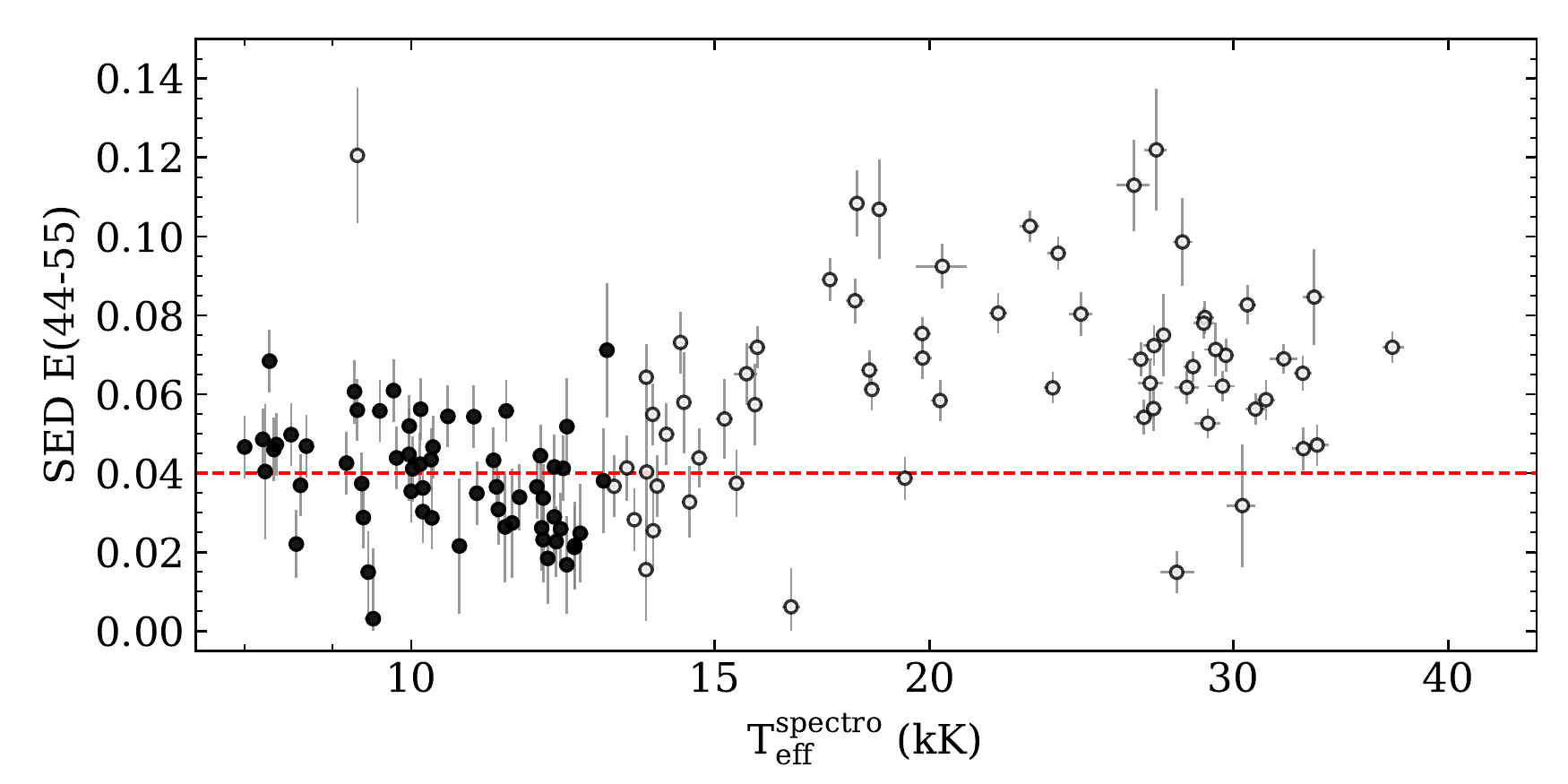}
   }
     \caption{Reddening versus effective temperature for the stars in \ngc. Stars with \teff\ below 13~kK (filled symbols) were used to derive the average E(44-55) (dashed line). }
     \label{pic:reddening}
\end{figure}

We first attempted a simultaneous fit of \teff\ and $\Theta$ in \ngc\ while keeping the reddening fixed to \ebmv\ = 0.046\,mag \citep{gratton2005} but we found a clear trend between \teff$^{\rm spec}$ and \teff$^{\rm SED}$ among the cold stars. This is because both the reddening and \teff\ affect the shape of the SED in a similar way for BHB stars. Thus, we fixed \teff\ to its spectroscopic value and left the reddening (and $\Theta$) as free parameters. We obtained results as seen in Fig.~\ref{pic:reddening}. In stars hotter than $\sim$15~kK, we notice a clear increase in the reddening value with \teff. This behavior is related to the discrepancy between \teff$^{\rm spec}$ and \teff$^{\rm SED}$ in the hot stars, which is discussed in the following subsection.
Because the temperatures obtained from the spectra and the SED fits are in good agreement for stars cooler than 13~kK, we use the reddening values obtained from these stars to compute the average reddening. We proceeded in the same way for \Cen.

The average reddening values obtained for both clusters, \emono\ $=0.04\pm0.01$\,mag for \ngc\ and \emono\ $=0.12\pm0.02$\,mag for \Cen, are in perfect agreement with the expected values from the literature (\citeauthor[2010 edition]{harris1996}).
We note here that, for \teff\,=\,10\,000\,K, the extinction conversion \ebmv/\emono\ = 0.976 \citep{2019ApJ...886..108F}.
For \Cen, we used the reddening map of \citet{bellini2017b} to take into account differential reddening and thus applied a correction to the reddening value of each star. However, these corrections are within $\pm$0.01\,mag, thus relatively small.

\subsection{Spectroscopic \teff\ versus photometric \teff}

\begin{figure}
\resizebox{\hsize}{!}{
   \includegraphics{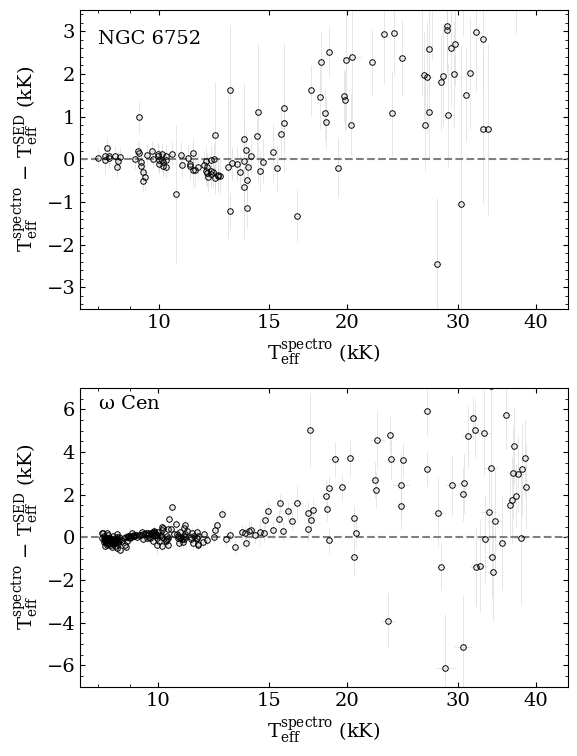}
   }
     \caption{Difference in effective temperature between the spectroscopic and photometric fits for \ngc\ (upper panel) and \Cen\ (lower panel).}
     \label{pic:delta_teff}
\end{figure}

With the reddening fixed, we then performed a second run of SED fits where \teff\ and $\Theta$ were the two free parameters as explained in Sect.~\ref{subsec:mrl}. Here we compared the effective temperatures determined from the spectral fits with those obtained from SED fits. The results are shown in Fig.~\ref{pic:delta_teff}. The temperatures derived from both methods are in good agreement up to $\sim$15 kK. Above this temperature, the spectroscopic \teff\ are generally hotter than the SED \teff\ by two to three thousand Kelvin. For stars hotter than $\sim$30~kK, neither method is expected to provide robust values. 
In the case of MUSE spectroscopy, the Paschen series disappears and few spectral lines are left to constrain the atmospheric parameters (see Figs.~\ref{pic:spectral_fits} and \ref{pic:spectral_fits2}). 
In the SED's case, the peak of the emitted flux moves to the far-UV in the hottest stars, and the flux slope in the optical range loses its sensitivity to temperature changes. We note that the known reddening and F225W magnitudes in \Cen\ are decisive for constraining \teff\ from photometry in the hot objects.

For stars between $\sim$15-30~kK, it is not clear which method provides the best \teff. 
In Sect.~\ref{subsec:ccp}, we saw that the spectroscopic \teff\ provide realistic temperatures for the M-jump, namely 18.6~kK and 19.5~kK in \ngc\ and \Cen\ respectively. 
If we do the same procedure using this time the photometric effective temperatures, we obtain 17.2~kK and 17.1~kK for \ngc\ and \Cen. Such temperature is lower than expected from evolutionary models, which put the M-jump at 20$-$18 kK. 
The temperatures at the G-jump remain unchanged.

\begin{figure*}
\resizebox{\hsize}{!}{
  \includegraphics{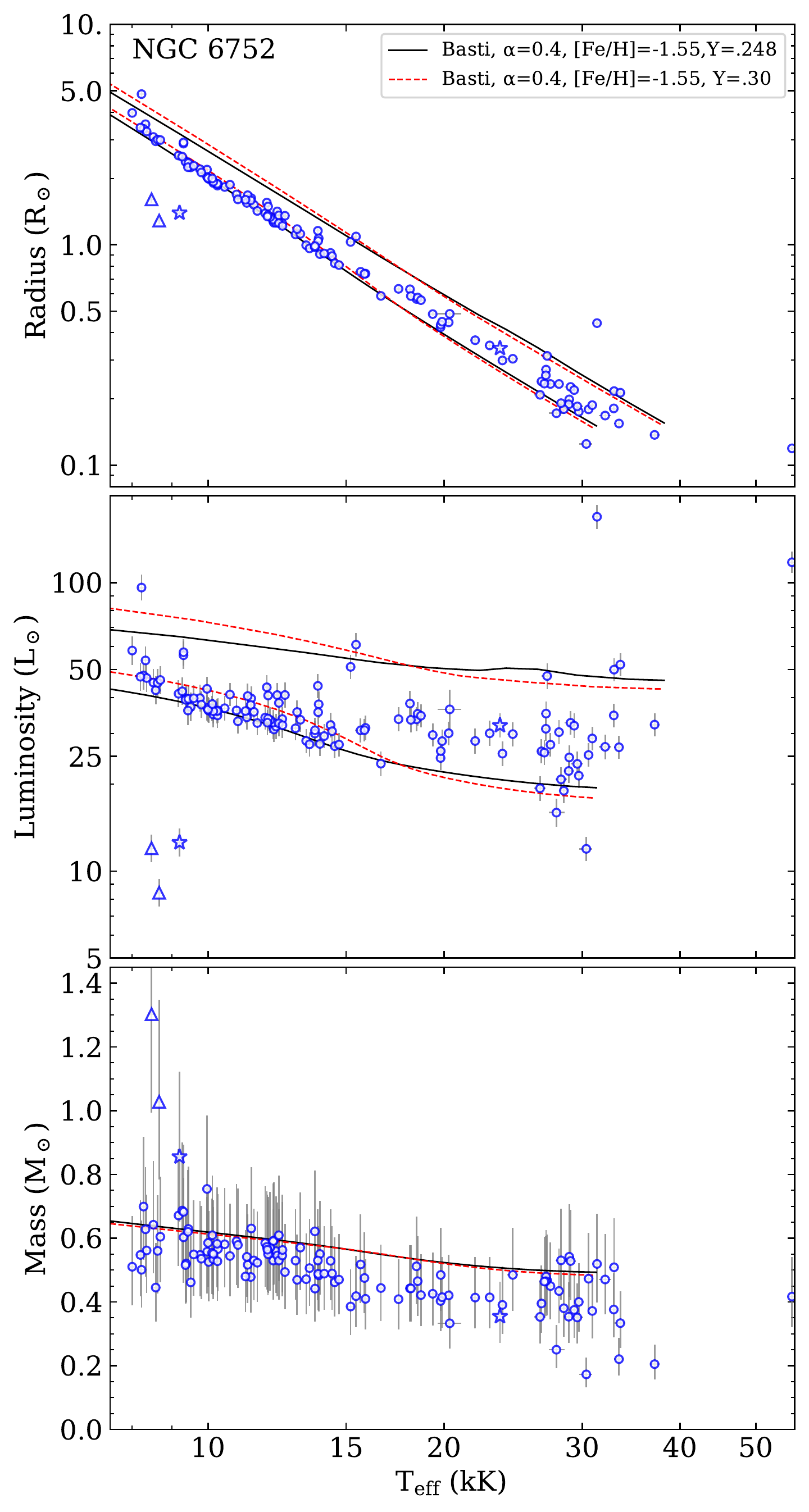}
\includegraphics{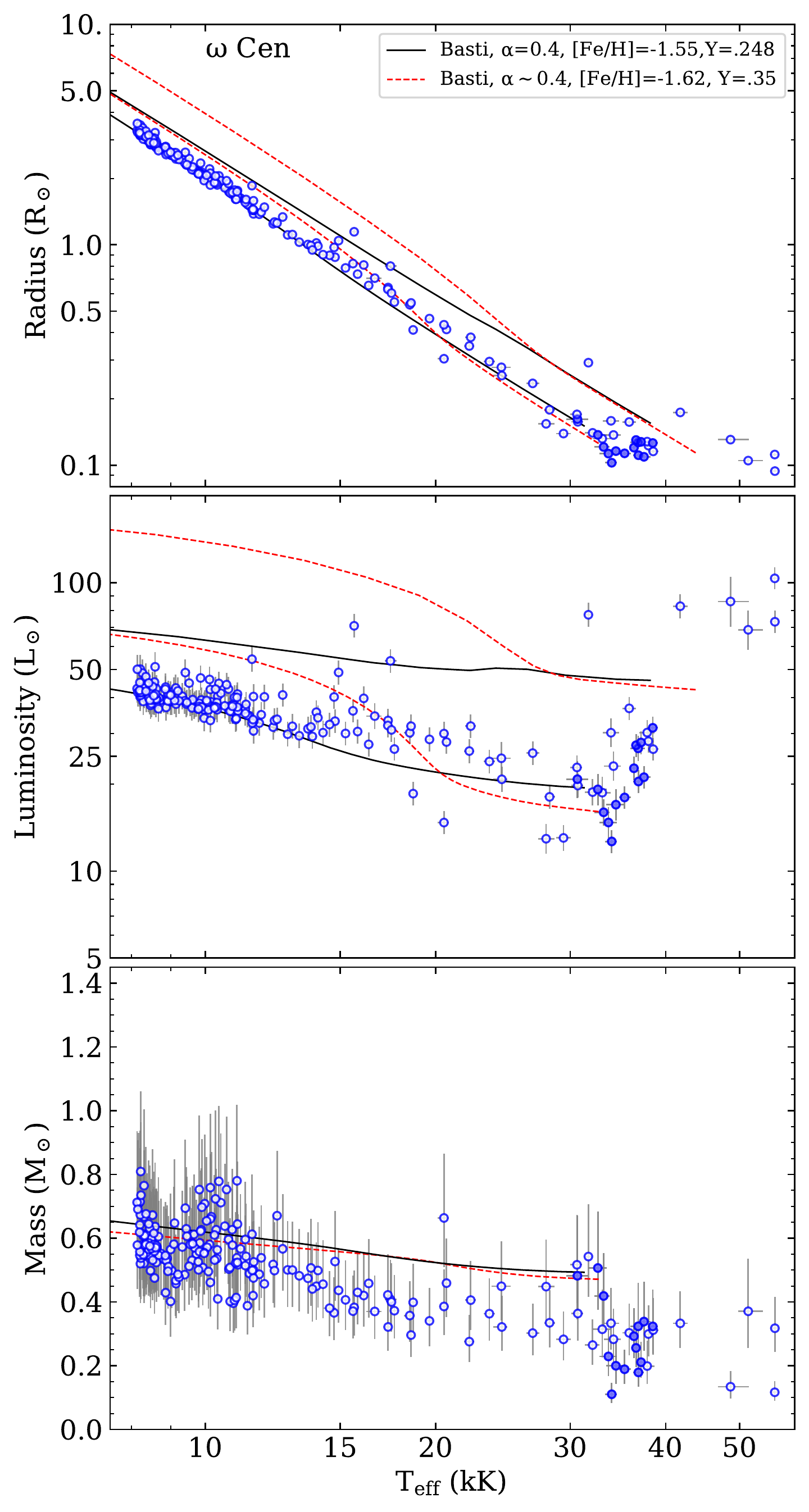}
   }
     \caption{Radius, luminosity, and mass versus \teff\ for the stars in \ngc\ (left panels) and \Cen\ (right panels). \teff\ is obtained from the spectral fits and fixed to this value in the SED fitting process. The filled symbols in \Cen\ indicate the He-rich (blue hook) stars. We show the theoretical ZAHB and TAHB for He-normal and He-enriched models.
     }
     \label{pic:app_rlm}
\end{figure*}

We also performed the SED fits for both clusters with \teff\ fixed to the spectroscopic value, thus leaving $\Theta$ as the only free parameter. The  resulting radii, luminosities, and masses are shown in Fig.~\ref{pic:app_rlm}. In \ngc, we see that the stars hotter than 15~kK start to deviate from the predicted tracks; they are larger, more luminous, and less massive than expected. In \Cen, the discrepancies in terms of radius and luminosity are less pronounced than for \ngc, but the mass discrepancy is stronger.

It remains unclear to us why the spectroscopic \teff\ of the hot stars (\teff\ $\gtrsim$ 15~kK) do not agree with that of their SED. At this point, we cannot say which method provides the most accurate \teff. However, for the cold stars, we have consistent results from both photometric and spectral fits, as well as a good agreement with theoretical prescriptions. This demonstrates that the SED fitting method is a powerful tool to analyze A-BHB and B-BHB stars.

\section{Description of the online tables}
The results of our spectroscopic analysis of the MUSE spectra are only available online at CDS as Table B.1 and B.2 for \ngc\ and \Cen, respectively. We include in these tables columns with star identification numbers, coordinates, atmospheric parameters derived, and their uncertainties as described in Sect.~\ref{subsec:spec_fit}, the number of individual spectra combined, and the $S/N$ of the resulting spectra. We also add a column with alternative star names from previous studies. 

The results of the SED fits are presented in four tables. The first two (Table B.3 and B.4, for \ngc\ and \Cen, respectively) present the results obtained when fitting simultaneously $\theta$ and \teff.  Along with the stars identification numbers, coordinates, and parameters obtained (\teff\, $\theta$, radius, luminosity, and mass), we also list the magnitudes collected from the three catalogs we used \citep{nardiello2018,bellini2017a,anderson10}. 
Finally, Tables B.5 and B.6 (for \ngc\ and \Cen, respectively) include the results of the SED fits when $\theta$ is the only free parameter and \teff\ is fixed to the spectroscopic value.

\end{appendix}
\end{document}